\documentclass[10pt,journal,compsoc]{IEEEtran}
\IEEEoverridecommandlockouts
\usepackage{ragged2e}
\usepackage{cite}
\usepackage{amsmath,amssymb,amsfonts}
\usepackage{algorithm}
\usepackage[noend]{algpseudocode}
\usepackage{graphicx}
\usepackage{textcomp}
\usepackage{xcolor}
\usepackage[pscoord]{eso-pic}
\usepackage{lipsum}
\def\BibTeX{{\rm B\kern-.05em{\sc i\kern-.025em b}\kern-.08em T\kern-.1667em\lower.7ex\hbox{E}\kern-.125emX}}
\usepackage{amsmath,amssymb,amsfonts}
\usepackage{textcomp}
\usepackage{booktabs} % For formal tables
\usepackage{enumitem}
\usepackage{colortbl} % Suresh Addition
\usepackage{graphicx,multirow}
\usepackage{epstopdf}
\usepackage{diagbox}
\usepackage{amsmath}
\usepackage{svg}
\usepackage{subfig}		%Subfigures inside a figure
\usepackage{pgfplots}
\pgfplotsset{compat=1.15}
\usepackage{verbatim} 
\usepackage{multicol}
\usepackage{hyphenat}
\usepackage{lipsum}
\usepackage{enumitem}
\usepackage{etoolbox}
\usepackage{euscript}
\usepackage{eufrak}
\usepackage{textcomp}
\usepackage{mathrsfs}
\usepackage{bm}
\usepackage[english]{babel}
\usepackage{blindtext}
\usepackage{textcase}
\usepackage{color}
\usepackage[tablename=TABLE, tableposition=top]{caption}
\usepackage{pifont}
\usepackage{makecell}
\usepackage[T1]{fontenc}
\usepackage{hyperref}
\usepackage[acronym]{glossaries}
\DeclareMathAlphabet{\mathpzc}{OT1}{pzc}{m}{it}

\usepackage{fancyhdr}
\usepackage{draftwatermark}
\SetWatermarkText{\textit{This article is currently under peer-review.}}
\SetWatermarkFontSize{0.6cm}
\SetWatermarkAngle{0}
\SetWatermarkHorCenter{0.6 \textwidth }
\SetWatermarkVerCenter{0.05 \textwidth }
\renewcommand{\baselinestretch}{1.0}
\addto\extrasenglish{%

}
\newcommand{\cmark}{\ding{51}}%
\newcommand{\xmark}{\ding{55}}%
\definecolor{mycolor}{rgb}{0.0,0.0,0.0}
\usepackage{tikz}
\newcommand*\circled[1]{\tikz[baseline=(char.base)]{
            \node[shape=circle,draw,inner sep=2pt] (char) {#1};}}
\algnewcommand{\LeftComment}[1]{\Statex \(\triangleright\) #1}
\algnewcommand{\HeadComment}[1]{\Statex \(\blacktriangleright\) #1}
\newcommand{\algrule}[1][.2pt]{\par\vskip.5\baselineskip\hrule height #1\par\vskip.5\baselineskip}
\definecolor{add}{rgb}{1, 0, 0}
\definecolor{rev}{rgb}{0, 0, 1}
\definecolor{reg}{rgb}{0, 0, 0}
\newcommand{\suresh}[1]{\textcolor{black}{#1}}

\newcommand{\siva}[1]{\textcolor{black}{#1}}
\newcommand{\salim}[1]{\textcolor{black}{#1}}
\newcommand{\sam}[1]{\textcolor{black}{#1}}
\newcommand{\su}[1]{\textcolor{mycolor}{#1}}
\makeglossaries
\newacronym{vlsi}{VLSI}{\text{V}ery \text{L}arge \text{S}cale \text{I}ntegration}
\newacronym{mttf}{MTTF}{\text{M}ean \text{T}ime \text{T}o \text{F}ailure}
\newacronym{mttc}{MTTC}{\text{M}ean \text{T}ime \text{T}o \text{C}rash}
\newacronym{noc}{NoC}{\text{N}etwork-on-\text{C}hip}
\newacronym{pr}{PR}{\text{P}artially \text{R}econfigurable}
\newacronym{cots}{COTS}{commercial-off-the-shelf}
\newacronym{nre}{NRE}{\text{N}on \text{R}ecurring \text{E}ngineering}
\newacronym{rtl}{RTL}{\text{R}egister \text{T}ransfer \text{L}evel}
\newacronym{vcu}{VCU}{\text{V}ideo \text{C}odec \text{U}nit}
\newacronym{apu}{APU}{\text{A}pplication \text{P}rocessing \text{U}nit}
\newacronym{gpu}{GPU}{\text{G}raphics \text{P}rocessing \text{U}nit}
\newacronym{rpu}{RPU}{\text{R}eal-time \text{P}rocessing \text{U}nit}
\newacronym{gps}{GPS}{\text{G}lobal \text{P}ositioning \text{S}ystem}
\newacronym{ai}{AI}{\text{A}rtificial \text{I}ntelligence}
\newacronym{iot}{IoT}{\text{I}nternet of \text{T}hings}
\newacronym{ic}{ICs}{\text{i}ntegrated \text{c}ircuits}
\newacronym{esl}{ESL} {\text{E}lectronic \text{S}ystem \text{L}evel}
\newacronym{eda}{EDA} {\text{E}lectronic \text{D}esign \text{A}utomation}
\newacronym{clr}{CLR} {\text{C}ross\hyp\text{l}ayer \text{R}eliability}
\newacronym{qos}{QoS} {\text{Q}uality of \text{S}ervice}
\newacronym{hmpsoc}{HMPSoC} {\text{H}eterogeneous \text{M}ulti-\text{P}rocessor \text{S}ystem\hyp on\hyp\text{C}hip}
\newacronym{mpsoc}{MPSoC} {\text{M}ulti-\text{P}rocessor \text{S}ystem\hyp on\hyp\text{C}hip}
\newacronym{soc}{SoC} {\text{S}ystem\hyp \text{o}n\hyp\text{C}hip}
\newacronym{fpga}{FPGA} {\text{F}ield \text{P}rogrammable \text{G}ate \text{A}rray}
\newacronym{dpr}{DPR} {\text{D}ynamic \text{P}artial \text{R}econfiguration}
\newacronym{prr}{PRR} {\text{P}artially \text{R}econfigurable \text{R}egion}
\newacronym{prm}{PRM} {\text{P}artially \text{R}econfigurable \text{M}odule}
\newacronym{pe}{PE} {\text{P}rocessing \text{E}lement}
\newacronym{dse}{DSE} {\text{d}esign \text{s}pace \text{e}xploration}
\newacronym{ga}{GA} {\text{G}enetic \text{A}lgorithms}
\newacronym{bti}{BTI} {\text{B}ias \text{T}emperature \text{I}nstability}
\newacronym{nbti}{NBTI} {\text{N}egative \text{B}ias \text{T}emperature \text{I}nstability}
\newacronym{pbti}{PBTI} {\text{P}ositive \text{B}ias \text{T}emperature \text{I}nstability}
\newacronym{em}{EM} {\text{E}lectro\text{m}igration}
\newacronym{gob}{GOB} {\text{G}ate \text{O}xide \text{B}reakdown}
\newacronym{hci}{HCI} {\text{H}ot \text{C}arrier \text{I}njection}
\newacronym{tddb}{TDDB}{\text{T}ime \text{D}ependent \text{D}ielectric \text{B}reakdown}
\newacronym{seu}{SEU} {\text{S}ingle \text{E}vent \text{U}pset}
\newacronym{ser}{SER} {\text{S}oft \text{E}rror \text{R}ate}
\newacronym{gdb}{GDB} {\text{G}ate \text{D}ielectric \text{B}reakdown}
\newacronym{tmr}{TMR} {\text{T}riple \text{M}odular \text{R}edundancy}
\newacronym{dmr}{DMR} {\text{D}ual \text{M}odular \text{R}edundancy}
\newacronym{ecc}{ECC}{\text{E}rror \text{C}hecking and \text{C}orrecting}
\newacronym{sram}{SRAM}{\text{S}tatic \text{R}andom \text{A}ccess \text{M}emory}
\newacronym{dram}{DRAM}{\text{D}ynamic \text{R}andom \text{A}ccess \text{M}emory}
\newacronym{llc}{LLC}{\text{L}ast \text{L}evel \text{C}ache}
\newacronym{l1}{L1}{\text{L}evel \text{1}}
\newacronym{dimm}{DIMM}{\text{D}ual \text{i}n-line-\text{M}emory \text{M}odule}
\newacronym{snc}{SNC}{\text{S}ingle-\text{N}ibble-error-\text{C}orrecting}
\newacronym{dnd}{DND}{\text{D}ouble-\text{N}ibble-error-\text{D}etecting}
\newacronym{sec}{SEC}{\text{S}ingle-bit-\text{E}rror-\text{C}orrecting}
\newacronym{ded}{DED}{\text{D}ouble-bit-\text{E}rror-\text{D}etecting}
\newacronym{dec}{DEC}{\text{D}ouble-bit-\text{E}rror-\text{C}orrecting}
\newacronym{ted}{TED}{\text{T}riple-bit-\text{E}rror-\text{D}etecting}
\newacronym{ivi}{IVI}{\text{I}nstruction \text{V}ulnerability \text{I}ndex}
\newacronym{fvi}{FVI}{\text{F}unction \text{V}ulnerability \text{I}ndex}
\newacronym{sed}{SED}{\text{S}obel \text{E}dge \text{D}etection}
\newacronym{cnn}{CNN}{\text{C}onvolutional \text{N}eural \text{N}etworks}
\newacronym{dnn}{DNN}{\text{D}eep \text{N}eural \text{N}etworks}
\newacronym{os}{OS}{\text{O}perating \text{S}ystem}
\newacronym{avf}{AVF}{\text{A}rchitectural \text{V}ulnerability \text{F}actor}
\newacronym{milp}{MILP}{\text{M}ixed \text{I}nteger \text{L}inear \text{P}rogramming}
\newacronym{sofr}{SOFR}{\text{S}um-\text{o}f-\text{F}ailure \text{R}ate}
\newacronym{clb}{CLB}{\text{C}onfigurable \text{L}ogic \text{B}locks}
\newacronym{bram}{BRAM}{\text{B}lock \text{RAM}}
\newacronym{dsps}{DSPs}{\text{D}igital \text{S}ignal \text{P}rocessing blocks}
\newacronym{mcts}{MCTS}{\text{M}onte \text{C}arlo \text{T}ree \text{S}earch}
\newacronym{ttp}{TTP} {\text{T}ree \text{T}raversal \text{P}roblem}
\newacronym{fir}{FIR} {\text{F}inite \text{I}mpluse \text{R}esponse}
\newacronym{mtbf}{MTBF}{Mean Time between Failures}
\newacronym{ura}{\textit{uRA}}{User-modulated Run-time Adaptation}
\newacronym{aura}{\textit{AuRA}}{Agent-based User-modulated Run-time Adaptation}
\newacronym{moea}{MOEA}{\text{M}ulti-\text{O}bjective \text{E}volutionary \text{A}lgorithms}
\newacronym{dvfs}{DVFS}{\text{D}ynamic \text{V}oltage and \text{F}requncy \text{S}caling}
 \newacronym{icap}{ICAP}{\text{I}nternal \text{C}onfiguration \text{A}ccess \text{P}ort}
\newacronym{rl}{RL}{\text{R}einforcement \text{L}earning}
\newacronym{pvt}{PVT}{\text{P}rocess, \text{V}oltage, and \text{T}emperature}
\newacronym{nhpp}{NHPP}{\text{N}on-\text{H}omogeneous \text{P}oisson \text{P}rocess}
\newacronym{fit}{FIT}{\text{F}ailures \text{I}n \text{T}ime}
\newacronym{mosfet}{MOSFET}{\text{M}etal \text{O}xide \text{S}emiconductor \text{F}ield \text{E}ffect \text{T}ransistor }
\newacronym{nmos}{NMOS}{\text{N}egative channel \text{M}etal \text{O}xide \text{S}emiconductor}
\newacronym{pmos}{PMOS}{\text{P}ositive channel \text{M}etal \text{O}xide \text{S}emiconductor}
\newacronym{osi}{OSI}{\text{O}pen \text{S}ystems \text{I}nterconnection}
\newacronym{bist}{BIST}{\text{B}uilt-\text{i}n \text{S}elf \text{T}est}
\newacronym{fxp}{FxP}{\text{F}ixed \text{P}oint}
\newacronym{flp}{FP32}{IEEE-754 Floating Point}
\newacronym{ann}{ANN}{\text{\text{A}}rtificial \text{\text{N}}eural \text{\text{N}}etworks}
\newacronym{pofx}{PoFx}{\text{Po}sit to \text{F}i\text{x}ed Point}
\newacronym{nan}{NAN}{\text{N}ot \text{A} \text{N}umber}
\newacronym{unum}{Unum}{\text{U}niversal \text{num}ber}
\newacronym{emac}{EMAC}{\text{E}xact \text{M}ultiply and \text{AC}cumulate}
\newacronym{mac}{MAC}{ \text{M}ultiply and \text{AC}cumulate}
\newacronym{hls}{HLS}{\text{H}igh \text{L}evel \text{S}ynthesis}
\begin{document}
%     \onecolumn
%     	\input{00_coverLetter}
%     \pagebreak
% 	\twocolumn
% \title{\huge{$\mathcal{J}$\textit{-ExPAN(N)D}: \textit{Ex}ploring \textit{P}OSITs for \textit{Energy} Efficient \textit{A}rtificial \textit{N}eural \textit{N}etwork \textit{D}esign}}
% \title{{\textit{ExPAN(N)D}: \textit{Ex}ploring \textit{P}osits for Efficient \textit{A}rtificial \textit{N}eural \textit{N}etwork \textit{D}esign} in FPGA-based Edge Processing}

\title{\LARGE{{\textit{ExPAN(N)D}: \underline{Ex}ploring \underline{P}osits for Efficient \underline{A}rtificial \underline{N}eural \underline{N}etwork \underline{D}esign} in FPGA-based Systems}}

\author{Suresh Nambi, Salim Ullah, Aditya Lohana, Siva Satyendra Sahoo, Farhad Merchant and Akash Kumar
% \IEEEmembership{Member, IEEE}
% \thanks{The manuscript was first submitted for review on 09 August 2020. The project was supported partially from the funding of ...}
\thanks{S. Nambi, S. Ullah, A. Lohana, S. S. Sahoo and A.Kumar are with The Chair for Processor Design, TU Dresden, Germany (emails: \{\textit{suresh.nambi;aditya.lohana}\}@mailbox.tu-dresden.de, \{\textit{salim.ullah;siva\_satyendra.sahoo;akash.kumar}\}@tu-dresden.de)}
\thanks{F. Merchant is with The Institute for Communication Technologies and Embedded Systems, RWTH Aachen, Germany (email:farhad.merchant@ice.rwth-aachen.de) }
% \thanks{\textcolor{blue}{\textbf{This article is under peer-review with IEEE TETC} }}
}

\IEEEtitleabstractindextext{\justify \begin{abstract}% \justify
\su{The recent advances in machine learning, in general, and Artificial Neural Networks (ANN), in particular, has made smart embedded systems an attractive option for a larger number of application areas. However, the high computational complexity, memory footprints, and energy requirements of machine learning models hinder their deployment on resource-constrained embedded systems. Most state-of-the-art works have considered this problem by proposing various low bit-width data representation schemes, optimized arithmetic operators' implementations, and different complexity reduction techniques such as network pruning. To further elevate the implementation gains offered by these individual techniques, there is a need to cross-examine and combine these techniques' unique features. This paper presents ExPAN(N)D, a framework to analyze and ingather the efficacy of the \textit{Posit} number representation scheme and the efficiency of \textit{fixed-point} arithmetic implementations for ANNs. The Posit scheme offers a better dynamic range and higher precision for various applications than IEEE $754$ single-precision floating-point format. However, due to the dynamic nature of the various fields of the Posit scheme, the corresponding arithmetic circuits have higher critical path delay and resource requirements than the single-precision-based arithmetic units. Towards this end, we propose a novel \textit{Posit to fixed-point converter} for enabling high-performance and energy-efficient hardware implementations for ANNs with minimal drop in the output accuracy. We also propose a modified Posit-based representation to store the trained parameters of a network. Compared to an $8$-bit fixed-point-based inference accelerator, our proposed implementation offers $\approx46\%$ and $\approx18\%$ reductions in the storage requirements of the parameters and energy consumption of the MAC units, respectively.}

%Today embedded systems are being used across a large variety of applications. The increasing advances in machine learning, in general, and Artificial Neural Networks (ANN), in particular, has made smart embedded systems an attractive option for an even larger number of application areas. However, the high computational cost of ANNs makes them harder to be implemented on resource-constrained embedded systems. The main obstacles to implementing energy-efficient ANNs on embedded hardware platforms are -- (1) The high power dissipation of the massively parallel ANN implementations and, (2) The high communication overheads and storage requirements to support such massive parallelism. Most state-of-the-art works overcome these issues by methods such as \textit{pruning} and \textit{scheduling}, and \textit{in-memory} and \textit{near-memory} computing. In this article, we propose methods that are orthogonal to such techniques and can complement them towards reducing the power, energy and storage requirements of embedded machine learning. Specifically, we propose using the Posit number representation to store the massive number of weights resulting in reduced storage and communication overhead. Further, we propose low-cost converters to enable low-powered fixed-point arithmetic operations, resulting in drastically reduced computational costs. Using the proposed methods, we report considerable reduction in the power and energy cost for ANNs implemented on FPGAs.
\end{abstract}

% \begin{IEEEkeywords}
% At least four keywords or phrases in 
% alphabetical order, separated by commas. For a list of suggested keywords, 
% send a blank e-mail to \href{mailto:keywords@ieee.org}{mailto:keywords@ieee.org} or visit 
% \href{http://www.ieee.org/organizations/pubs/ani_prod/keywrd98.txt}{http://www.ieee.org/organizations/pubs/ani\_prod/keywrd98.txt}
% \end{IEEEkeywords}
% Note: There should no nonstandard abbreviations, acknowledgments of support, 
% references or footnotes in in the abstract.
}

\newcounter{tblEqCounter} %create a counter
\maketitle
% \begin{abstract}
% \input{abstract}
% \end{abstract}
\glsresetall
\section{Introduction}
\label{sec:intro}
% \par {\color{blue} 1.1. Increasing Complexity of \gls{dnn} \& Shift towards Edge Computing and associated resource constraints }
\IEEEPARstart{M}{achine} \su{learning algorithms have become an essential factor in various modern applications, such as scene perception and image classification~\cite{DBLP:journals/corr/abs-1708-02709,DBLP:journals/corr/HeZR015,6639344}. Over the past few years, these algorithms have mainly relied on the performance of modern computing systems to support the increasing complexity of the algorithms. For example, the massively parallel architectures, such as Graphics Processing Units (GPUs), and cloud-based computing have been traditionally used to train these algorithms. However, to utilize these trained machine learning models on resource-constrained embedded systems, the computational complexity and storage requirements of these algorithms must be reduced.}

\begin{figure}[t] 
   \centering
   	\subfloat[FP32\label{sub_fig1}]{%
       \includegraphics[width=0.32\columnwidth]{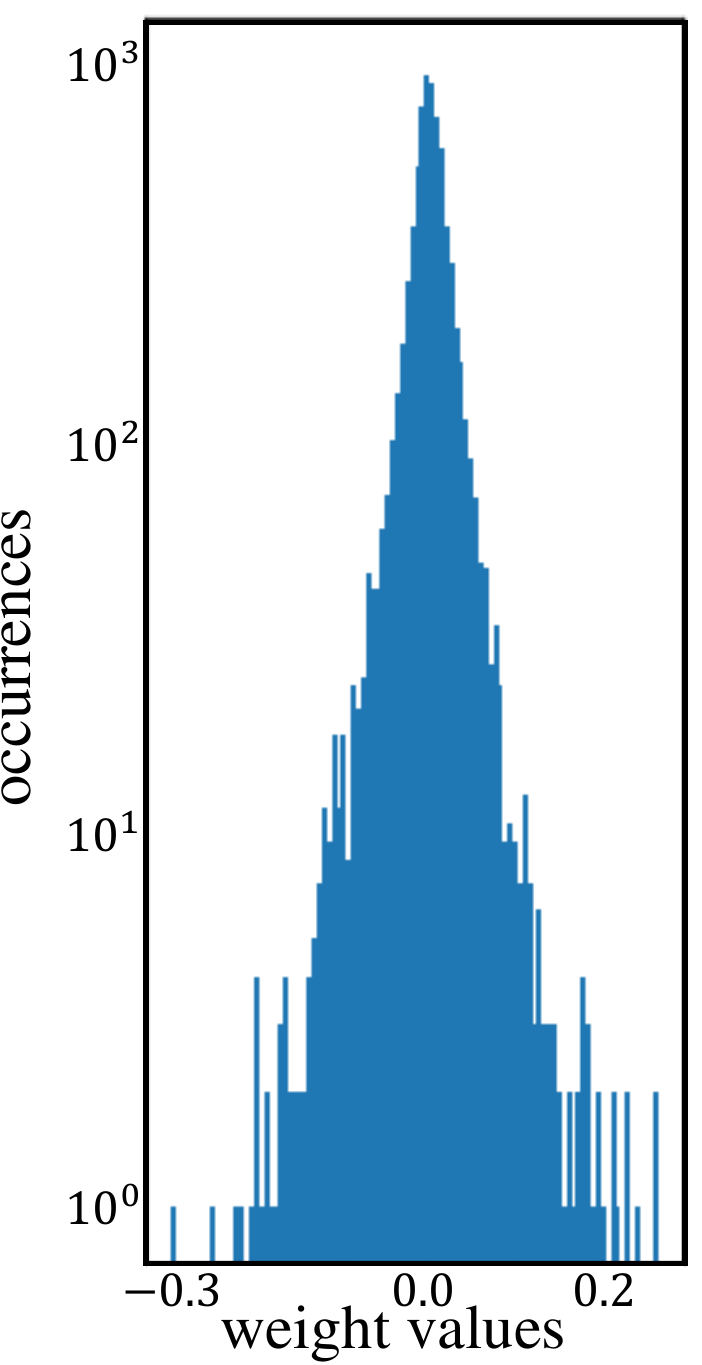}}
    \hspace*{0.2em}
    \subfloat[FxP-8\label{sub_fig2}]{%
        \includegraphics[width=0.32\columnwidth]{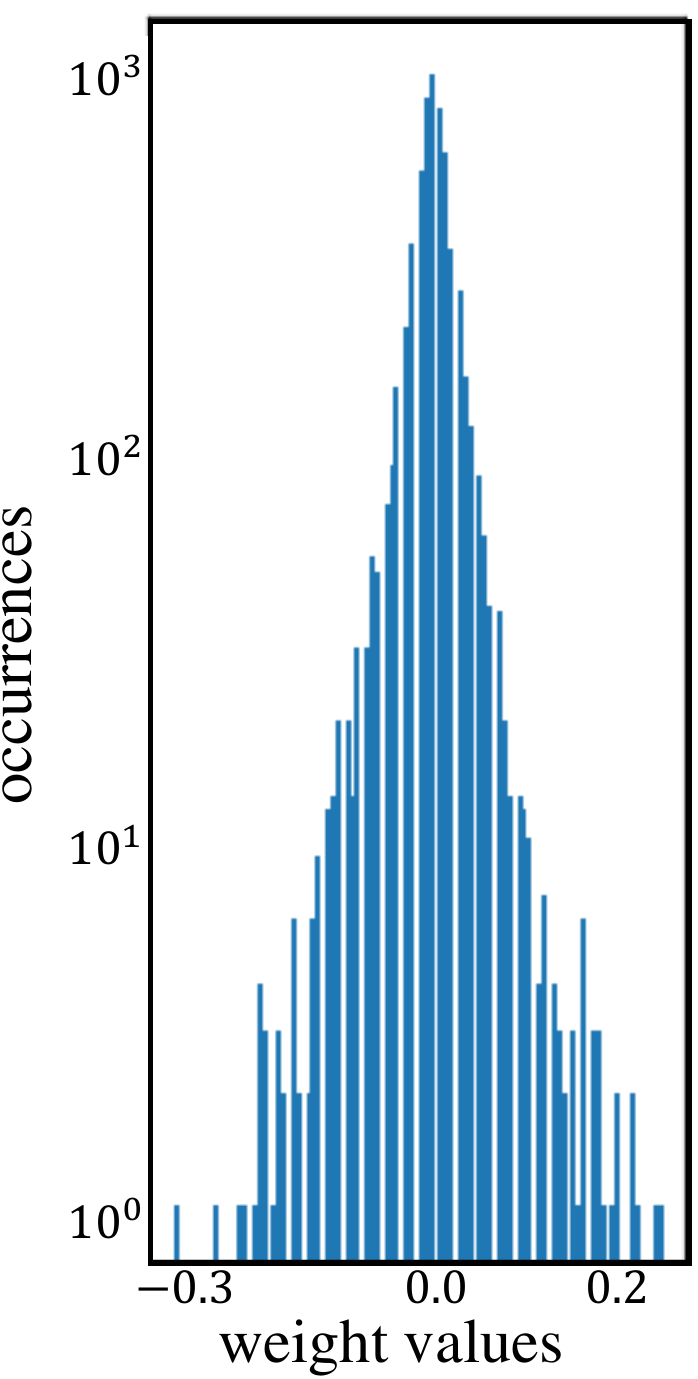}}
\hspace*{0.2em}
    \subfloat[Posit-8 \label{sub_fig2}]{%
        \includegraphics[width=0.32\columnwidth]{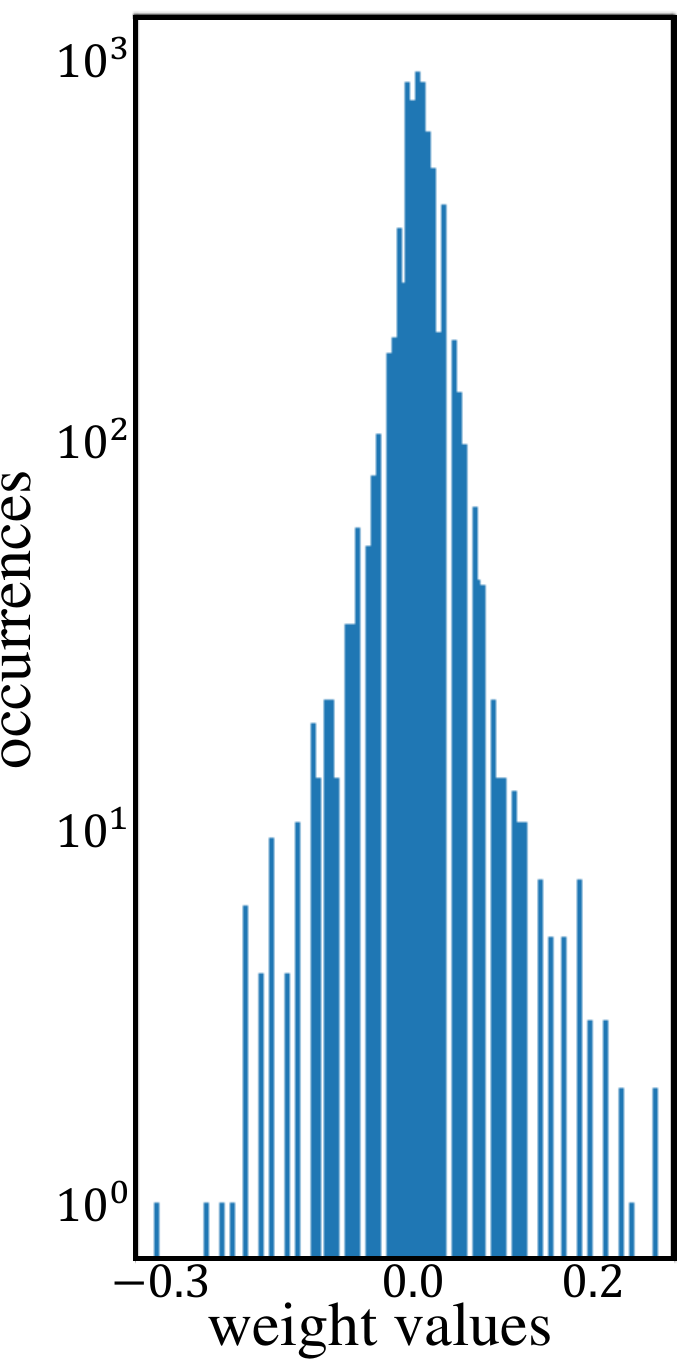}}
  \caption{Distribution of pre-trained weights of Conv2\_1 layer of VGG16~\cite{simonyan2014deep}. (a) Single-precision floating-point, (b)  8-bit linear fixed-point quantization: average absolute relative quantization-induced error=0.295 (c) Posit~(8, 2)-based quantization: average absolute relative quantization-induced error=0.052.}
  \label{fig:motiv1} 
\end{figure}
\su{Many recent works have considered this problem to define various optimization techniques to reduce the complexity of machine learning models, such as \gls{ann}. For example, the techniques used in\cite{han2015deep} and~\cite{Liu_2015_CVPR} have employed the sparsity of \gls{dnn} to reduce the total number of trained parameters. The works in~\cite{8877390}, \cite{langroudi2019cheetah} and~\cite{10.5555/3045390.3045690} have explored other number representation  techniques, such as \textit{bfloat16}, \textit{Posit} and \textit{\gls{fxp}}, to overcome the storage requirements of single-precision \gls{flp}.}
\salim{Depending on the configuration used, each of these number representation techniques provides different dynamic range to represent the parameters (weights and biases) of a network. For example, \autoref{fig:motiv1}(a) shows the \gls{flp}-based distribution of the pre-trained weights of the \emph{Conv2\_1} layer of VGG16 \gls{dnn}~\cite{simonyan2014deep}. The pre-trained weights have a dynamic range between $-0.3$ to $+0.3$, with most of the weights clustered around `$0$'. To reduce the memory footprint of the weights and associated computational complexity, \autoref{fig:motiv1}(b) represents the distribution using an 8-bit fixed point linear quantization scheme, referred to as \emph{FxP8}. The FxP8 provides a set of $256$ uniformly distributed discrete values, which generates an average relative error of $0.295$ in the quantized weights. To reduce the quantization-induced errors, ~\autoref{fig:motiv1}(c)  shows the trained parameters
%of the single-precision-based distribution 
using an 8-bit Posit scheme. The Posit technique covers more values around $0$, resulting in an average relative error of $0.052$ in the quantized weights. \su{Therefore, it is imperative to define number representation schemes (or quantization methods), which can significantly maintain \gls{flp}-based machine learning models' accuracy and reduce their corresponding computational complexity and storage requirements.}}
\par
\begin{figure}[t] 
   \centering
   	\subfloat[Accuracy \label{mot2:sub_fig1}]{%
       \includegraphics[width=0.32\columnwidth]{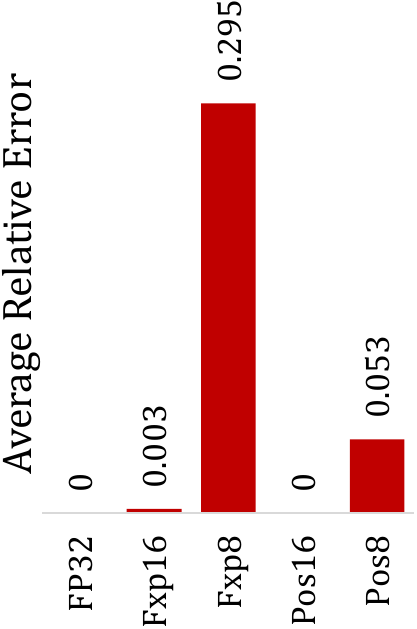}}
\hspace*{0.2em}
    \subfloat[Computation \label{mot2:sub_fig2}]{%
        \includegraphics[width=0.32\columnwidth]{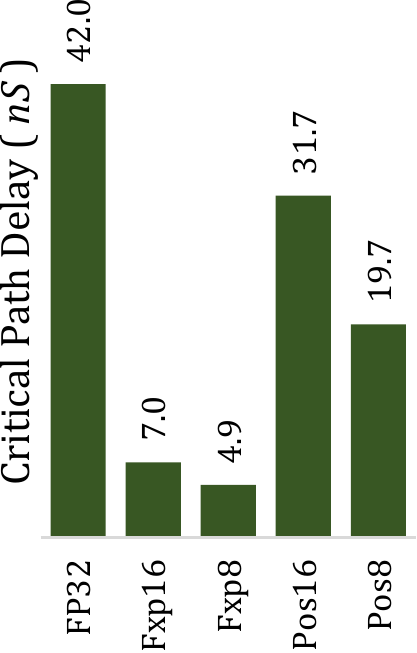}}
\hspace*{0.2em}
    \subfloat[Storage \label{mot2:sub_fig3}]{%
        \includegraphics[width=0.32\columnwidth]{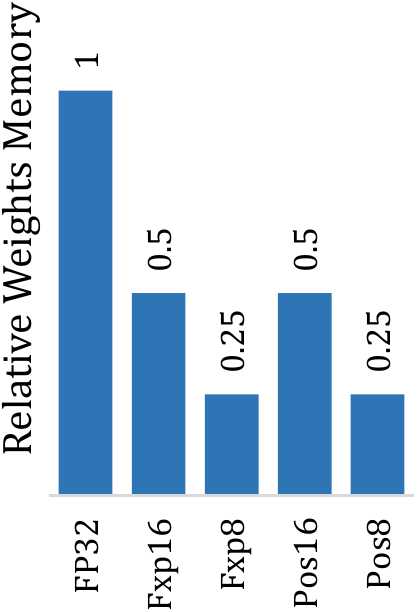}}
  \caption{Accuracy and performance comparison of various schemes for numbers representation for the Conv2\_1 layer of pre-trained VGG16~\cite{simonyan2014deep} : (a) Average absolute relative error with respect to FP32-based parameters, (b) critical path delay, (c) normalized memory footprints. }
  \label{fig:MetricsMot} 
\end{figure}

{
\salim{The various number representation schemes (quantization methods) 
result in varying performance overheads of their associated arithmetic hardware. For example, \autoref{fig:MetricsMot} shows the comparison of the effect of using different quantization methods across multiple performance aspects -- behavioral (error in the quantization of weights), computational (critical path delay of a \gls{mac} unit), and memory requirements (weights' storage) in the Conv2\_1 layer of pre-trained VGG16. The hardware implementation results have been obtained by implementing each technique on the Xilinx UltraScale \gls{fpga} using Vivado HLS 2018.2. For a fair comparison, the critical path delay (CPD) is obtained from \gls{mac} units implemented using 6-input lookup tables (LUTs) and with a latency of a single cycle. As shown by the results, higher bit-widths for the quantization schemes significantly reduce quantization-induced errors.
The FP32 implementation has the highest memory footprint with the worst CPD of $42ns$. The Posit schemes provide better coverage of the FP32-based pre-trained parameters than the corresponding FxP-based schemes. However, the FxP-based arithmetics' simplicity results in significantly reducing the CPD of the \gls{mac} units when compared with the corresponding Posit schemes. 
}
}
% This is just a placeholder !! : We have a lot of data and can expand this figure to contain multiple windows to further drive home the point !! (Note also not valid for all layers of VGG 16)

% In this paper we aim to improve \gls{dnn} inference by  adapting Posit representation to act as a storage format~\cite{dedinechin:hal-01959581} for \gls{dnn} weights but perform computation on Fixed point numbers after optimized conversion. Our contributions to achieve the former goal are as follows  
% \begin{itemize}[leftmargin=*]
% 	\item Propose a \gls{pofx} Conversion Algorithm
% 	\item Develop Normalized Posit Representation to take advantage of the limited dynamic range of FP32 weights  
% 	\item Synthesize parameterized \gls{pofx} conversion hardware modules with Posit and Fixed Point configuration as parameters
% 	\item Implement a Software tool to ascertain the most suitable \gls{pofx} configuration for a particular \gls{dnn} layer and calculate the associated classification accuracy
% 	\item Compare the resource utilization and associated speedup of our proposed design with existing implementations
	
% \end{itemize}
% \par {\color{blue} 1.4. Motivation for PoFx Conversion [Include Quantization Error Results as Figure]}

%-----------------------------------------
% Contributions
%-----------------------------------------
%%%%%%%%%%%%%%%%%%%%%%%%%%%%%%%%%%%%%%%%%%%%%%%%%%%%
%%%%%%%%%%%%%%%%% Contributions of the paper %%%%%%%
%%%%%%%%%%%%%%%%%%%%%%%%%%%%%%%%%%%%%%%%%%%%%%%%%%%%
\par{
Most state-of-the-art works do not consider application-specific optimizations to the quantization methods. For instance the Posit related works focus on representing the whole range of real numbers, $(-\infty, \infty)$, rather than the actual range of the parameters in the application. Similarly, many related works consider each quantization method in isolation and do not attempt to leverage the best features of multiple methods. \su{To this end, we propose \textit{ExPAN(N)D} framework for \underline{Ex}ploring the joint use of \underline{P}osit and \gls{fxp} representations for \underline{D}esigning efficient \gls{ann}s. The major contributions in this paper are as follows.}\\
\textbf{Contributions:}
% \begin{enumerate}[wide, labelwidth=!, labelindent=0pt]
\begin{enumerate}[wide]
% 	%Framework for QoS-aware platform-aware accelerators for RL
% 	\item \textcolor{blue}{Framework for QoS-aware platform-aware accelerators for RL\\
% 	Quote results }.
	%%%%%%%%%%%%%%%%%%%%%%%%%%%%%%%%%%%%%%%%%%%%%%%%%%%%
	%Hardware Design.
	\item \su{We propose a reduced bit-length Posit-based representation that improves the encoding efficiency to reduce the communication and storage costs in \gls{ann}s. Using our proposed representation for each $N$-bit Posit number, we only store $N-1$ bits.}
	\item %\suresh{Hardware Design. Quote results}.
	\su{We propose a novel arithmetic hardware design, referred to as \gls{pofx}, that aims to combine the best of both Posit and \gls{fxp} number representations. The proposed hardware unit offers resource-efficient and low-latency conversion of Posit-based numbers to \gls{fxp}-based numbers to leverage the lower computation overheads of fixed-point arithmetic.For example, an 8-bit \gls{pofx}-based MAC provides up to $15\%$ resource overhead with a corresponding $46\%$ reduction in the storage requirement of a network's parameters.}
	%\siva{We propose a novel arithmetic hardware design that aims to combine the best of both Posit and \gls{fxp} number representations. Specifically, We also leverage the lower computation overheads of fixed-point arithmetic.}
	
	%%%%%%%%%%%%%%%%%%%%%%%%%%%%%%%%%%%%%%%%%%%%%%%%%%%%
	%Behavioral Analysis.
	\item \salim{Framework for Behavioral Analysis: We provide a high-level framework for the efficient and thorough exploration of various quantization schemes to satisfy the accuracy constraints of a DNN. The proposed framework explores the limitations and the interplay of various quantization schemes, such as \emph{FxP to Posit to FxP}, to minimize the quantization-induced errors. The framework prunes the non-optimal quantization configurations by analyzing the quantization induced-errors in (a) parameters of individual layers, (b) output activations of each layer using quantized weights, and (c) final output of the network. \su{For example, our framework explores various $N$-bit Posit configurations to achieve output accuracy comparable to an $M$-bit \gls{fxp}-based quantization, where $N < M$. }}

	%%%%%%%%%%%%%%%%%%%%%%%%%%%%%%%%%%%%%%%%%%%%%%%%%%%%
	%Accelerator DSE.
	\item %\siva{Accelerator DSE. Quote results}.
	\siva{We explore the impact of using the proposed hardware designs in a fully-connected layer. Specifically we use an automated design flow, using state-of-the-art \gls{hls} tools, to explore storage-computation trade-offs in the design of FPGA-based accelerators for \gls{ann}s.} \su{For example, compared to an \gls{fxp}-based accelerator, the \gls{pofx}-based accelerator provides up to $46\%$ and $18\%$ reductions in the storage and energy requirements of an accelerator. } 
	%%%%%%%%%%%%%%%%%%%%%%%%%%%%%%%%%%%%%%%%%%%%%%%%%%%%
	%Open source framework for estimation and DSE for DVFS and CLR integrated system design 
% 	\item The proposed optimizations for designing accelerators for Q-learning will be available as an open source tool at \textit{\url{https://blinded_for_peer_review}}. 
\end{enumerate}
}
%-----------------------------------------
% Paper Organization
%-----------------------------------------
%%%%%%%%%%%%%%%%%%%%%%%%%%%%%%%%%%%%%%%%%%%%%%%%%%%%
%%%%%%%%%%%%%%%%%% Organization of the paper %%%%%%%
%%%%%%%%%%%%%%%%%%%%%%%%%%%%%%%%%%%%%%%%%%%%%%%%%%%%
\par{
The rest of the paper is organized as follows.
In~\autoref{sec:bckgrndRel}, we provide the relevant background and brief overview of related work.
The system model used for the evaluation of the proposed methods is presented in~\autoref{sec:sysModel}.
In~\autoref{sec:dsgnMeth}, we explain the proposed methodology for exploring the use of Posit representation for \gls{ann}s, along with the proposed hardware designs.
In~\autoref{sec:expRes}, we discuss the results from the experimental evaluation of the different components of the proposed methodology. 
Finally, we conclude the article in~\autoref{sec:conc} with a summary and a discussion on the scope for related future research.
}

\section{Background and Related Works}
\label{sec:bckgrndRel}
\subsection{Posit Number System}
\label{subsec:posRep}
% ############################# Subsection Description ##################################

% Brief description of Redundancies in Floating point arithmetic
% Advantages of Posit representation over Floating point
% Decoding a Posit representation to its decimal value -> introduce "variable bit-field length" concept
% Outcome : Variable Bit Field length provides flexibility/ better range .. but HW expensive not compute eff.
% ########################################################################################

\begin{figure}[t]
	\centering
	\scalebox{1}{\includegraphics[width=0.95 \columnwidth]{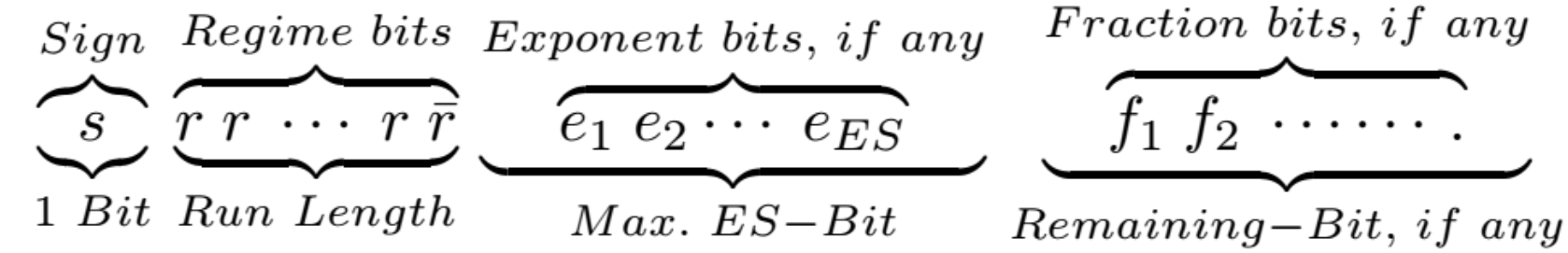}}
	\caption{Posit number representation}
	\label{fig:posit_format}
\end{figure}

\su{The IEEE 754-2008 compliant floating-point (floats)-based arithmetic has become ubiquitous in modern-day computing and is deeply embedded in compilers and low-level software routines. However, the floats have several limitations, such as non-identical results across systems, redundant/wasted bit patterns, and a limited dynamic range. 
The Posit number scheme overcomes these limitations by offering a better dynamic range and portability across various computing platforms. \autoref{fig:posit_format} shows the various fields~(\textit{sign}, \textit{regime}, \textit{exponent} and \textit{fraction}) of the Posit number scheme. A Posit configuration is characterized by its total bit length ($N$) and the number of bits reserved for exponent~($ES$). Utilizing the four fields of the Posit scheme, Eq.~\ref{eq2} defines the computation of a Posit value.  The \textit{regime} field, in \autoref{fig:posit_format}, is utilized to compute the value of $k$ in Eq.~\ref{eq2}. The \textit{regime} field is terminated when an inverted bit ($\bar{r}$) is encountered, and the associated value of $k$ is determined by the number of identical bits ($m$); if the identical bits are a string of $0s$, then $k = - m$; if they are a string of $1s$, then $k = m - 1$. Next, the \textit{exponent} ($e$) and \textit{fraction} values ($f$) are determined using the remaining bits. The utilization of \textit{regime} field provides a better dynamic range to Posit number scheme. For example, the authors in~\cite{PositHardwareGenerator} have reported that for some applications, the $n$-bit floats can be replaced by $m$-bit Posit-based numbers (where $m$ < $n$) to achieve comparable output accuracy.}

\begin{table*}[ht]
\caption{Posit-based hardware developments at a glance }
\def\arraystretch{1.05}
\centering
\resizebox{1.9 \columnwidth}{!}{
\begin{tabular}{|l|c|l|c|c|c|c|c|}
\hline
\multicolumn{1}{|c|}{\textbf{Related Work}} & \textbf{Main Objective} & \multicolumn{1}{c|}{\textbf{\begin{tabular}[c]{@{}c@{}}Degrees of \\ Freedom\end{tabular}}} & \textbf{\begin{tabular}[c]{@{}c@{}}Posit-based \\ Arithmetic\end{tabular}} &
\textbf{\begin{tabular}[c]{@{}c@{}}\gls{fxp}-based \\ Arithmetic\end{tabular}} &
\textbf{\begin{tabular}[c]{@{}c@{}}ANN-specific \\ Optimizations\end{tabular}} & \textbf{Energy-Aware} & \textbf{Open Source} \\ \hline
Chaurasiya, et.al~\cite{PositHardwareGenerator} & \begin{tabular}[c]{@{}c@{}}Posit Arithmetic \\ Unit Generator\end{tabular} & Computation & \cmark & \xmark & \xmark & \xmark &\xmark\\ \hline
Jaiswal, et. al~\cite{PositArithmetic} & \begin{tabular}[c]{@{}c@{}}Posit Arithmetic \\ on FPGA\end{tabular} & Computation & \cmark &\xmark & \xmark & \xmark &\cmark\\ \hline
Jaiswal, et. al~\cite{PACoGen} & \begin{tabular}[c]{@{}c@{}}Posit Arithmetic \\ Architectures\end{tabular} & Computation & \cmark &\xmark & \xmark & \xmark &\cmark\\ \hline Podobas, et.al~
\cite{PositsFPGA} & \begin{tabular}[c]{@{}c@{}}Posit-based  \\ hardware implementation\end{tabular} & Computation & \cmark &\xmark & \xmark & \xmark &\xmark\\ \hline
Carmichael, et.al~\cite{DeepPositron} & \begin{tabular}[c]{@{}c@{}}Posit-based DNNs\end{tabular} & \begin{tabular}[c]{@{}l@{}}Computation, \\ Communication\end{tabular} & \cmark &\xmark & \cmark & \xmark &\xmark\\ \hline
Langroudi, et.al~\cite{AdaptvePosit} & \begin{tabular}[c]{@{}c@{}}Posit-based DNNs with\\adpative dynamic range\end{tabular} & \begin{tabular}[c]{@{}l@{}}Computation, \\ Communication\end{tabular} & \cmark &\xmark & \cmark & \xmark &\xmark\\ \hline Cococcioni, et.al
~\cite{FastDNNusingPosit} & \begin{tabular}[c]{@{}c@{}}Posit-based DNNs for \\ image processing\end{tabular} & Computation & \cmark &\xmark & \cmark & \xmark &\xmark\\ \hline Murillo, et.al
~\cite{DeepPeNSieve} & \begin{tabular}[c]{@{}c@{}}Posit-based Deep \\ Learning Framework\end{tabular} & Computation & \cmark &\xmark & \cmark & \xmark &\xmark\\ \hline Langroudi, et.al~
\cite{PositNNFramework} & \begin{tabular}[c]{@{}c@{}}Posit-based Deep \\ Neural Networks Inference\end{tabular} & Computation & \cmark &\xmark & \textcolor{black}{\cmark} & \xmark &\xmark\\ \hline Langroudi, et.al~
\cite{InferenceOnEmbeddedDevices} & \begin{tabular}[c]{@{}c@{}}Posit vs Fixed point for \\  Deep Learning Inference \end{tabular} & Computation & \cmark &\cmark &  \textcolor{black}{\cmark} & \xmark &\xmark \\ \hline Zhang, et.al~
\cite{zangPositMAC} & \begin{tabular}[c]{@{}c@{}}Posit-based Multiply \\ and Accumulate Unit\end{tabular} & Computation & \cmark &\xmark & \xmark & \xmark &\xmark\\ \hline
\textit{ExPAN(N)D} & \begin{tabular}[c]{@{}c@{}}Posit based DNNs \end{tabular} & \begin{tabular}[c]{@{}l@{}}Storage, \\ Computation, \\ Communication\end{tabular} & \cmark &\cmark & \cmark & \cmark &\textcolor{black}{\cmark}\\ \hline
\end{tabular}}
\label{tab:PositHardwareDevelopment}
\end{table*}

\begin{equation} \label{eq2}
% Decimal\:Value =
Posit~value = s * (2^{2^{ES}})^{k} * 2^{e} * 1.f
\end{equation}

% The Posit representation is characterized by its bit length ($N$) and exponent size ($ES$). There can be a maximum of four fields: sign, regime, exponent and fraction as illustrated in Eq.~\eqref{eq1}. The regime field is terminated when an inverted bit ($\bar{r}$) is encountered and the associated value is determined by the number of identical bits in the run (m); if the identical bits are 0, then $k = - m$; if they are 1, then $k = m - 1$. Next, the exponent (e) and fraction values (f) are determined using the remaining bits. 

% The formula that allows us to retrieve the decimal value of the Posit number is specified in~\ref{eq2}. 
\su{Compared to the floats and fixed-point number representation schemes, Posit requires more computational resources. In the following section, we summarize the state-of-the-art works related to hardware implementation of Posit-based arithmetic circuits.}

% Posits numbers require more extensive and sequential hardware circuits for decoding and encoding their associated values due to variable bit length in comparison to their Floating point and Fixed point counterparts. They also lack intuitive Posit specific operator optimizations as they are still a relatively new system~\cite{dedinechin:hal-01959581}, hence its counterparts are more compute efficient. 

% \begin{figure}[t]
% 	\centering
% 	\scalebox{1}{\includegraphics[width=1 \columnwidth]{figures/Posit_Visualization.jpg}}
% 	\caption{Posit Numbers Visualized on Circular Number Lines}
% 	\label{fig:POSITNumSys}
% \end{figure}
% Develop own figure or Cite Stanford Posit Lecture
% \par{One of the key aspects of the Posit Numerical Representation is the variable bit field sizes of its various components. There are four major components Regime, Exponent and Mantissa. As is evident from the figure below the Regime bit length is bit sequence dependent and terminates only when encountering a flipped bit.}

\subsection {Posit Arithmetic Hardware}
\label{subsec:posArithHW}
% ############################# Subsection Description ##################################

% Challenges : Developing effective Hardware implementation due to variable bit-field length
% Highlight existing Optimizations and strategies for processing Posit Numbers
% Focus on the lack of Posit Specific Operations and conversion back to FltPt like format for core compute
% Outcome : Highlight HW expensive nature of Posit arithmetic & need to always decode Posit for Computation

% ########################################################################################

\par{ \suresh{The major challenges faced while developing an efficient hardware implementation for Posit arithmetic involve-- (1)~handling run-time length variation in individual Posit fields, (2)~extraction of Posit components to facilitate further manipulation and, (3)~implementation of rounding algorithms as proposed in the Posit standard. \su{\autoref{tab:PositHardwareDevelopment} presents an overview of the state-of-the-art work related to Posit-based arithmetic and highlights our proposed framework's key focus. These works are summarized below.}}}
\par{\suresh{The authors in~\cite{PositHardwareGenerator} tackle run-time varying field length by developing hardware arithmetic architectures for conversion from Posit to floating point and vice-versa. The work in~\cite{PositsFPGA} proposes a tool to generate pipelined Posit operators to be used as a drop-in replacement in processing units. In~\cite{PositArithmetic}, authors present the architecture of a parameterized Posit arithmetic unit to generate posit adders and multipliers of any bit-width. Similarly, PACoGen~\cite{PACoGen} employs a three-stage process which involves Posit data extraction, core arithmetic processing and Posit construction to perform parameterized Posit arithmetic including multiplication and division. It proposes improvements in Posit data extraction methodology and a pipelined architecture for Posit (N=32, ES=6). Posit arithmetic has also been integrated into Clarinet~\cite{jain2020clarinet} which is a RISC-V ISA based processor that supports the use of a Posit arithmetic core. However, the RISC-V implementations are not capable of handling large-scale applications.}}
\subsection{Arithmetic Hardware for ANN Inference}
\label{subsec:posANN}
% ############################# Subsection Description ##################################

% Highlight the compatibility of Posits with DNN inference due to tapered accuracy
% Summarize existing work (Posit for DNN Inference) -> Deep Positron  
% Drawbacks and questionable scalability in existing designs
% Introduce Fixed Point representation based for inference (Compute Efficient)
% Outcome : Need for custom Posit representation modified for DNN inference

%##########################################################################################
\par{\salim{A plethora of recent works have considered different quantization schemes to reduce the memory footprints and computational complexity of \gls{dnn}s for resource-constrained embedded systems and edge devices for IoT. These techniques can be categorized into (a) in-training quantization, and (b) post-training quantization schemes. For example, the techniques proposed in~\cite{DBLP:journals/corr/ZhouNZWWZ16, 8318896, DBLP:journals/corr/RastegariORF16,DBLP:journals/corr/CourbariauxBD15} have considered various fixed-point schemes for in-training quantization. The in-training quantization schemes can overcome most of the quantization-induced errors. However, these techniques cannot be utilized for the quantization of the parameters of pre-trained DNNs.  For example, for the quantization of pre-trained DNNs, \cite{10.5555/3045390.3045690,DBLP:journals/corr/abs-1811-05896, 9126777, 8587722} have proposed different schemes. The techniques presented in~\cite{9126777, 8587722} have focused on the utilization of logarithmic data representations to avoid the computationally expensive multiplication operations. However, some recent works, such as~\cite{9116476, 10.1145/2966986.2967021, 8863138} have utilized fixed-point quantization schemes to employ the well-explored high-performance and energy-efficient approximate adders and multipliers. The utilization of approximate arithmetic units~\cite{8342140, ullah2018smapproxlib, ullah2020area, 10.1109/ASP-DAC47756.2020.9045171} provides another degree of freedom for achieving the accuracy, performance, and energy constraints of DNNs for IoT. For example, the authors of~\cite{9116476} have utilized the library of approximate multipliers~\cite{ullah2018smapproxlib} to provide approximate accelerators for reduced-precision DNNs.  }}

\su{Some recent works have also explored the utilization of Posit numbers for training and inference phases of \gls{ann}. For example, the work in~\cite{FastDNNusingPosit} has used ARM scalable vector extension SIMD engine to present vectorized extensions for the cppPosit C++ posit arithmetic library. The authors of~\cite{DeepPositron} have proposed an exact multiply and accumulate (EMAC) for implementing the \gls{mac} operations in \gls{ann}. Their results show that the Posit-based representation of networks' parameters performs better than fixed-point-based representation in retaining the output accuracy of~\gls{ann}. However, the Posit-based EMACs have significantly higher resource utilization and energy-delay product (EDP) than the fixed-point-based \gls{mac} operations. In~\cite{zangPositMAC}, the authors have also proposed a parametrized Posit MAC generator to produce the HDL code of a Posit MAC unit. However, they do not present the efficacy of their proposed design in any real-world application. In~\cite{PositNNFramework}, the authors have also used the EDP metric to compare their proposed Posit-based framework with the \gls{flp}- and \gls{fxp}-based implementations; the \gls{fxp}-based implementations always produce lower EDP values than the corresponding Posit-based designs. Further, they do not report the overall resource utilization of their presented designs. The work in \cite{InferenceOnEmbeddedDevices} and~\cite{AdaptvePosit} have considered Posits for storing the trained weights of \gls{ann} and then utilizing the \gls{flp}-based operations to compute output values.}

\su{Currently, the Posit numerical scheme's utilization in implementing accelerators for various applications is hampered by the unavailability of resource-optimized and energy-efficient Posit arithmetic units. In our proposed work, we aim to leverage the useful storage capability of Posit by modifying the Posit number representation to store numbers within the sub-normal region and the compute efficiency of \gls{fxp}-based arithmetic by implementing a \gls{pofx} converter.}

\section{System Model}
\label{sec:sysModel}
\begin{figure*}[t]
	\centering
	\subfloat[Application \label{fig:appModel}]{
			\includegraphics[width=0.8 \columnwidth]{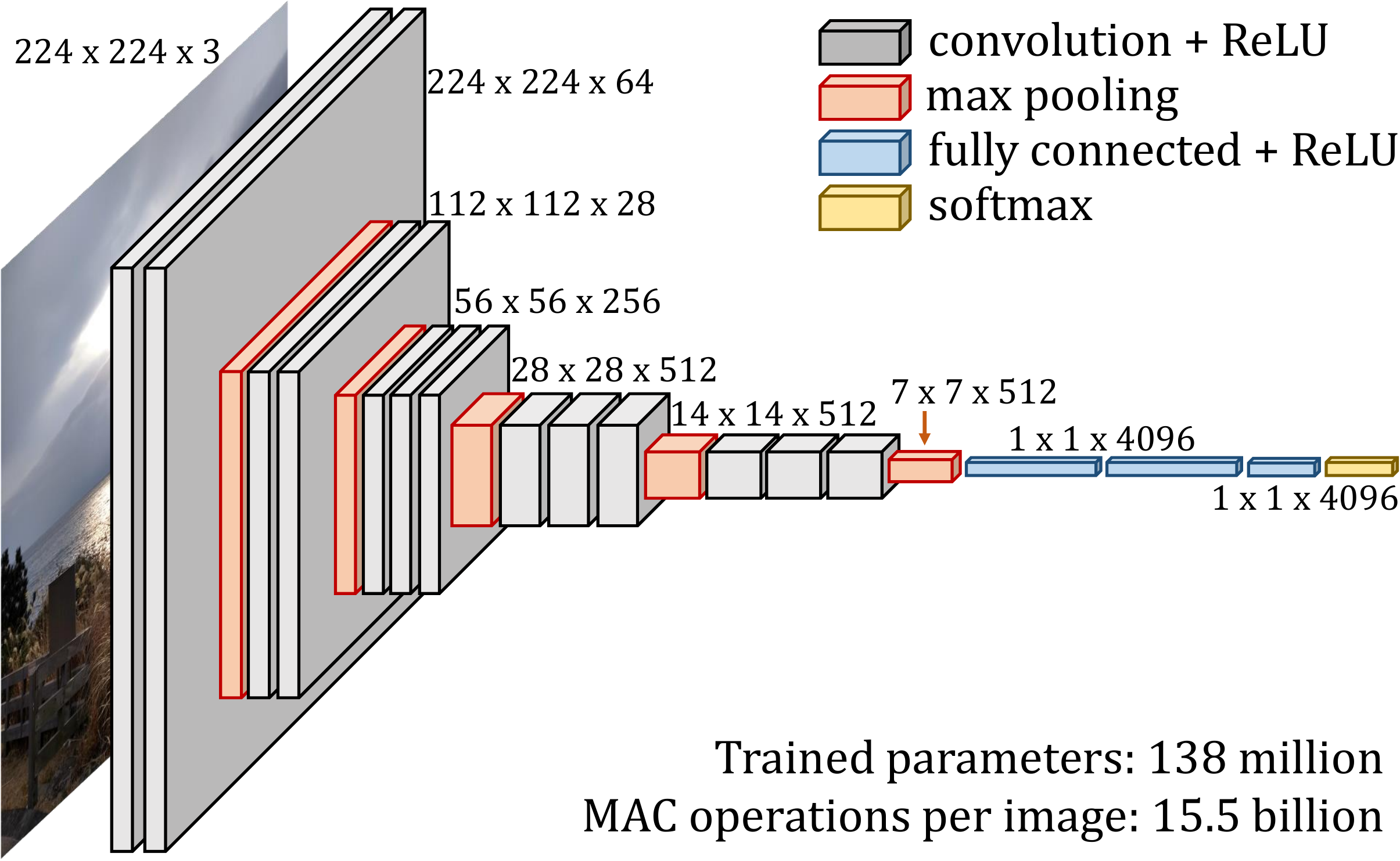}
        	} \hspace{10pt}
	\subfloat[Architecture \label{fig:archModel}]{
    	\includegraphics[width=0.8 \columnwidth]{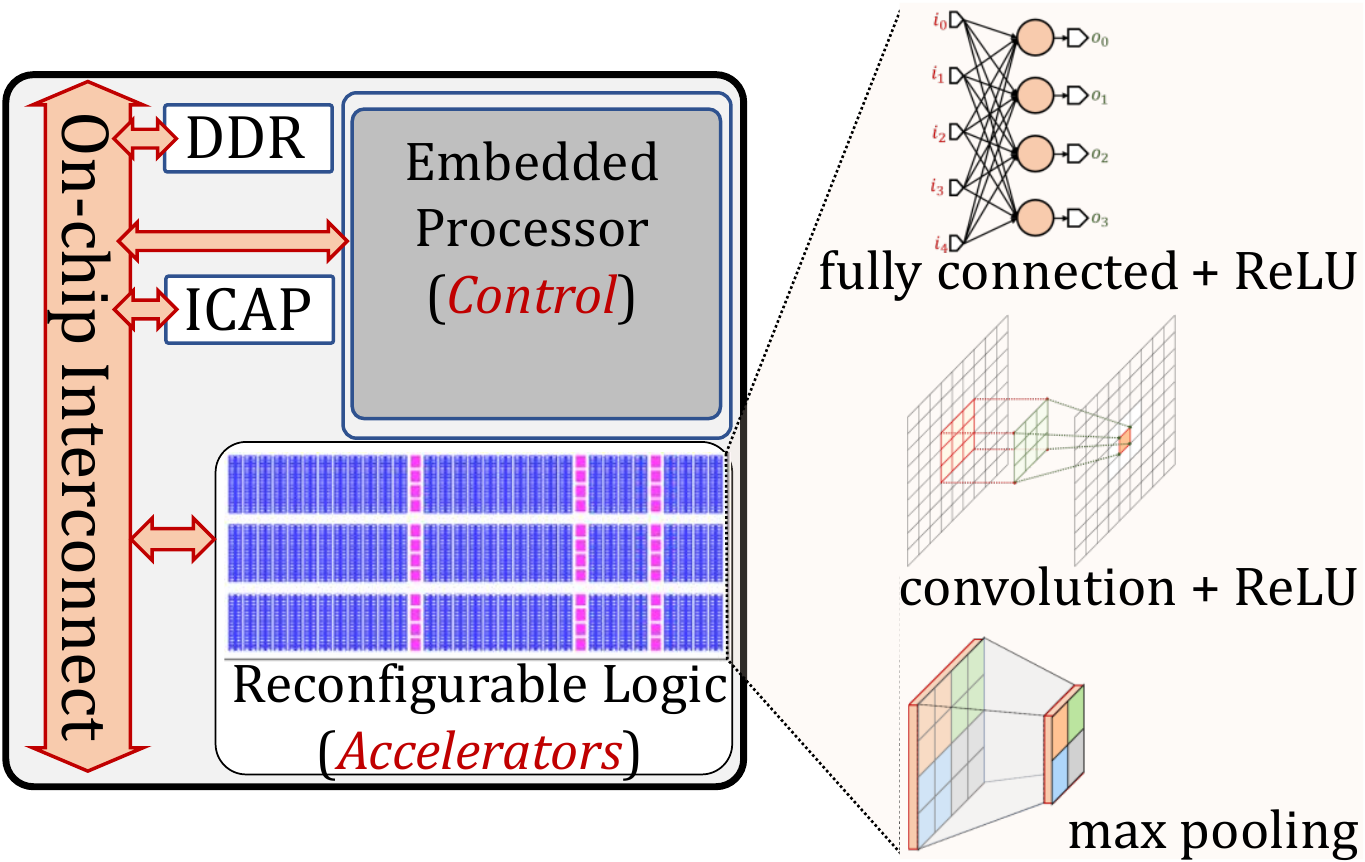}
	   } %\hspace{10pt}
	\caption{System Model}	
	\label{fig:systemModel}
\end{figure*}
\subsection{Application Model}
\label{subsec:appModel}
% \begin{figure}[t]
% 	\centering
% 	\scalebox{1}{\includegraphics[width=1 \columnwidth]{figures/appModel.pdf}}
% 	\caption{Application Model (VGG16~\cite{simonyan2014deep})}
% 	\label{fig:appModel}
% \end{figure}
\siva{The hardware designs proposed in our current work can be used for any arbitrary application that needs to communicate and/or store a large number of parameters. However, in this article, we limit our exploration to \gls{ann}s. \autoref{fig:systemModel}(a) shows one of the more widely used \gls{ann}---the VGG16~\cite{simonyan2014deep}---in research. As shown in the figure, VGG16 is composed of 16 layers of 4 different types---\textit{convolutional}, \textit{max pooling}, \textit{fully connected} and \textit{softmax}. Although we use the VGG16 as the application for evaluating our proposed methodology, the methods are applicable to any arbitrary \gls{ann} as most networks are composed of a subset of these types of layers. \autoref{fig:systemModel}(a) also shows the dimension of the parameters that are used in each of the layers. Using accelerators for inference usually involves communicating and storing these large number of trained parameters---138 million for VGG16. Consequently, the quantization methods used for the parameters can influence the corresponding storage and communication overheads. Similarly, given the large number of \gls{mac} operations involved in the inference of a single input---15.5 billion for VGG16---the speed and power dissipation of the \gls{mac} unit determines the throughput and energy consumption of \gls{ann} inference.} 

\subsection{Architecture Model}
\label{subsec:archModel}
% \begin{figure}[t]
% 	\centering
% 	\scalebox{0.98}{\includegraphics[width=1 \columnwidth]{figures/archModel.pdf}}
% 	\caption{Architecture Model}
% 	\label{fig:archModel}
% \end{figure}
\siva{
\autoref{fig:systemModel}(b) shows the architecture model used in this article. As shown in the figure, we assume an \gls{fpga}-based \gls{soc} as the hardware platform. It contains an embedded processor along with reconfigurable logic similar to the Zynq EPP~\cite{rajagopalan2011xilinx}. 
We assume that the accelerators for different types of layers of an \gls{ann} are executed on the reconfigurable logic and can implement the proposed hardware designs. For any accelerator, we assume that the parameters of the corresponding layer are fetched from the main memory through streaming interfaces with the on-chip AXI interconnect~\cite{logicore2017axi}. Similarly the input and output activations are transferred from and to the main memory using AXI streaming interfaces as well. Hardware platforms based on the Zynq EPP, such as the Ultra96-V2~\cite{ultra96}, are being widely marketed as edge processing devices for \gls{iot}.
}

% \caption{Application Model (VGG16~\cite{simonyan2014deep})
% \caption{Architecture Model}
% \subsection{Approximation Model}
% \label{subsec:approxModel}
\section{Design Methodology}
\label{sec:dsgnMeth}

\begin{figure}[t]
	\centering
% 	\scalebox{1}{\includegraphics[width=1 \columnwidth]{figures/Methodology.pdf}}
	\scalebox{1}{\includegraphics[width=0.99 \columnwidth]{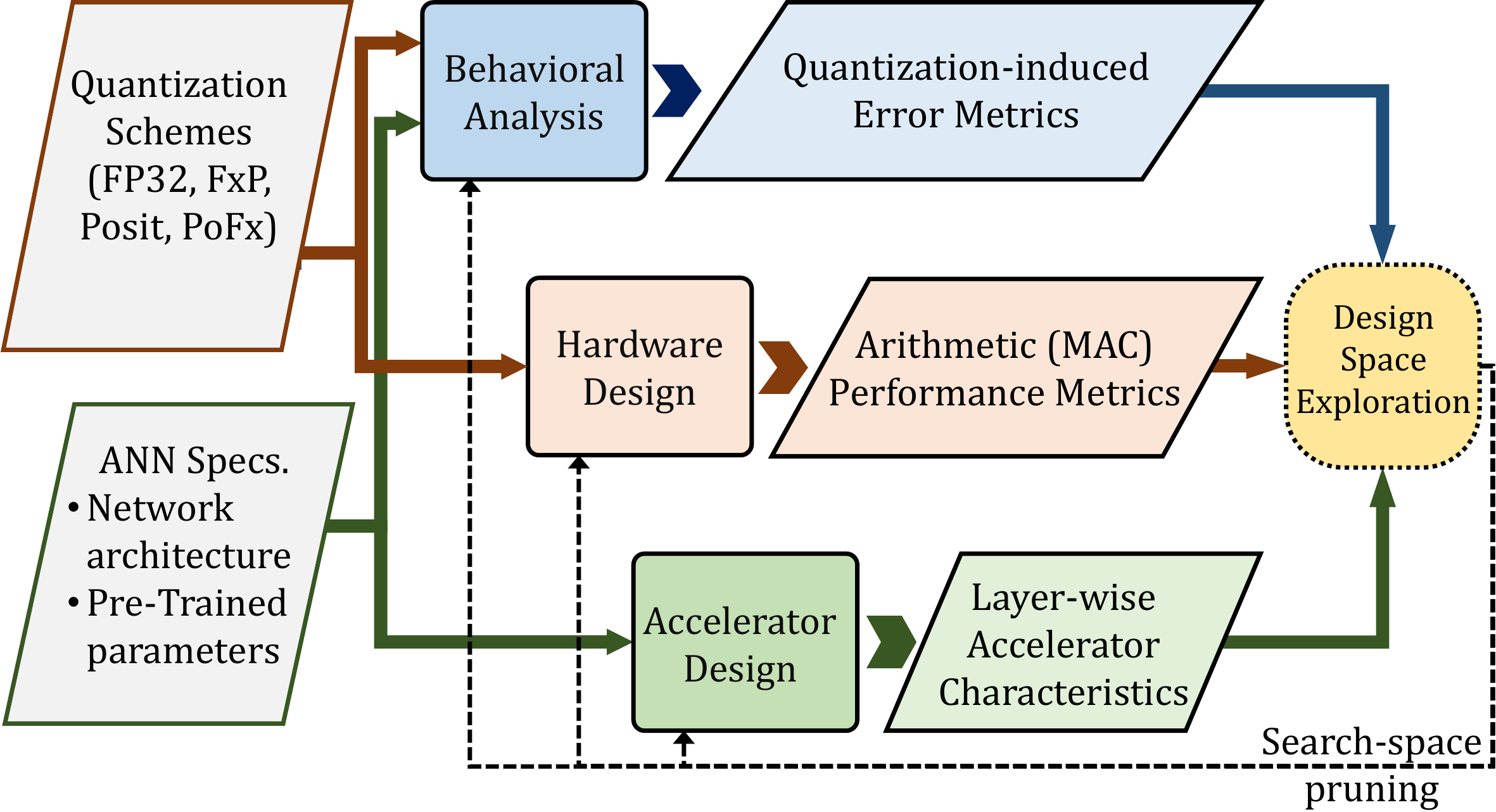}}
	\caption{Proposed design methodology}
	\label{fig:DsgnMeth}
\end{figure}

\siva{
The top-level view of ExPAN(N)D is shown in \autoref{fig:DsgnMeth}. The \textit{Hardware design} and characterization of the \gls{mac} units for various quantization schemes forms the central theme around which the other two methods---\textit{Behavioral analysis} and \textit{Accelerator design}---are implemented. \textit{Behavioral analysis} enables the estimation of quantization-induced errors in a given \gls{ann} using the proposed hardware designs. Similarly, \textit{Accelerator design} allows the designer to estimate the performance-resource trade-offs resulting from implementing various quantization schemes in an accelerator for a given layer of the \gls{ann}. The results from each of the three methods can be used to constraint the search space in the design of an efficient \gls{ann} using successive design space pruning. However, the implementation of an effective \gls{dse} methodology is beyond the scope of this article.
}
\subsection{Hardware Design}
\label{subsec:hwDesignMeth}
% \par{In this section we will look at hardware optimizations specific to \gls{pofx} conversion, develop the Normalized posit representation and implement a parameterizable compute efficient \gls{fxp} MAC which will operate on data extracted from a Posit representation based storage format.  }

% Elaborate on Normalized Posit Representation 
\subsubsection{Normalized Posit Representation}
\label{subsubsec:NormPoFx}
% MOTIVATION : Need for Normalized representation 
% DEPENDENCY : Posit N,ES is better than linear fixed quantization
\par{The Posit representation is inherently designed to encode numbers 
%spread across the number line from $+\infty$ to $-\infty$ 
in the range $(-\infty, \infty)$.
However, due to their tapered accuracy, numbers near $\pm1$ have better accuracy in comparison to extremely small or large numbers \cite{PositsPrimer}. Thus, low-precision Posit numbers perform better than an equivalent linear fixed-point representation during the quantization of normalized \gls{ann} weights.
%though the former was designed to store a wider range of numbers.
While processing normalized numbers, sub-optimal utilization of all possible Posit bit-patterns  leads to half of them being unused. This can translate to communication and storage overheads, as more than required bits are being transferred around. Similarly, a higher number of bits, than that required for storing the information, are processed during each computation. Hence, we propose \textit{normalized Posit}---an alternative representation based on Posits which preserves its encoding efficiency, hardware realization and tapered accuracy while doubling the usable bit patterns within the normalized range. This normalized Posit representation is a logical subset of Posits that is customized for the representation of normalized numbers. 
%To understand our thought process during its development, look at~
For example, \autoref{tab: PositLookup} shows all the possible bit-patterns and their equivalent real values for a Posit configuration of $N=4, ES=0$. The highlighted rows in the table show the bit-patterns which represent normalized numbers. It is evident that the two leading bits of the Posit representation are identical when the bit pattern denotes a normalized number; we leverage this finding to drop the leading Posit bit in our proposed normalized Posit representation.}

\par{This Posit representation helps us encode $N$-bit Posit functionality within the normalized range with $N-1$ bits. This leads to a reduction in storage requirement while still being able to reuse existing Posit arithmetic hardware by replicating the leading bit near the processing unit. However, existing hardware implementations are not optimized to perform normalized Posit-only arithmetic. Existing implementations do not take complete advantage of the benefits arising as a consequence of the potentially unidirectional nature of bit shifts required to extract normalized Posits. To this end, we propose a novel parameterized Posit-to-\gls{fxp} converter, \textit{\gls{pofx}}, that implements an optimized extraction for normalized Posit numbers.
%to highlight its low resource utilization.      
}

\begin{table}[t]
\caption{Posit(N=4, ES=0) to normalized Posit representation}
\centering
\def\arraystretch{1.0}
\resizebox{0.75 \columnwidth}{!}{
\begin{tabular}{|c|c|c|c|c|c|}
\hline
\textbf{Posit}         & \textbf{s}       & \textbf{k}      & \textbf{f}     & \textbf{Value}   & \textbf{ExPAN(N)D}       \\ \hline
\rowcolor{lightgray}
    \textbf{00}00 & 0 & -3 & 0   & 0  & \textbf{0}00   \\ \hline
\rowcolor{lightgray}
\textbf{00}01 & 0 & -2 & 0   & 0.25 & \textbf{0}01 \\ \hline
\rowcolor{lightgray}
\textbf{00}10 & 0 & -1 & 0   & 0.5 & \textbf{0}10  \\ \hline
\rowcolor{lightgray}
\textbf{00}11 & 0 & -1 & 0.5 & 0.75 & \textbf{0}11 \\ \hline
0100          & 0          & 0           & 0            & 1     & -         \\ \hline
0101          & 0          & 0           & 0.5          & 1.5     & -       \\ \hline
0110          & 0          & 1           & 0            & 2     & -         \\ \hline
0111          & 0          & 2           & 0            & 4       & -       \\ \hline
1000          & 1          & -3          & 0            & NaR   & -         \\ \hline
1001          & 1          & 2           & 0            & -4       & -      \\ \hline
1010          & 1          & 1           & 0            & -2    & -         \\ \hline
1011          & 1          & 0           & 0.5          & -1.5     & -      \\ \hline
\rowcolor{lightgray}
\textbf{11}00 & 1 & 0 & 0   & -1 & \textbf{1}00   \\ \hline
\rowcolor{lightgray}
\textbf{11}01 & 1 & -1 & 0.5 & -0.75 & \textbf{1}01\\ \hline
\rowcolor{lightgray}
\textbf{11}10 & 1 & -1 & 0   & -0.5 & \textbf{1}10  \\ \hline
\rowcolor{lightgray}
\textbf{11}11 & 1 & -2 & 0   & -0.25 & \textbf{1}11\\ \hline
\end{tabular}
}
\label{tab: PositLookup}
\end{table}

\subsubsection{\gls{pofx}: Normalized Posit to Fixed-Point Converter}
\label{subsubsec:pos2fxp}
% JUSTIFICATION : for the use of PoFx Converter
% DEPENDENCY : Fixed Point Operations are more compute efficient 
\par{
%Further, due to the lack of intuitive Posit operations, 
\begin{algorithm}[h]
\caption{\textit{Posit (N,ES) to \gls{fxp} (M,F)  }}
\label{alg:convPOSIT2Fxd}
% \begin{multicols}{2}
\small{
\begin{algorithmic}[1]
	\Require{$N,~ES,~M,~F$}
	\LeftComment{$N$:~Input Posit Bit Length}
	\LeftComment{$ES$:~Maximum Exponent Bit Length}
	\LeftComment{$M$:~\gls{fxp} Output Length}
	\LeftComment{$F$:~Fraction length in \gls{fxp} Output}
	\algrule[2pt]
	\HeadComment{\textbf{A1: Extract Sign Component to \gls{fxp} Output}}
	\State{$S   = POSIT~[N-1]$}
	\State{$MAG [F] = 1$} \Comment{Set Leading Bit}
	\HeadComment{\textbf{A2: Implement conditional Two's Complement}}
	\If{$POSIT[N-1] ==1$}
	    \State{$POSIT[N-2:0] = !~POSIT [N-2:0] + 1$}
	\EndIf
	\HeadComment{\textbf{A3: Implement Modified Leading Zero Detector}}
    \If{$POSIT[N-2] == 0$}  \label{algo1_mark_line1}
        \State{$P[N-2:0]= !~POSIT[N-2:0]$} 
    % \Else
        % \State{$P[N-2:0]= POSIT[N-2:0]$}
    \EndIf 
    \LeftComment{To avoid LOD by inversion of bit sequence}
    \label{algo1_mark_line2}
    % \LeftComment{Implement Modified LZD Detector}
    \State{$LZD [N-2] = P[N-2]$} \Comment{Always $1$ }
    \For{$(i=N-3;~i>=0;~i--)$}
        \State{$LZD[i] = LZD [i+1] ~\&~ P [i]$} 
    \EndFor
    \algrule[2pt]
  	\HeadComment{\textbf{B1: Evaluate Regime Value }}
  \State{$V = ~\#~1's~In~LZD$}
 % \LeftComment{XOR continuous 1s sequence from LHS based on place value of count to be extracted}
  \If{$POSIT[N-2]==0$}
  \State{$K = - V$}
  \Else
  \State{$K = V-1 $}
  \EndIf
    \HeadComment{\textbf{B2: Extract Exponent and Fraction Fields}}
    \State{$E:[e_{ES-1},...,e_1,e_0] = \vec{0}$}
    \For {$(i=N-4;~i>=0;~i--)$}
        \State{$EXT [i] =~!(LZD [i+1] ~|~ LZD [i])$}
    \EndFor
    
    \State{$ST [N-4] = EXT[N-4]$}
    \For {$(i=N-5;~i>=0;~i=i--)$}
        \State{$ST[i]=EXT[i+1] \oplus EXT[i]$}
    \EndFor
    \LeftComment{To Generate Silhouette ST for Extraction}
    % \LeftComment{Extract Exponent and Fractional Fields based on silhouette}
    \State{$switch = N-4-ES$}
    \For {$(i=0;~i<=N-4;~i++)$}
        \State{$set = 0$}
        \For{$(j=0;~j<=i;~j++)$}
        \State{$set=set | (ST[N-4-i+j] ~\&~ POSIT [j]) $}
        \EndFor
        \If{$i<=switch$}
            \State{$MAG[F-1- switch+i] = set$} 
        \Else
          \State{$E[i-1-switch] = set$}
        \EndIf
    \EndFor
    \algrule[2pt]
    \HeadComment{\textbf{C: Shift Calculation}}
    \State{$SHIFT = 2^{ES}*K+E$ }
    \LeftComment{\textcolor{black}{$SHIFT$ register size = $\left \lceil{log_2(M)}\right \rceil$}}
    \algrule[2pt]
    \HeadComment{\textbf{D: Bit Shift Implementation}}
    \State{MAG $<<$ SHIFT}\Comment{-ve Value = Right Shift}
    % \LeftComment{Combinational MUX based Implementation}
    \algrule[2pt]
    \HeadComment{\textbf{E: Sign Magnitude to Two's Complement Block}}
% \algstore{genPoFX_1}
% HAVING TO SPLIT AS characters overflow and algorithm is not displayed
\end{algorithmic}
}
% \end{multicols}
\end{algorithm}
\begin{figure}[t]
	\centering
	\scalebox{1}{\includegraphics[width=0.999 \columnwidth]{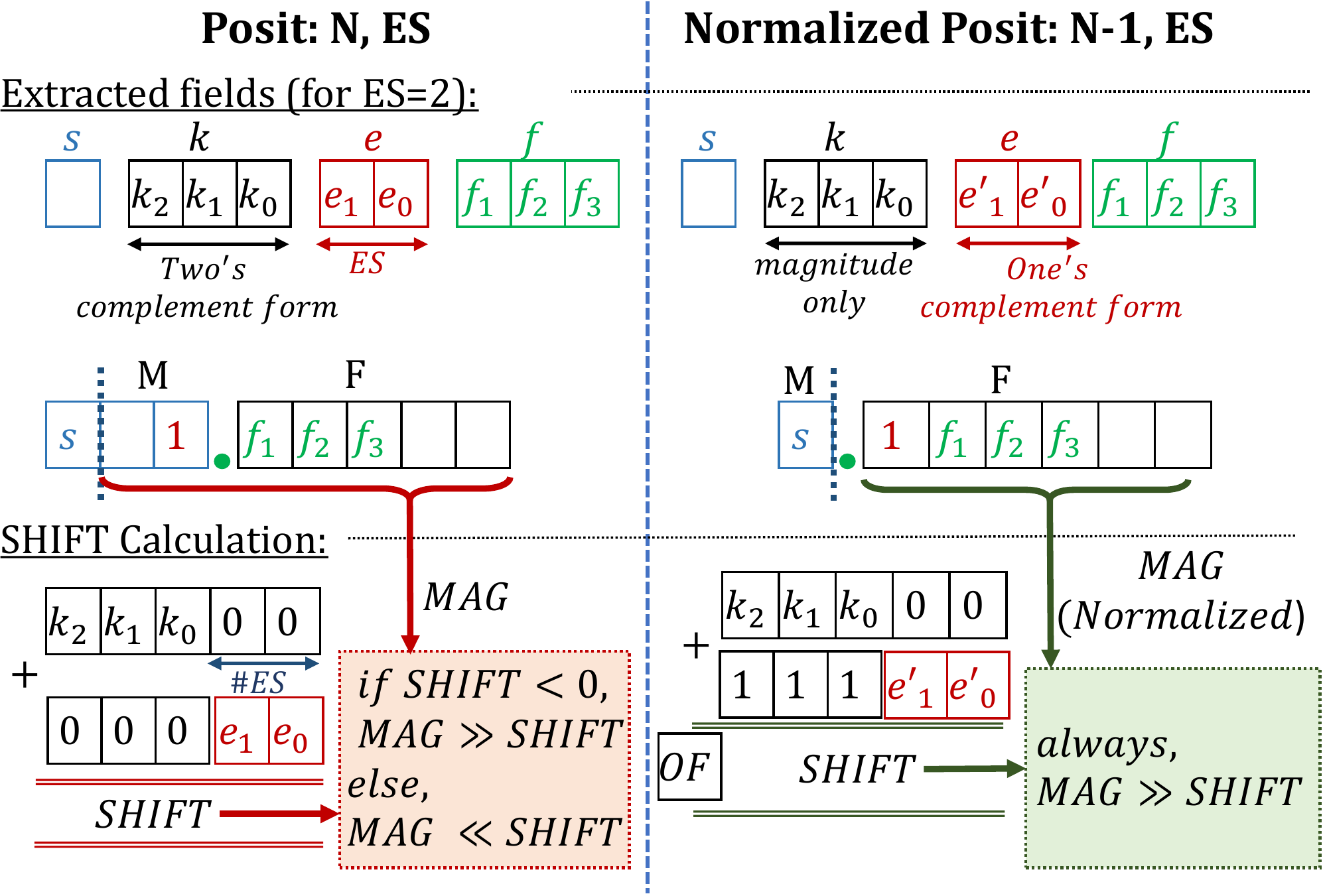}}
	\caption{Comparing shifting operations in Posit and normalized Posit representations}
	\label{fig:methHW}
\end{figure}
Most Posit-based computations require a decode stage to extract the value before arithmetic operations. Currently, Posit-based arithmetic relies  heavily on extraction of Posit numbers to a floating point like representation before operating on them, which leads to increased resource utilization. Instead, we design a novel resource-efficient parameterized \gls{pofx} converter which facilitates the use of existing resource-efficient \gls{fxp} arithmetic optimizations.}

\par{The proposed~\gls{pofx} conversion algorithm is an intuitive technique which effectively converts a Posit to a \gls{fxp} number developed from the way Posits are decoded. 
Taking the example of the Posit($N=4, ES=0$) bit-patterns in~\autoref{tab: PositLookup}, we illustrate how we use minimal resources during conversion to a \gls{fxp} format after Posit field-extraction by working at the bit level. 
The key to developing this algorithm rests on recognizing that the fraction field extracted from the Posit representation is identical to that required in the \gls{fxp} output. Thus, once the data in the Posit bit pattern is extracted into its components $s$, $k$, $e$ and $f$; the posit value 
%is defined by~Eq.~\eqref{eq2} which 
only requires us to set a bit and store the extracted fraction bits to its right followed by a final bit shift determined by the equation $2^{ES}*k+e$. This equation can be implemented by adding the $e$ value to the bit-sequence obtained by appending $k$ to $ES$ number of zero bits as illustrated in~\autoref{fig:methHW}. The sign-bit along with the shifted bit sequence gives us the Posit representation in sign-magnitude \gls{fxp} format.} 

\par{\gls{pofx} conversion algorithm for manipulation at the bit-level which converts Posit representation, Posit($N,ES$) to fixed-point representation \gls{fxp}($M,F$) is summarized in Algorithm~\ref{alg:convPOSIT2Fxd}. \textit{Stage A} (comprising of \textit{Stages} \textit{A1}, \textit{A2} and \textit{A3}) stores the sign bit and prepares the Posit bits for subsequent extraction. \textit{Stage B1} implements 
%evaluates the value of V using 
an optimized algorithm to evaluate the number of contiguous $1$'s. \textit{Stage B2} performs bit manipulations to ascertain location of exponent and fraction bits and subsequently extracts them. All the loop indices are carefully evaluated based on the constraints arising from the Posit representation. \textit{Stage C} performs the bit shift calculation and \textit{Stage D} implements the bit shifts. The final \textit{Stage E} is optional depending on the application and involves the conversion of sign-magnitude format to two's complement.}

\par{The proposed~\gls{pofx} can be adapted to perform normalized~\gls{pofx} conversion which leads to lower resource utilization and improved performance in~\gls{ann}s. This is primarily due to the drastic simplification of \textit{Stage C} and \textit{Stage D} as in this case the shifts are unidirectional, that is towards the right, making the value smaller. For normalized Posits we set $F=M-1$ as all but one bit would be used for the sign. The first bit is replicated within \textit{Stage A} followed by simplified extraction in \textit{Stage B1} as the regime bit would always begin with zero thus $K$ would store only magnitude. We use an optimized algorithm to evaluate the modified shift equation $2^{ES}*K-E$ in \textit{Stage C} which is illustrated in~\autoref{fig:methHW}.
\suresh{We store $1$ after the assumed decimal point in normalized \gls{pofx} extraction and thus always need to right shift one time less. This is achieved implicitly by adding the one's complement of $E$ to $2^{ES}*K$; further we will set the overflow flag~(OF) if the required number of shifts exceeds the width of the $MAG$ field. \textit{Stage D} is replaced with a standalone right bit-shifter while \textit{Stage E} remains unchanged.}}

\par{The five stages in our proposed design can be pipelined to further improve the throughput of the \gls{pofx} converter as there are no feedback paths between the stages, thus eliminating data hazards. We note that though normalized Posit representation can represent the value $-1$, the normalized~\gls{pofx} cannot extract the same due to its implicit storage in sign-magnitude format. For the rest of the article, the term \gls{pofx} will be used to denote the normalized \gls{pofx}. Similarly, Posit($N,ES$) and Posit($N-1,ES$) will be used to denote Posit and normalized Posit respectively.}

\subsubsection{MAC Unit with PoFx Converter}
\label{subsubsec:macPoFx}
% JUSTIFICATION : Why PoFx based MAC ?
\par
\begin{figure}[t]
	\centering
	\scalebox{1}{\includegraphics[width=0.9 \columnwidth]{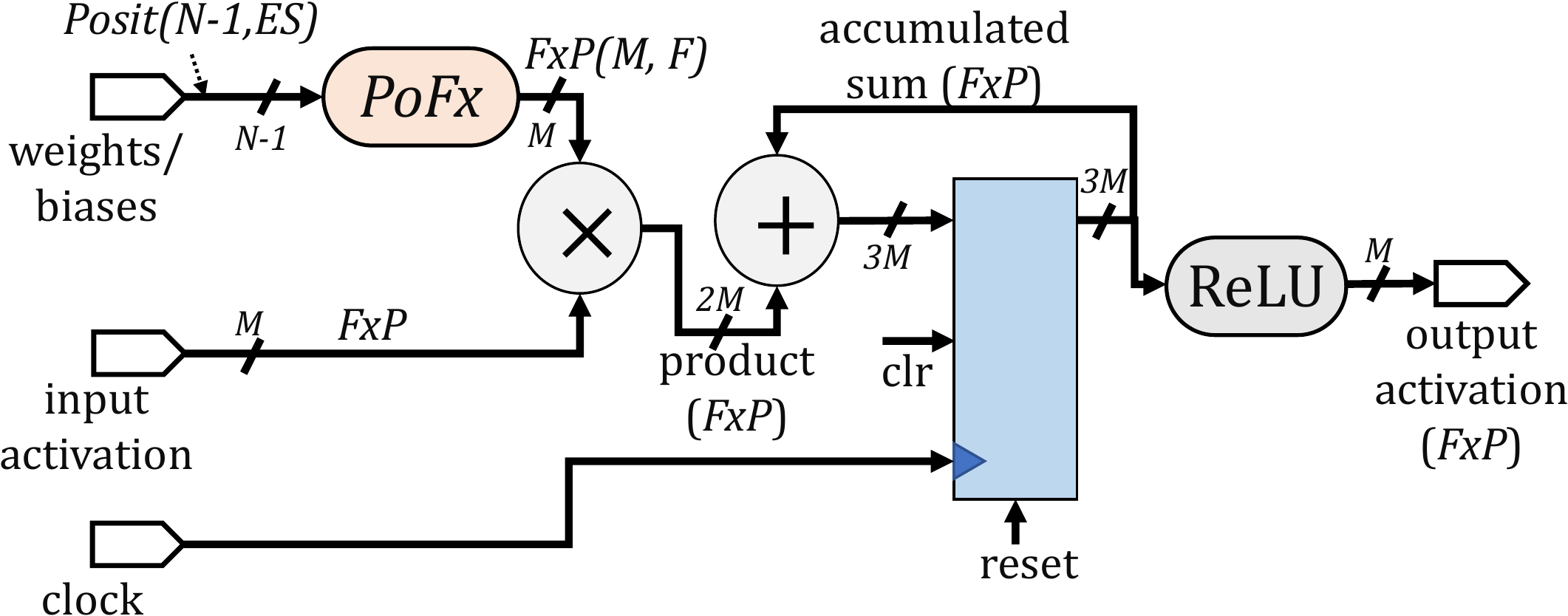}}
	\caption{\gls{mac} unit (with ReLU activation) using \gls{pofx}}
	\label{fig:methMAC}
\end{figure}

{
\siva{The \gls{pofx} converter can be used for any application that can benefit from storing a large number of parameters efficiently. As a special case for \gls{ann}s, we integrate the normalized \gls{pofx} into \gls{mac} units to facilitate the use of our proposed optimizations for improving low-precision \gls{ann} inference. \autoref{fig:methMAC} shows the schematic of a parameterized \gls{pofx} converter based \gls{mac} along with \textit{ReLU} activation function. As shown in the figure, the weights/biases are assumed to be stored/communicated as Posit($N-1,ES$) numbers. These values are then converted to their corresponding $M$-bit~\gls{fxp} representations and multiplied with the $M$-bit input activation values.} \su{To accommodate the overflows resulting from the accumulation of a large number of $2M$-bit values, we propose to use a $3M$-bit adder. After accumulating all the values, for a single node in a layer of an ANN, we pass the $3M$-bit result to the activation function.}
}
\par{\siva{It can be noted that the \gls{pofx}-based \gls{mac} unit allows the designer to represent the weights/biases with a fewer number of bits while still being able to implement different kinds of \gls{fxp}-based arithmetic optimizations, such as precision-scaling, approximations, etc. However, the effect of such a reduced bit representation on the \gls{ann}'s behavior, and the corresponding reduction in the compute and communication/storage overheads of the associated accelerators for each layer needs to be estimated. The next two sub-sections provide the details of our contributions regarding these aspects of designing a \gls{pofx}-based \gls{ann}.}
}
%  However, the effect of such a reduced-bit representation on the \gls{ann} accuracy parameters needs to be evaluated to determine the appropriate number representation for each layer of the network. Similarly, the trade-offs between reduced communication/storage due to reduced bits and the overheads of the \gls{pofx} needs to be evaluated for the design of accelerators for each layer. The next two sub-sections provide the details of our contribution regarding these two aspects of designing a \gls{pofx}-based \gls{ann}.
\subsection{Behavioral Analysis}
\label{subsec:behavMeth}
\begin{figure}[t]
	\centering
	\scalebox{1}{\includegraphics[width=1 \columnwidth]{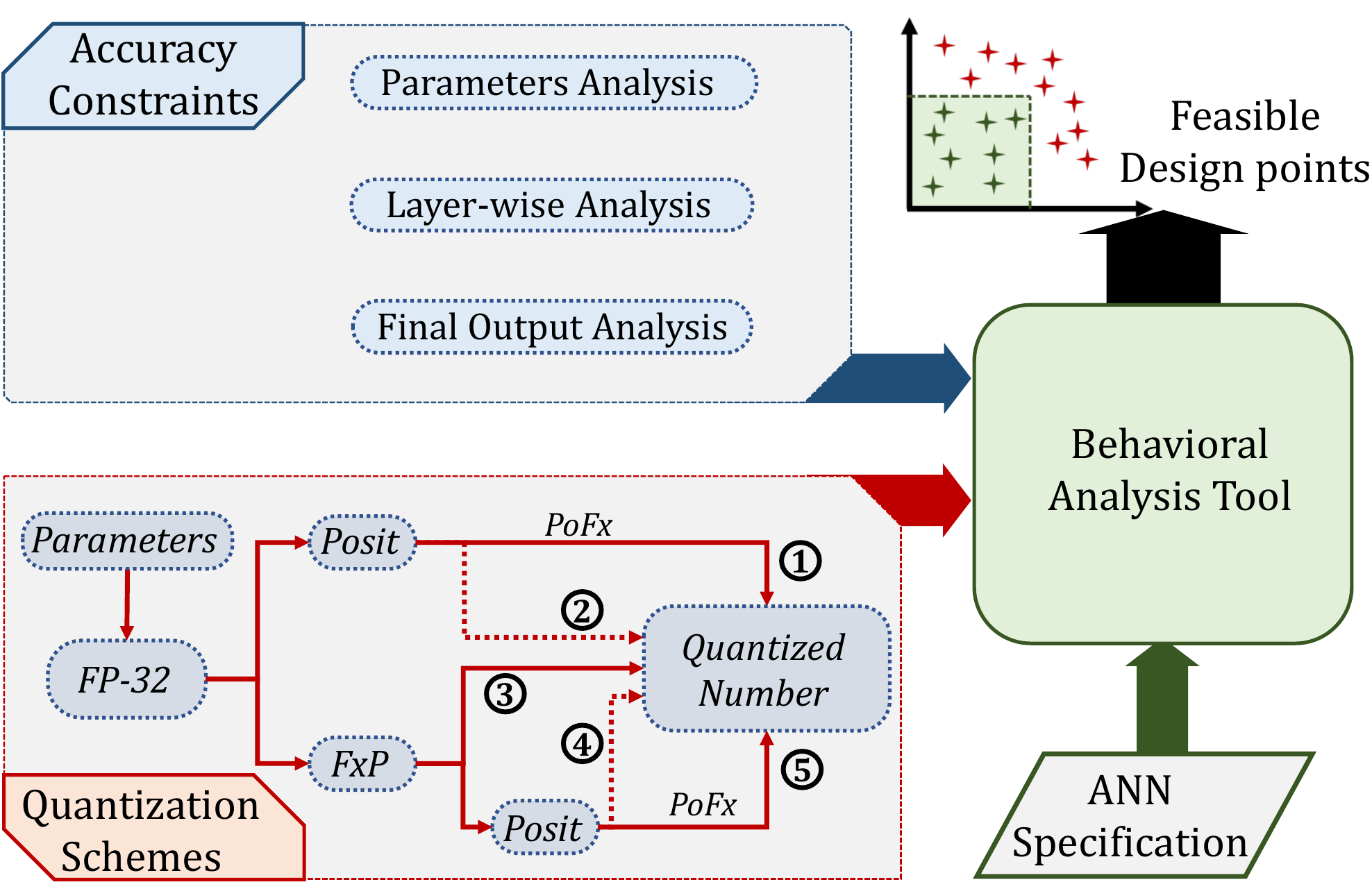}}
	\caption{Framework for behavioral analysis}
	\label{fig:behavMeth}
\end{figure}
\salim{To evaluate the impact of various quantization schemes on the output accuracy of a DNN, we have utilized TensorFlow for implementing a high-level behavioral framework, as shown in~\autoref{fig:behavMeth}. 
%To satisfy the output accuracy constraints of a DNN, 
It evaluates each quantization scheme's efficacy by analyzing its impact on (a) accuracy of the quantized parameters, (b) errors generated in the output activations of each layer due to quantized parameters, and (c) the accuracy of the final output of the quantized DNN compared to \gls{flp}-based output. The multi-level analysis of the quantization induced errors helps in the early elimination of the infeasible configurations. For this work, we have considered various configurations of the \gls{fxp}-based linear quantization and Posit-based representations, denoted by the \emph{Quantization Schemes} in~\autoref{fig:behavMeth}. However, our proposed framework is generic and allows the integration of other types of quantization schemes. \su{Our proposed workflow performs a thorough analysis of the inter-conversions of these schemes to evaluate the impact of the available quantization step sizes and the dynamic ranges offered by each scheme. For example, the \gls{fxp}-based representation of an \gls{flp}-based parameter can be achieved as shown by \circled{1}, \circled{3} and \circled{5} paths in the figure.}
% \begin{itemize}
%     \item \gls{flp} $\rightarrow$ $\underset{Storage}{FxP}$ \gls{fxp}$_{_{{storage}}}$ 
%     \item \gls{flp} $\rightarrow$ Posit $\rightarrow$ \gls{fxp} conversion using \gls{pofx}
%     \item \gls{flp} $\rightarrow$ \gls{fxp} $\rightarrow$ Posit $\rightarrow$ \gls{fxp} conversion using \gls{pofx}
% \end{itemize}
}
%\sam{To emphasize the importance of the interplay of these quantization schemes, \autoref{fig:motiv1} has highlighted the \gls{fxp}-based quantized weights of the Conv2\_1 layer of a pre-trained VGG16. Each configuration has a different impact on quantization-induced errors. }
\noindent\su{As shown by the classification accuracy results in Section~\ref{sec:expRes}, the utilization of each of these schemes has a distinct impact on the final output accuracy.}
\salim{%The \emph{Computation Mode} in~\autoref{fig:behavMeth} allows different storage options for quantized parameters. For example, an FxP quantized parameter can be stored as either an FxP number or as a Posit number. 
After providing the description of an \gls{ann} and the various quantization schemes, the proposed framework provides quantization configurations fulfilling the desired accuracy constraints. These selected configurations are then used by our proposed \emph{Accelerator Design} tool flow to compute their respective performance metrics.
}
% \begin{figure}[t] 
%   \centering
%   	\subfloat[Floating-point-32\label{sub_fig1}]{%
%       \includegraphics[width=0.16\textwidth]{figures/posit_fxp_trimmed.pdf}}
%     \hspace*{0.2em}
%     \subfloat[Posit\label{sub_fig2}]{%
%         \includegraphics[width=0.16\textwidth]{figures/fxp_trimmed.pdf}}
% \hspace*{0.2em}
%     \subfloat[Linear \label{sub_fig2}]{%
%         \includegraphics[width=0.155\textwidth]{figures/fxp_posit_fxp_trimmed.pdf}}
%   \caption{Distribution of 8-bit \gls{fxp} quantized weights of Conv2\_1 layer of VGG16~\cite{simonyan2014deep}. (a) Posit (8, 2) to \gls{fxp}: average absolute error=0.003  (b) \gls{fxp}: average absolute error=0.002 (c) \gls{fxp} to Posit~(8,2) to \gls{fxp}: average absolute error=0.002.}
%   \label{fig:motiv_behav} 
% \end{figure}

% \begin{figure}[t]
% 	\centering
% 	\subfloat[Framework for behavioral analysis \label{fig:behavMeth}]{
% 			\includegraphics[width=0.51 \columnwidth]{figures/methBehav_2.pdf}
%         	} %\hspace{1pt}
% 	\subfloat[HLS-based Accelerator Design \label{fig:hlsMeth}]{
%     	\includegraphics[width=0.48 \columnwidth]{figures/methHLS_2.pdf}
% 	   } %\hspace{10pt}
% 	\caption{Behavioral Analysis and Accelerator Design}	
% 	\label{fig:behavAcc}
% \end{figure}
\subsection{Accelerator Design}
\label{subsec:rsrcMeth}
%\lipsum[10]
% \begin{figure}[t]
% 	\centering
% 	\scalebox{1}{\includegraphics[width=1 \columnwidth]{figures/methHLS_2.pdf}}
% 	\caption{HLS-based Accelerator Design}
% 	\label{fig:hlsMeth}
% \end{figure}
\par
\begin{figure}[t]
	\centering
	\scalebox{1}{\includegraphics[width=1 \columnwidth]{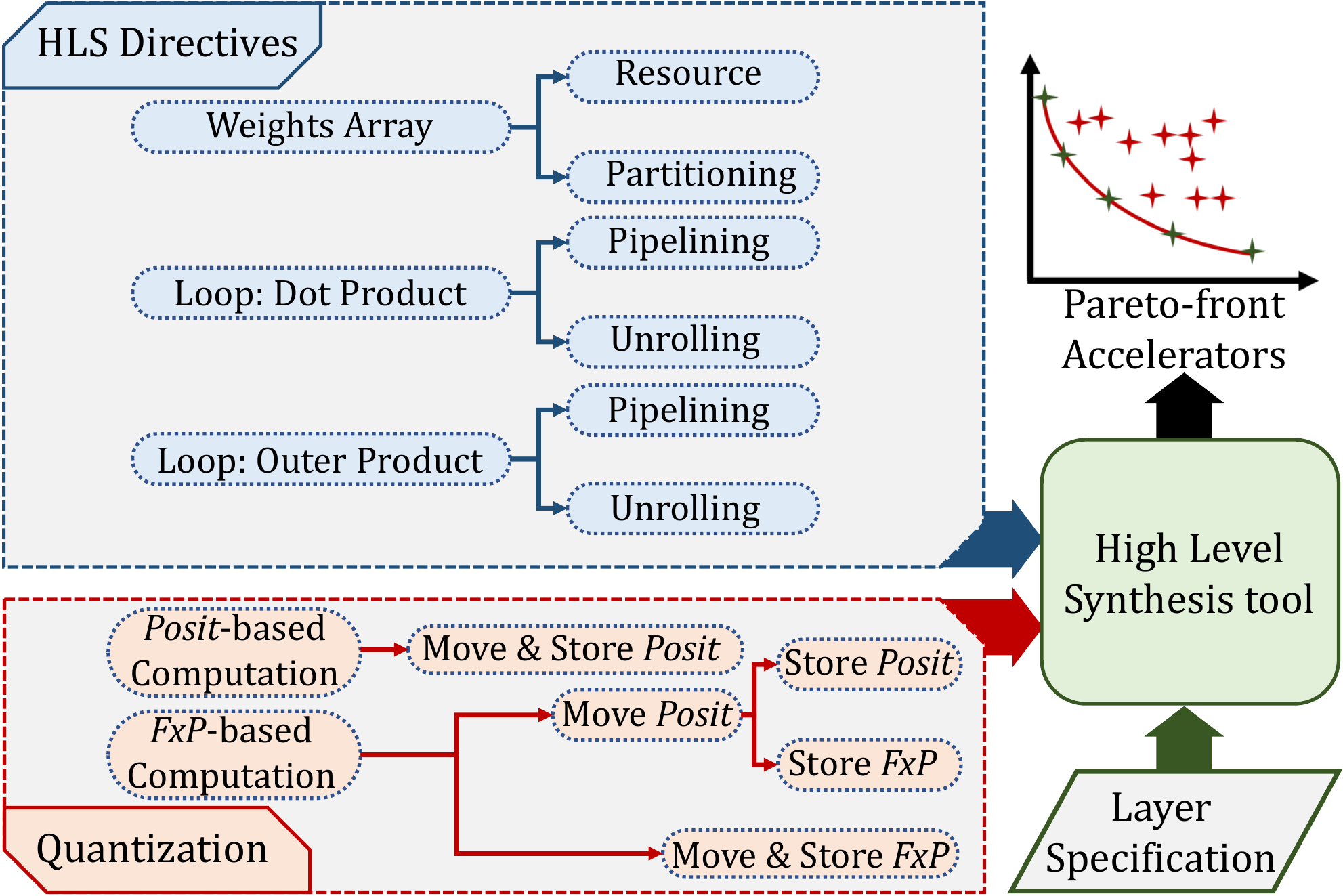}}
	\caption{HLS-based Accelerator Design}
	\label{fig:hlsMeth}
\end{figure}

{
\siva{
%Using the \gls{pofx}-based \gls{mac} unit for the accelerators of an \gls{ann} can result in reduced communication and storage overheads due to the reduced bit representation of normalized Posits. However, the \gls{pofx} converter introduces additional resource and timing overheads compared to \gls{fxp}-only computation schemes. Therefore, the associated trade-offs between computation and communication/storage needs to be estimated. \autoref{fig:hlsMeth} shows the \gls{hls}-based design flow used for this estimation.
\su{ The HLS-based design flow, shown in~\autoref{fig:hlsMeth}, is used for evaluating the associated trade-offs between computation overhead and communication/storage gains offered by the \gls{pofx}-based \gls{mac} units.}
The design choices tree originating from \gls{hls} directives shows the various degrees of freedom (not exhaustive) associated with the design of an accelerator for a fully-connected layer. We assume a weight-stationary~\cite{eyeriss2016} design, where a set of weights for a subset of the artificial neurons in the layer are transferred once to the hardware accelerator. Subsequently, each input activation vector is transferred and the corresponding output activation of each neuron is computed. Therefore, the computation of each output activation vector can be seen as the multiplication of a matrix (\textit{weights}) by a vector (\textit{input activations}). Consequently, \gls{hls} directives of \textit{pipelining} and \textit{loop unrolling} can be applied to the computation of the \textit{Dot Product} (evaluation of the output activation of each node) and the \textit{Outer Product} (evaluation of all output activations) for obtaining designs with varying performance and resource utilization. Similarly, the type of resources allocated for the weights matrix, BRAMs or LUTRAMs, and the associated array-partitioning choices can affect the accelerator characteristics.}
}
\par{
\siva{The design decisions associated with the quantization schemes are integrated into the \gls{hls}-based flow. The computation mode, Posit- or \gls{fxp}-based, and the associated bit-widths impact the accelerator performance considerably. The proposed accelerator allows the designer to send and store the weights in Posit($N-1,ES$) or \gls{fxp} format. If the weights are moved and stored as Posit($N-1,ES$), the \gls{mac} units need to have the \gls{pofx} unit integrated into it (similar to~\autoref{fig:methMAC}). 
%However, if the weights are moved as Posit($N-1,ES$) and stored as \gls{fxp}, the \gls{pofx} is used only during transferring the weights and need not have to be converted during each \gls{mac} operation.
\su{However, if the weights are moved as Posit($N-1,ES$) and stored as \gls{fxp} (using \gls{pofx}), the MAC units do not require the run-time conversion during each computation. %integration of the \gls{mac}.
}
However, this approach increases the storage requirements compared to storing as Posit($N-1,ES$). It must be noted that the joint exploration across \gls{hls} directives and quantization schemes is necessary for a good estimation of accelerator characteristics. Performance improvement using \gls{hls} directives usually involves replicating compute and memory resources which are in turn dependent upon the choices related to the quantization schemes.}
}

\section{Experiments and Results}
\label{sec:expRes}
\subsection{Experiment Setup} 
\label{subsec:expSetup}
\par{
\siva{The proposed \gls{pofx} converter and the associated computer arithmetic blocks were implemented using Verilog HDL. Python-based scripts were used for automating the generation of the parameterized designs. SmallPosit~HDL~\cite{SmallPositHDL} was used for generating the Posit-based arithmetic designs. The hardware designs were characterized using Xilinx Vivado Design Suite. For the calculation of the dynamic power of all implementations, Vivado Simulator and Power Analyzer tools have been utilized. All designs have been implemented on Xilinx Zynq UltraScale+ MPSoC (xczu3eg-sbva484-1-e device). The behavioral analysis was achieved using Python-based implementations and used TensorFlow~\cite{abadi2016tensorflow} for estimation of various quantization induced error metrics. Xilinx Vivado HLS 18.3 was used as the High-level Synthesis tool for accelerator design. While the results for the behavioral analysis correspond to the experiments using VGG16 as the test application, all the proposed methods can be used for any arbitrary application.}
% \suresh{Cite SmallPosit HDL ~\cite{SmallPositHDL}}
}

\subsection{Hardware Design}
\label{subsec:expHWDesign}

\subsubsection{Normalized \gls{pofx}}
\begin{figure}[t]
	\centering
	\scalebox{1}{\includegraphics[width=0.99 \columnwidth]{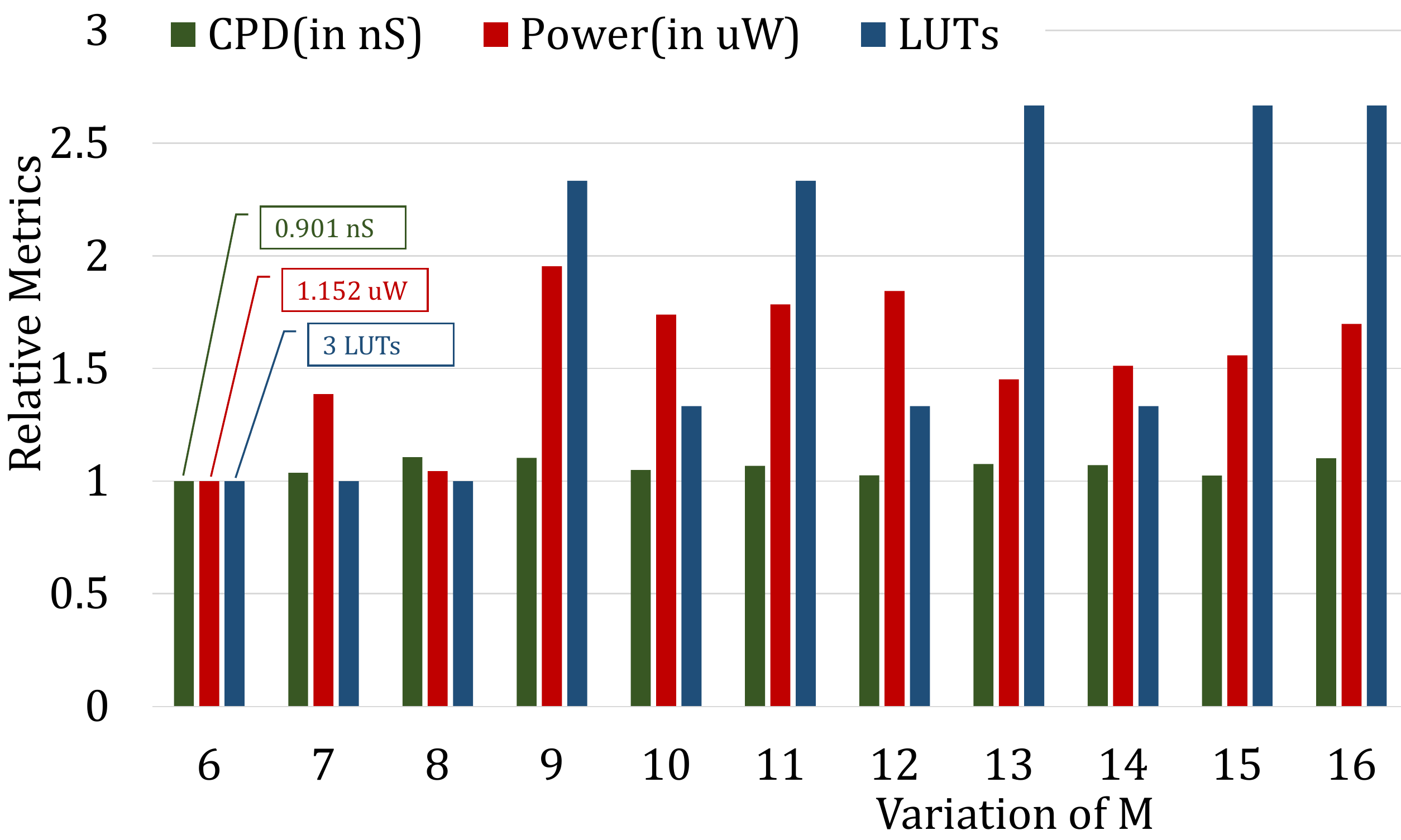}}
	\caption{\su{Performance metrics of \gls{pofx} with varying $M$ for Posit($N-1=5, ES=1$) for $2^N$ random input combinations}}
	\label{fig:exp_pofx_varM}
\end{figure}

\su{We analyze the impact of varying output bit-width ($M$) of \gls{pofx} converter on the overall performance of \gls{pofx} for a given configuration of Posit. \autoref{fig:exp_pofx_varM} presents the results of the analysis for Posit ($N-1=5$, $ES=1$) configuration\footnote{\su{Similar results are obtained for other Posit configurations.}}. The variation in $M$, for a \textit{fixed} Posit configuration, has an insignificant impact on the converter's CPD. For a specific value of $\left \lceil{log_2(M)}\right \rceil$, the overall LUT utilization also remains relatively unchanged\footnote{\su{As described in Algorithm~\ref{alg:convPOSIT2Fxd}, the $\left \lceil{log_2(M)}\right \rceil$ is used to calculate the size of the \textit{shift} register for computing the corresponding \gls{fxp} value.}}. For example, the total number of utilized LUTs by \gls{pofx} for $M=9$ is approximately $2.3$ times the total number of utilized LUTs for $M=8$. The total number of utilized LUTs also directly affects the dynamic power consumption of the \gls{pofx}. The minor variations in the Power metric of the \gls{pofx} is a result of the optimizations performed by the synthesis tool. Compared to resource utilization of Posit-based arithmetic units (discussed in the following sections), the \gls{pofx} has an insignificant contribution to the overall resource utilization of \gls{fxp}-based arithmetic units.}

%\par{\suresh{We first analyse the resource utilization and hardware performance of our proposed \gls{pofx} converter design. \autoref{fig:exp_pofx_varM} shows the variation of critical path delay, power dissipation and LUT utilization for increasing bit-width ($M$) of the \gls{fxp} output. It can be noted that the critical path delay remains remains relatively unchanged. Thsis is expected as the value of $ES$ is constant for all the designs shown in the figure.
%since Posit($N-1=5,ES=1$) is constant throughout. 
%As a result it requires identical hardware until the end of the extraction stage. Further, the penultimate stage which implements the bit shifts have similar critical path delays across variations in $M$. 
%The variation in the Power metrics are a result of the way the design is optimized by the tool and the effective number of bits which need to be switched depending on the input. The steep rise in LUT utilization from $M=8$ to $M=9$ is the result of switching from using $3$ bits to $4$ bits as per $log_{2}(M)$ to calculate shift values. The low resource utilization for standalone \gls{pofx} conversion highlights its effectiveness.}}

\begin{figure}
\centering
% \begin{minipage}{.49\textwidth}
%   \centering
%   \includegraphics[width=1 \columnwidth]{figures/exp_pofx_varM.pdf}
% 	\caption{Performance metrics of \gls{pofx} with varying $M$ for Posit($N-1=5, ES=1$)}
% 	\label{fig:exp_pofx_varM}
% \end{minipage}%
% \hspace*{0.2em}
% \begin{minipage}{.5\textwidth}
  \centering
   	\subfloat[CPD\label{exp_pofx_varNES_cpd}]{%
       \includegraphics[width=0.33 \columnwidth]{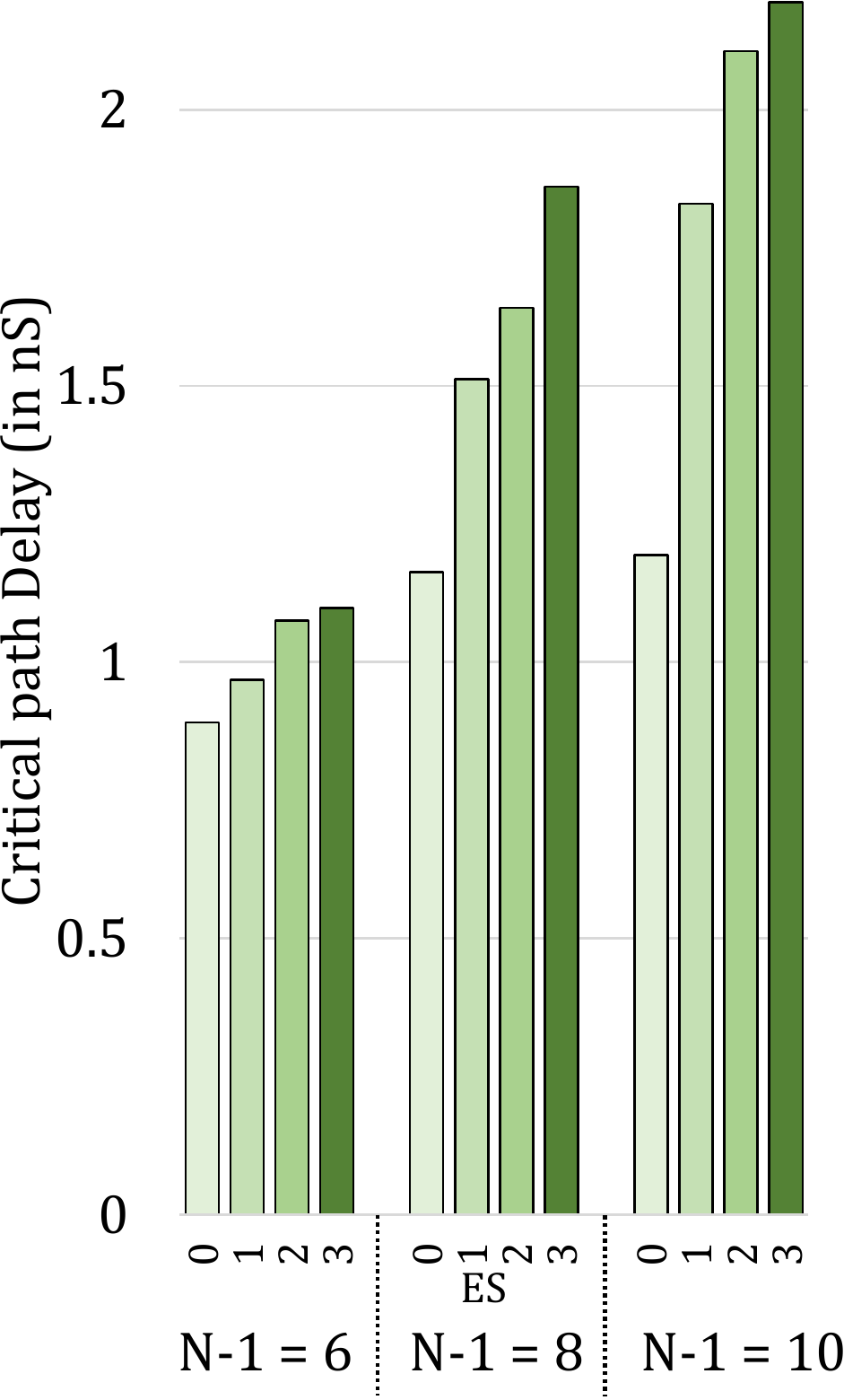}}
    % \hspace*{0.2em}
    \subfloat[Power\label{exp_pofx_varNES_pow}]{%
        \includegraphics[width=0.33 \columnwidth]{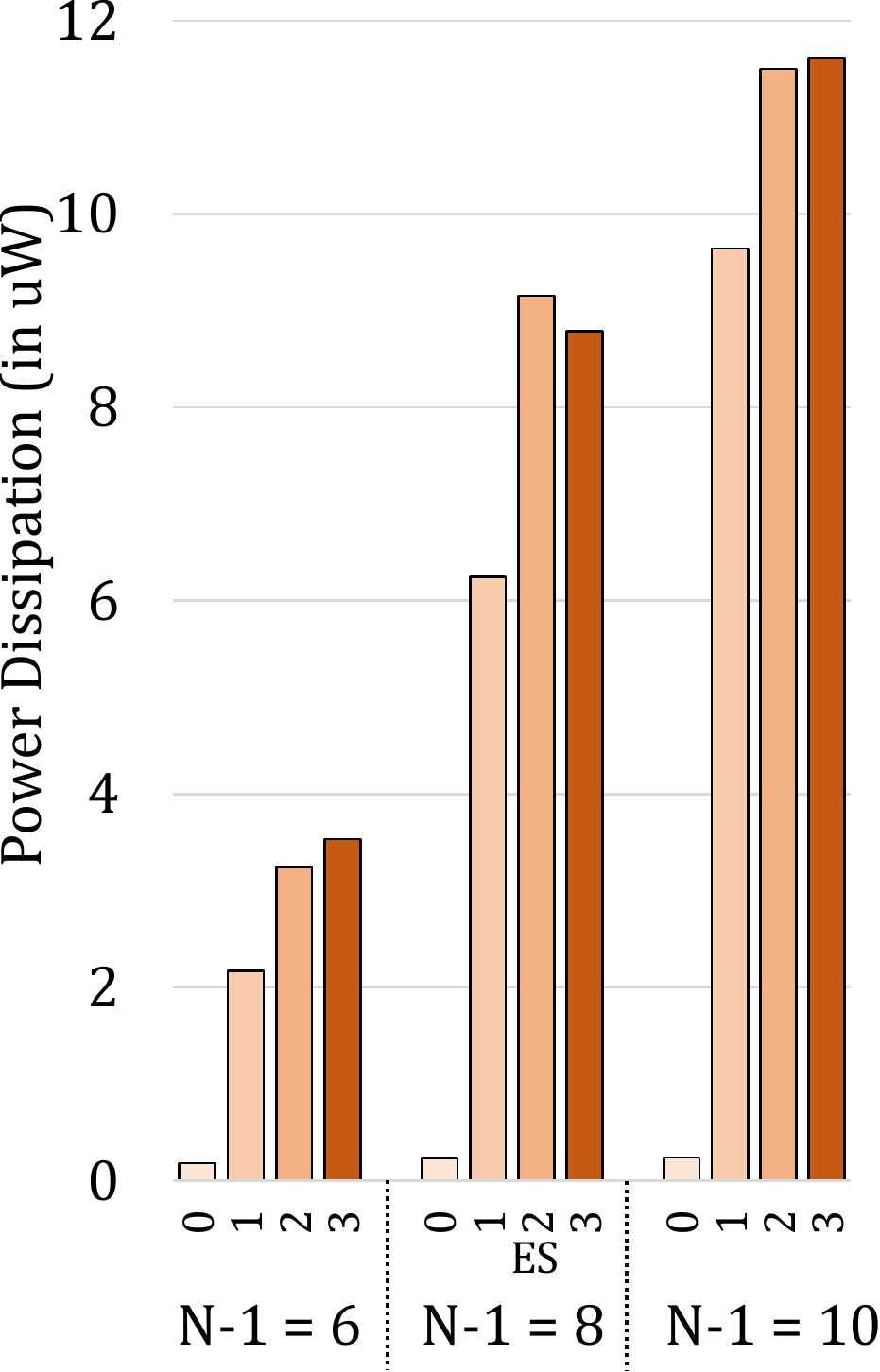}}
    % \hspace*{0.2em}
    \subfloat[Resources \label{exp_pofx_varNES_lut}]{%
        \includegraphics[width=0.33 \columnwidth]{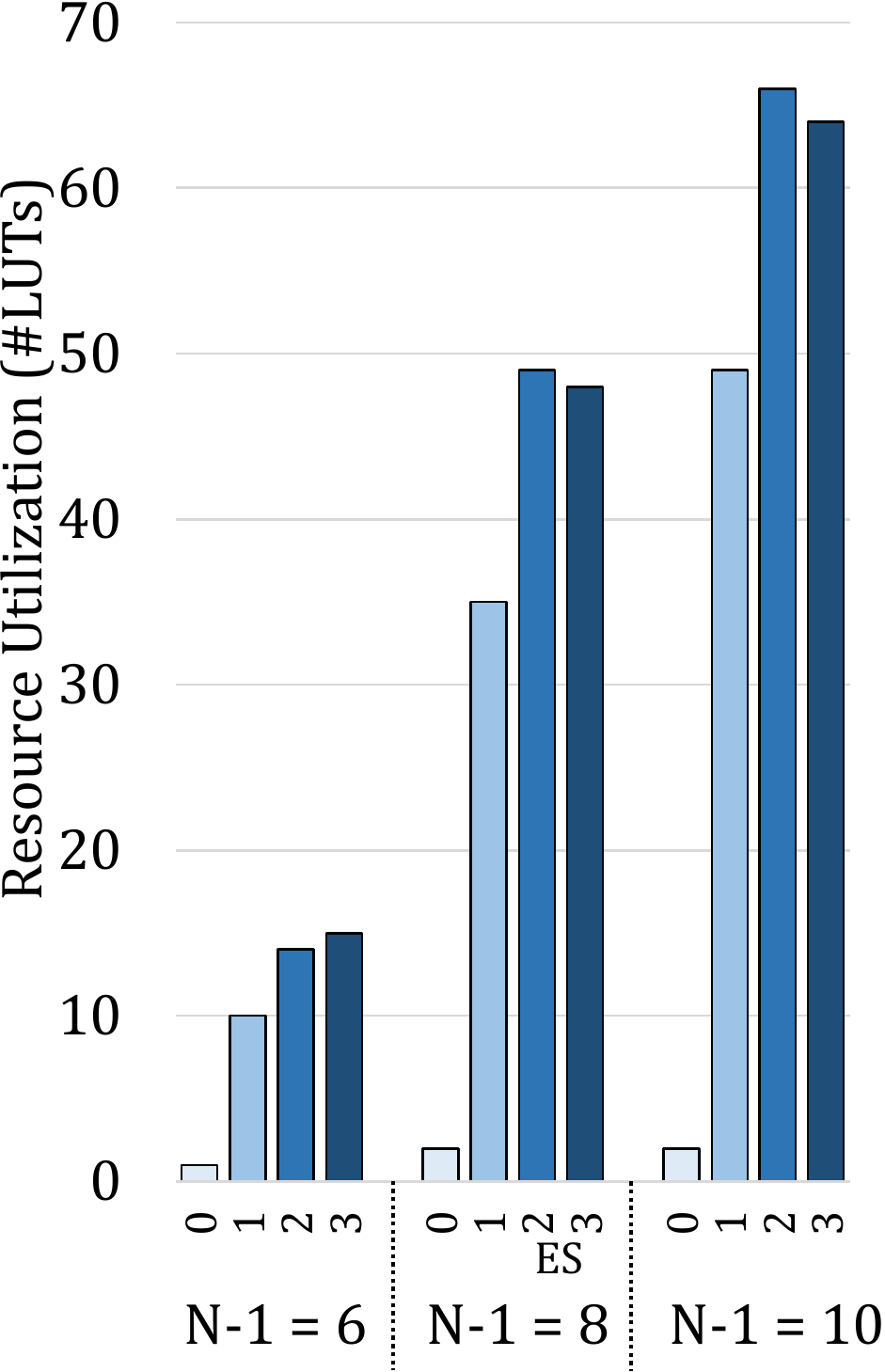}}
  \caption{Variation in the hardware performance metrics of \gls{pofx} with varying values of $ES$ and $N-1$ for Posit($N-1,ES$) representation being converted to \gls{fxp}($M=16, F=15$).}
  \label{fig:exp_pofx_varNES} 
% \end{minipage}
\end{figure}

\su{\autoref{fig:exp_pofx_varNES} compares the impact of various Posit configurations, varying $N-1$ and $ES$, on the performance metrics of \gls{pofx} for a fixed bit-width ($M$) of the output. The critical path delay follows an increasing trend with an increase in the values of $ES$ and $N-1$. This trend is primarily due to an increase in the logic required for Posit extraction due to increased variability in the individual field length. The designs with $ES = 0$ have minimum resource utilization. The absence of the exponent field results in significant simplification of the overall extraction circuit. However, the designs with $ES\in \{2,3\}$ have comparably higher resource utilization. A similar trend is also observed for the dynamic power consumption of the \gls{pofx} for various Posit configurations.}

\subsubsection{MAC Design Analysis}
\begin{figure}[t] 
   \centering
   	\subfloat[Critical Path Delay\label{exp_pofxMAC8_cpd}]{%
       \includegraphics[width=0.32\columnwidth]{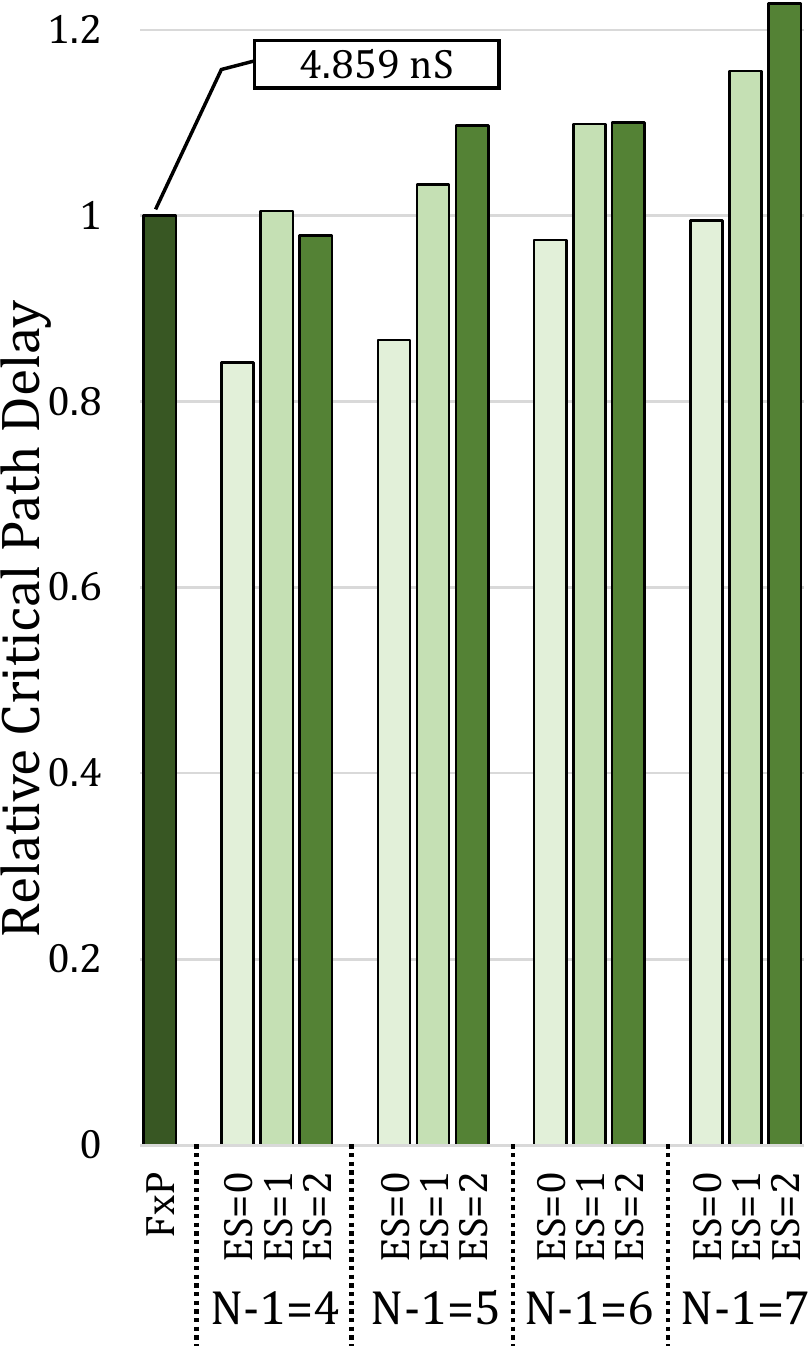}}
    \hspace*{0.2em}
    \subfloat[Power Dissipation\label{exp_pofxMAC8_pow}]{%
        \includegraphics[width=0.32\columnwidth]{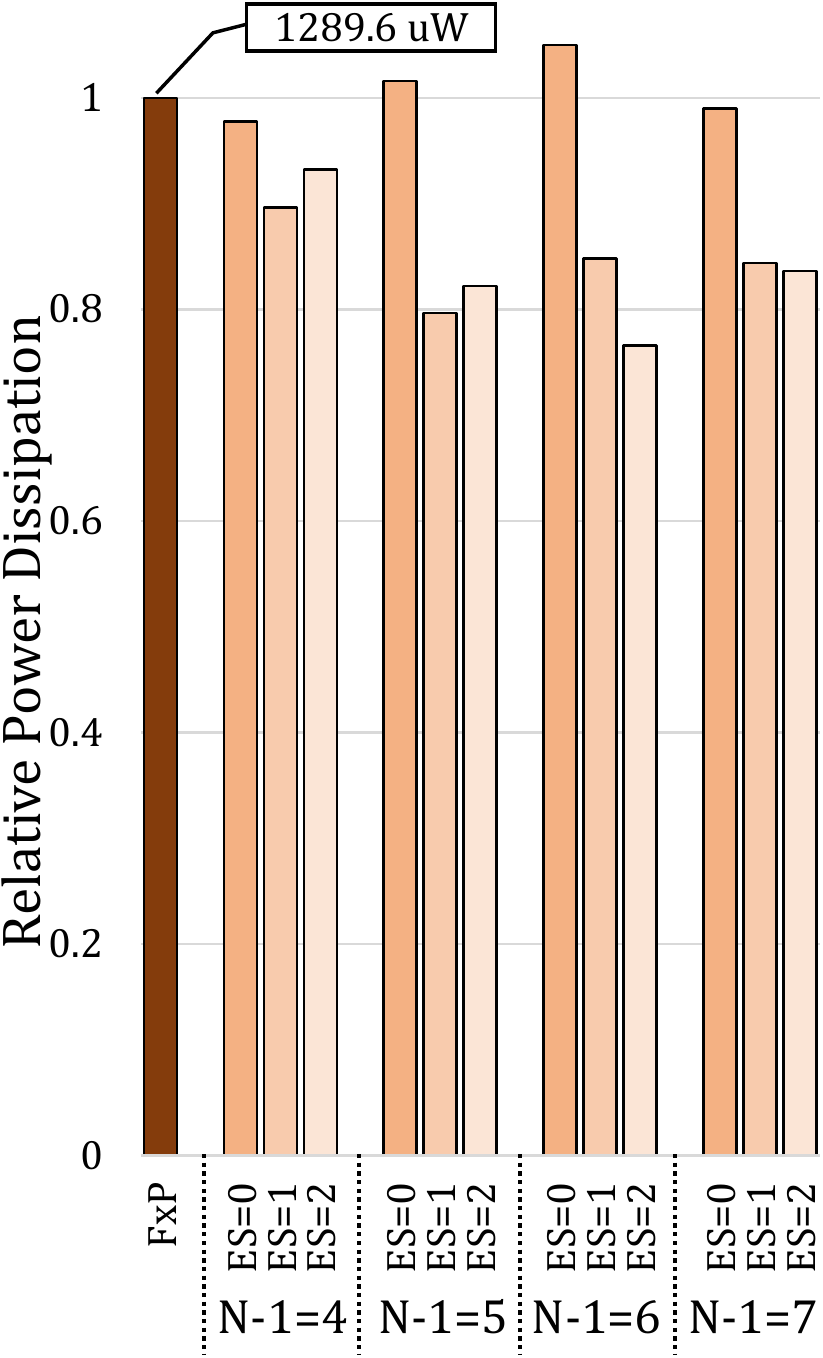}}
\hspace*{0.2em}
    \subfloat[Resources \label{exp_pofxMAC8_lut}]{%
        \includegraphics[width=0.32\columnwidth]{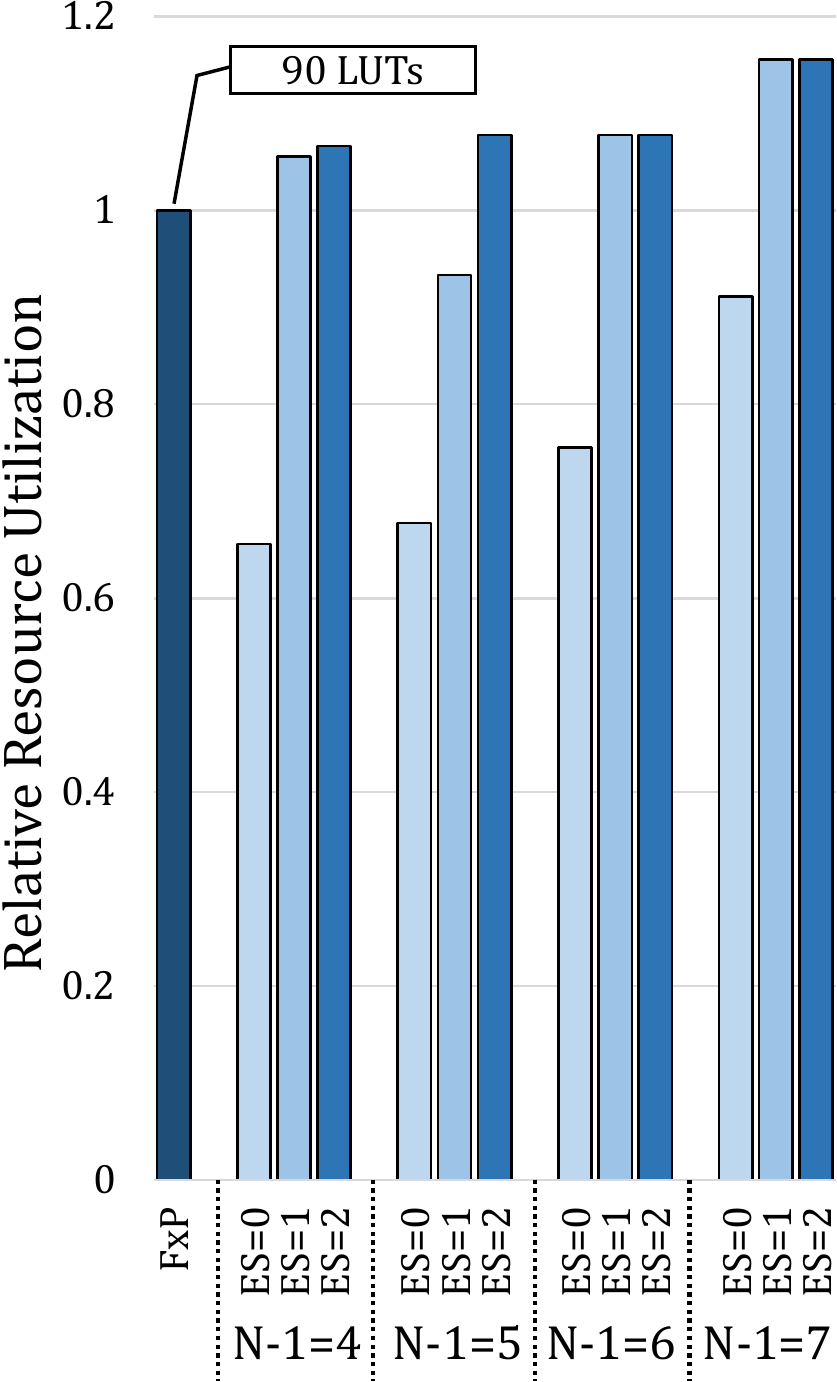}}
  \caption{Relative hardware performance metrics of \gls{pofx}-based \gls{mac} units with varying values of $ES$ and $N-1$ for Posit($N-1,ES$) inputs to 8-bit \gls{fxp} \gls{mac}.}
  \label{fig:exp_pofxMAC8} 
\end{figure}

\begin{figure}[t]
	\centering
	\scalebox{1}{\includegraphics[width=0.99 \columnwidth]{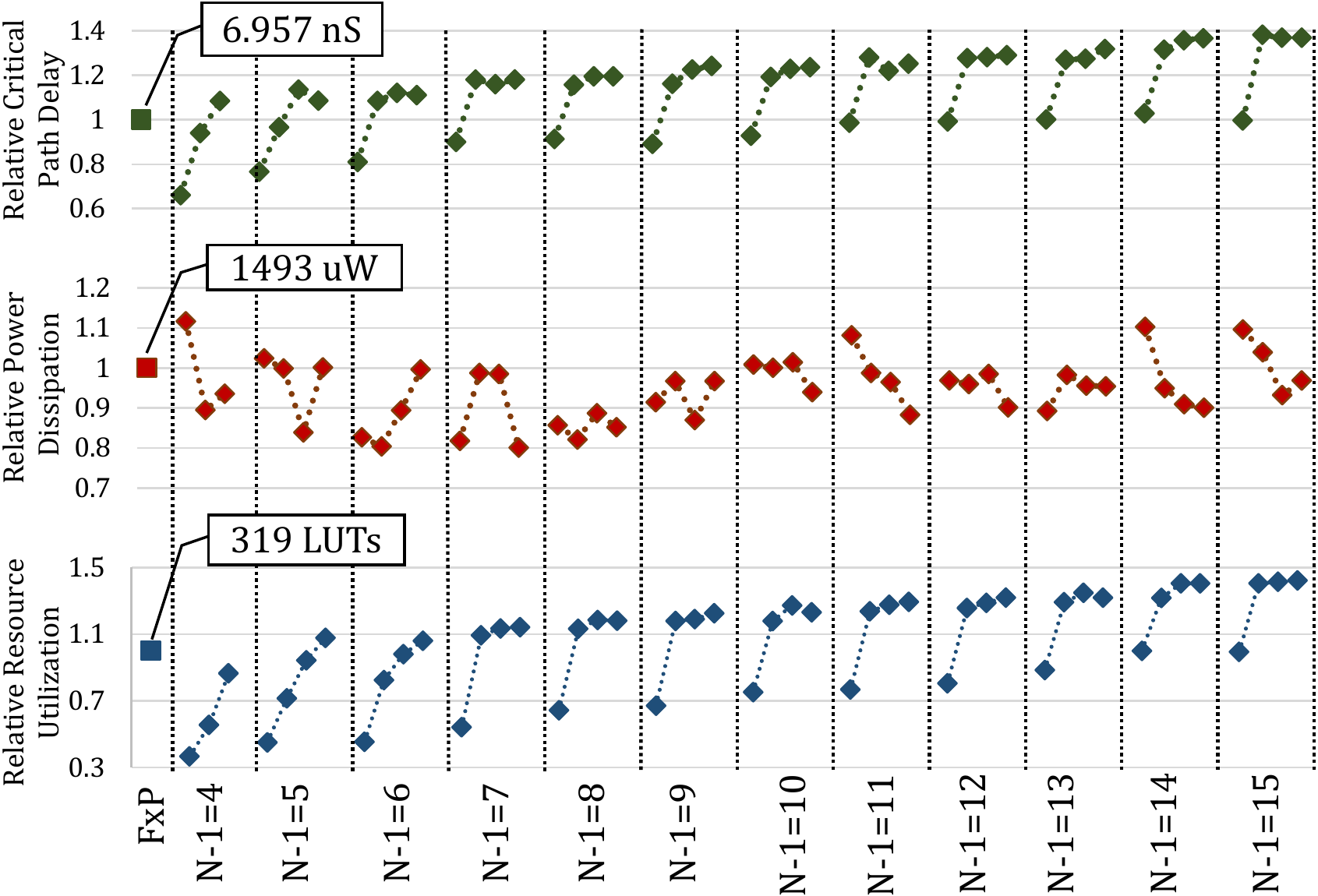}}
	\caption{Relative hardware performance metrics of \gls{pofx}-based \gls{mac} units with varying values of $ES$ and $N-1$ for Posit($N-1,ES$) inputs to 16-bit \gls{fxp} \gls{mac}. $ES\in \{0,1,2,3\}$ for all cases, except for $N-1=4$ where $ES\in \{0,1,2\}$.}
	\label{fig:exp_pofxMAC16}
\end{figure}

\su{The proposed \gls{pofx} allows the utilization of resource-efficient and high-performance \gls{fxp}-based arithmetic units for Posit number systems. To evaluate the efficacy of the proposed approach and estimate the associated overheads of the \gls{pofx}, we compare \gls{pofx}-based 8-bit MAC units\footnote{\su{As shown in~\autoref{fig:methMAC}, an M-bit FxP-based MAC includes a $M \times M$ multiplier and a $3M$-bit adder.}} with a traditional \gls{fxp}-based MAC unit. The results of these comparisons for various configurations of Posit are presented in~\autoref{fig:exp_pofxMAC8}. The critical path delay and resource utilization of the MAC follow a gradually rising trend with both $N$ and $ES$ values. It can be noted that in a few cases, especially for $ES= 0$, the \gls{pofx}-based MAC provides better performance across critical path delay, power dissipation, and LUT utilization than the \gls{fxp}-only MAC. 
For $ES=0$, the Posit scheme's dynamic range is limited, and the \gls{pofx} does not utilize the complete dynamic range of the \gls{fxp}. The limited number of unique \gls{fxp} values, after conversion, allows the synthesis tool to optimize the overall design of \gls{pofx}-based MAC to improve the associated performance metrics.}
{\suresh{
%We compare the hardware performance metrics of the 8-bit configuration proposed \gls{pofx}-based MAC with traditional fixed point counterpart in~\autoref{fig:exp_pofxMAC8}. 
%to highlight the hardware cost associated with the addition of the \gls{pofx} converter. 
%As shown earlier in~\autoref{fig:methMAC}, a $N$-bit \gls{fxp} \gls{mac} includes $2N$-bit multiplier output length and $3N$-bit accumulator stage. 
%The critical path delay and resource utilization follow a gradually rising trend with both $N$ and $ES$. It can be noted that in a few cases, especially for $ES=0$, the \gls{pofx}-based MAC provides better performance across critical path delay, power dissipation and LUT utilization than the \gls{fxp}-only \gls{mac}.
%which is its sub-block. 
%Although the \gls{pofx}-based MAC uses the \gls{fxp}-based multipliers and adders, the reduced  hardware metrics is possible as not all bit inputs of the internal \gls{fxp} MAC are addressed by the \gls{pofx} converter due to the tapered accuracy of Posits and limited dynamic range owing to the limited \gls{fxp} output. Thus, the synthesis tool can optimize the complete design based on the possible bit-level inputs leading to a reduction in the overall metrics. 
The power metrics do not follow a well defined trend as they are generated based on the bit switches required to obtain the correct bit-sequence as the output. Compared to the \gls{fxp}-only \gls{mac}, we report worst-case overheads of 22.8\%, 5.0\% and 15.5\% for critical path delay, power dissipation, and LUT Utilization, respectively.  
% Doesn't warrant a discussion and can't be explained with a one to one correlation
% Logical validity to be verified with Mr. Salim
%Similar trends are observed in~\autoref{fig:exp_pofxMAC16}, which plots the same  performance metrics for a larger ($16$-bit \gls{fxp} input) \gls{pofx}-based MAC.
\su{Similar trends are observed in~\autoref{fig:exp_pofxMAC16}, which compares the same performance metrics for a 16-bit \gls{fxp} MAC.}
}}

% \begin{figure}[t] 
%   \centering
%   	\subfloat[Critical Path Delay\label{exp_pofxMAC16_cpd}]{%
%       \includegraphics[width=1\columnwidth]{figures/exp_pofxMAC16_cpd.pdf}}\\
%     % \hspace*{0.2em}
%     \subfloat[Power Dissipation\label{exp_pofxMAC16_pow}]{%
%         \includegraphics[width=1\columnwidth]{figures/exp_pofxMAC16_pow.pdf}}\\
% % \hspace*{0.2em}
%     \subfloat[Resources \label{exp_pofxMAC16_lut}]{%
%         \includegraphics[width=1\columnwidth]{figures/exp_pofxMAC16_lut.pdf}}
%   \caption{Relative hardware performance metrics of \gls{pofx}-based \gls{mac} units with varying values of $ES$ and $N-1$ for Posit($N-1,ES$) inputs to 16-bit \gls{fxp} \gls{mac}.}
%   \label{fig:exp_pofxMAC16} 
% \end{figure}

\begin{figure}[t]
	\centering
% 	\scalebox{1}{\includegraphics[width=1 \columnwidth]{figures/exp_compare_MAC8.pdf}}
	\scalebox{1}{\includegraphics[width=1 \columnwidth]{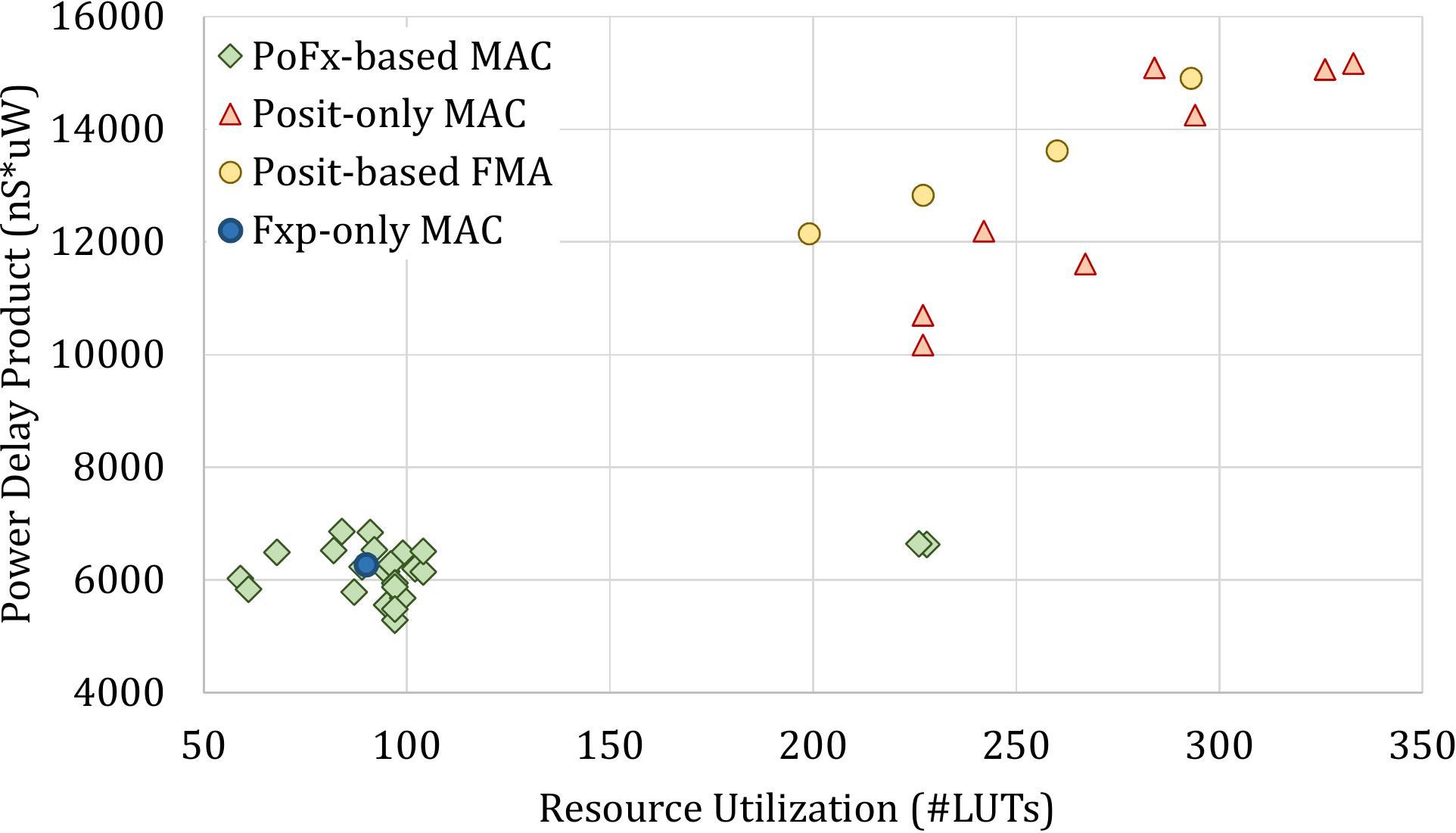}}
	\caption{Comparison of various 8-bit MAC implementations: \su{for Posit($N-1, ES$) $N-1 \in \left[4~..~7\right]~and~ES \in \{0,1,2\}$}}
	\label{fig:exp_compare_MAC8}
\end{figure}

\begin{figure}[t]
	\centering
% 	\scalebox{1}{\includegraphics[width=1 \columnwidth]{figures/exp_compare_MAC16.pdf}}
	\scalebox{1}{\includegraphics[width=1 \columnwidth]{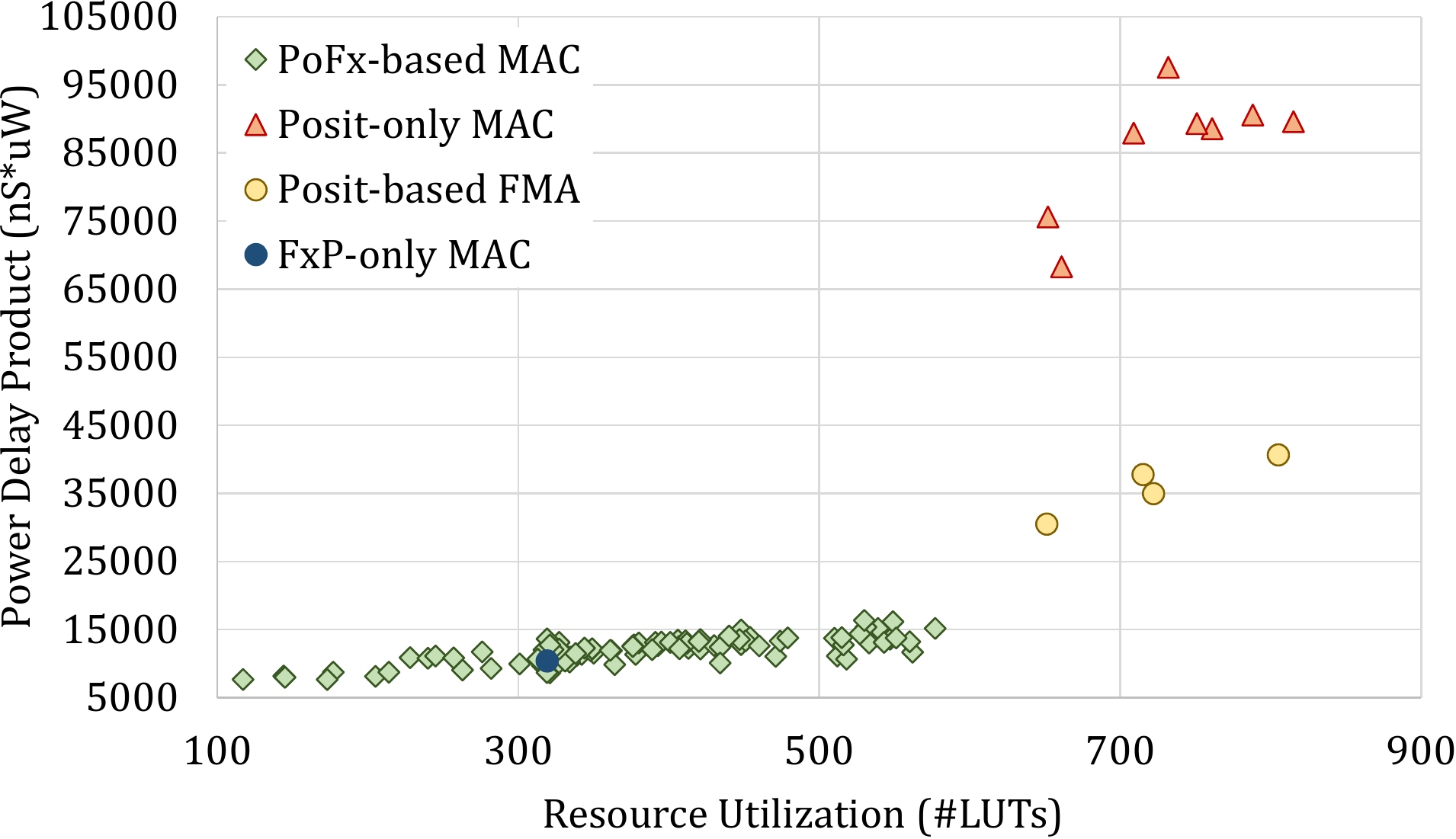}}
	\caption{Comparison of various 16-bit MAC implementations: \su{for Posit($N-1, ES$) $N-1 \in \left[4~..~15\right]~and~ES \in \{0,1,2,3\}$}}
	\label{fig:exp_compare_MAC16}
\end{figure}

\su{To further evaluate the efficacy of \gls{pofx}-based MAC design, we compare it with \gls{fxp}-only MAC, Posit-only MAC, and Posit-based 3-input Fused Multiply Add (FMA)~\cite{SmallPositHDL}. Moreover, for a thorough exploration of the \gls{pofx}-, \gls{fxp}-~, and Posit-based designs, we have synthesized two types of designs--- one that allows the synthesis tool to optimize across the constituent blocks (converters, multipliers, and adders) and the other that performs optimization for the constituent blocks separately.}
{\suresh{
%We compared the \gls{mac} designs from four different methods---\gls{pofx}-based \gls{mac}, \gls{fxp}-only \gls{mac}, Posit-only \gls{mac} and Posit 3-input Fused Multiply Add (FMA). For each of the first three types, we synthesized two types of designs--- one that allowed the synthesis tool to optimize across the constituent blocks (converters, multipliers and adders) and the other that performs optimization for the constituent blocks separately.
\autoref{fig:exp_compare_MAC8} and~\autoref{fig:exp_compare_MAC16} show the comparison of the power-delay-product (PDP) and the LUT utilization of these designs for $8$- and $16$-bit designs, respectively.  
%We analyse the distribution of the Power Delay Product (PDP) with respect to the resource utilization for four MAC designs including the proposed \gls{pofx}-based MAC, \gls{fxp} MAC, Posit MAC and Posit based three input Fused Multiply Add (FMA) for 8-bit MAC implementations in~\autoref{fig:exp_compare_MAC8} and 16-bit MAC implementations in~\autoref{fig:exp_compare_MAC16}. 
%Presence of multiple points for each design indicates the presence various configurations  with and without inter block optimizations. 
Posit-only \gls{mac}, which has been implemented by using a standalone $N$-bit Posit adder and $N$-bit Posit Multiplier, has significantly higher PDP and LUT utilization as a result of the extraction and packaging of Posits between stages. The Posit-based FMA, though optimized,  
requires more hardware resources for implementation. It can be observed that the \gls{pofx}-based \gls{mac} designs fall closely within the range of~\gls{fxp}-only \gls{mac}. Further, the Posit-only \gls{mac} and Posit-based FMA designs generate an $N$-bit output whereas, the proposed design generates a more precise $3N$-bit output once extracted. This can lead to lower inter-layer losses in \gls{ann}s as we can ascertain the type of rounding mechanism at the output based on the network to retain as much precision as possible before transferring the value to the next stage.}}

% \begin{figure}[t]
% 	\centering
% 	\subfloat[8-bit MAC implementations \label{fig:exp_compare_MAC8}]{
% 			\includegraphics[width=0.48 \columnwidth]{figures/exp_compare_MAC8_compensated.pdf}
%         	} \hspace{3pt}
% 	\subfloat[16-bit MAC implementations \label{fig:exp_compare_MAC16}]{
%     	\includegraphics[width=0.48 \columnwidth]{figures/exp_compare_MAC16_compensated.pdf}
% 	   } %\hspace{10pt}
% 	\caption{Comparison of various MAC implementations}	
% 	\label{fig:exp_compare_MAC}
% \end{figure}
\subsection{Behavioral Analysis}
\label{subsec:expBehav}
\salim{We have considered DNNs as a test case to show the impact of various number representation schemes on the output accuracy of high-level applications. For this work, we have used a pre-trained VGG16\cite{simonyan2014deep} network for the classification of the ImageNet dataset~\cite{10.1007/s11263-015-0816-y}. The VGG16 network mainly consists of 13 convolution layers and 3 fully connected layers. The very large number of the network's trained parameters, 138 million, makes it a sound candidate for evaluating efficiency of various quantization schemes. The single-precision \gls{flp}-based Top-1 and Top-5 percentage output classification accuracy of the 50000 validation images in the ImageNet dataset is 69.72\% and 89.09\%, respectively. Our proposed TensorFlow-based framework performs a multi-level analysis to identify possible quantization configurations fulfilling the output accuracy requirements of the network. 
%, as shown in~\autoref{tab:VGG_classification}.
}
\subsubsection{Weights Quantization Error Analysis}
\begin{figure}[ht]
	\centering
	\scalebox{1}{\includegraphics[width=0.99 \columnwidth]{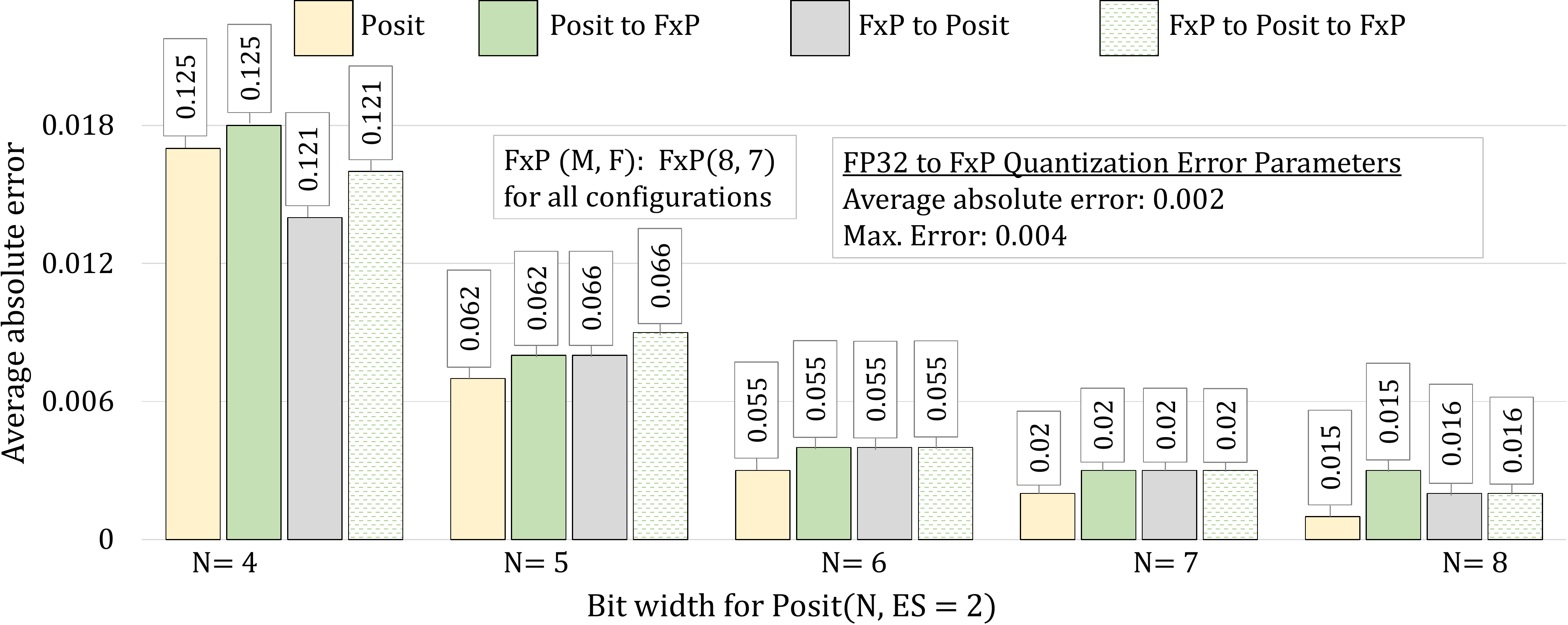}}
	\caption{Error analysis of various quantization schemes for Conv2\_1 layer of pre-trained VGG16\cite{simonyan2014deep}. $ES$ values are kept 2 for all configurations. Maximum absolute quantization-induced error of each configuration is shown above the corresponding average relative error bars.}
	\label{fig:exp_conv2_1_error}
\end{figure}

\salim{
%Our proposed TensorFlow-based framework performs a multi-level analysis to identify possible quantization configurations fulfilling the output accuracy requirements of the network. 
In the first step, our framework quantizes the parameters (weights and biases) of all layers and filters out the configurations having large quantization-induced errors. For example, \autoref{fig:exp_conv2_1_error} shows the average absolute and the maximum quantization-induced errors in the weights of the Conv2\_1 layer of the VGG16 network using different configurations of Posit and \gls{fxp} schemes. The 8-bit \gls{fxp} produces an average absolute error of $0.002$. For smaller values of $N$, Posit schemes produce more errors than the \gls{fxp}-based scheme in the quantized weights. However, for 7-bit and 8-bit Posit schemes, the average absolute errors are reduced to $0.002$ and $0.001$ only. We also evaluate the interconversions\footnote{As shown in~\autoref{fig:behavMeth}} of various schemes to identify feasible configurations for \gls{pofx}-based hardware. For example, the Posit($N-1=7,ES=2$)$\rightarrow$ $8$-bit \gls{fxp} scheme produces an average absolute error of $0.003$, whereas the 8-bit \gls{fxp}$\rightarrow$ Posit($N-1=7,ES=2$)$\rightarrow$ $8$-bit \gls{fxp} generates an average error of $0.002$ only. \autoref{fig:exp_conv2_1_error} also reveals that Posit($N-1=3,ES=2$)-based configurations can be eliminated in the first step due to large quantization induced-errors.}
\sam{ We have performed a similar analysis for all layers of the VGG-16 network by exploring all combinations of Posit($N, ES$) where $N \in \{4,5,6,7,8\}$ and $ES \in \{0,1,2,3\}$, and 8-bit \gls{fxp}. The analysis identifies the quantization schemes producing the minimum average absolute error and the maximum absolute error for each layer of the network. For each $N$-bit Posit scheme, the quantized parameters are analyzed to identify the values of $ES$ inducing minimum quantization errors.}

% \salim{
% \autoref{fig:exp_all_layers_error} shows a similar analysis for all layers of the VGG-16 network by exploring all combinations of Posit($N, ES$) where $N \in \{4,5,6,7,8\}$ and $ES \in \{0,1,2,3\}$. The figure shows the quantization schemes producing the minimum average absolute error and the maximum absolute error for each layer of the network. For each $N$-bit scheme, the quantized parameters are analyzed to identify the values of $ES$ inducing minimum quantization errors.     
% }
%\subsubsection{Layer-wise Quantization Error Analysis}
% \begin{figure}[t]
% 	\centering
% 	\scalebox{1}{\includegraphics[width=1 \columnwidth]{figures/conv2_1_error_analysis_trimmed.pdf}}
% 	\caption{Error analysis of different quantization schemes for Conv2\_1 layer of pre-trained VGG16. Maximum absolute quantization-induced error is shown by the dotted lines.}
% 	\label{fig:exp_conv2_1_error}
% \end{figure}
%\subsubsection{Layer-wise Quantization Error Analysis}
% \begin{figure*}[ht]
% 	\centering
% 	\scalebox{0.9}{\includegraphics[width=1 \textwidth]{figures/all_layers_trimmed.pdf}}
% 	\caption{Error analysis of different quantization schemes for all layers of pre-trained VGG16\cite{simonyan2014deep}. For each scheme, $ES$ value producing minimum error is selected.}
% 	\label{fig:exp_all_layers_error}
% \end{figure*}

\begin{table}[ht]
\caption{Pareto Analysis of \gls{mac} designs with weights quantization error. Objectives: PDP, Average Error, \#LUTs}
\centering
\def\arraystretch{1.0}
\resizebox{1.0 \columnwidth}{!}{
\begin{tabular}{|l|c|c|c|c|c|c|c|c|}
\hline
\multicolumn{1}{|c|}{\textbf{VGG16 Layer}} & \multicolumn{6}{c|}{\textbf{Number of points on   Pareto front}} & \multicolumn{2}{c|}{\multirow{3}{*}{\textbf{\begin{tabular}[c]{@{}c@{}}\%   Improvement in \\ hypevolume due to \\ PoFx-based \gls{mac}s\end{tabular}}}} \\ \cline{1-7}
\multirow{2}{*}{\textbf{MAC Type}} & \multicolumn{2}{c|}{\multirow{2}{*}{\textbf{PoFx-based}}} & \multicolumn{2}{c|}{\multirow{2}{*}{\textbf{Posit-based}}} & \multicolumn{2}{c|}{\multirow{2}{*}{\textbf{FxP-based}}} & \multicolumn{2}{c|}{} \\
 & \multicolumn{2}{c|}{} & \multicolumn{2}{c|}{} & \multicolumn{2}{c|}{} & \multicolumn{2}{c|}{} \\ \hline
\textbf{Max Bits} & \textbf{8} & \textbf{16} & \textbf{8} & \textbf{16} & \textbf{8} & \textbf{16} & \textbf{8} & \textbf{16} \\ \hline
% conv1\_1 & 9 & 7 & 1 & 12 & 1 & 0 & 100.4 & 82.0 \\ \hline
% conv1\_2 & 8 & 3 & 3 & 13 & 1 & 0 & 71.0 & 1.5 \\ \hline
% conv2\_1 & 8 & 3 & 3 & 12 & 1 & 0 & 68.4 & 1.3 \\ \hline
% conv2\_2 & 8 & 1 & 4 & 10 & 1 & 0 & 67.3 & 0.3 \\ \hline
% conv3\_1 & 8 & 1 & 4 & 10 & 1 & 0 & 65.0 & 0.3 \\ \hline
% conv3\_2 & 8 & 0 & 5 & 11 & 1 & 0 & 61.6 & 0.0 \\ \hline
% conv3\_3 & 8 & 0 & 5 & 11 & 1 & 0 & 61.9 & 0.0 \\ \hline
% conv4\_1 & 8 & 0 & 6 & 9 & 1 & 0 & 59.4 & 0.0 \\ \hline
% conv4\_2 & 7 & 0 & 6 & 9 & 1 & 0 & 56.2 & 0.0 \\ \hline
% conv4\_3 & 7 & 0 & 6 & 9 & 1 & 0 & 56.7 & 0.0 \\ \hline
% conv5\_1 & 7 & 0 & 6 & 9 & 1 & 0 & 58.6 & 0.0 \\ \hline
% conv5\_2 & 8 & 0 & 6 & 9 & 1 & 0 & 58.9 & 0.0 \\ \hline
% conv5\_3 & 7 & 0 & 6 & 9 & 1 & 0 & 58.6 & 0.0 \\ \hline
% fc6 & 7 & 0 & 7 & 10 & 1 & 0 & 40.9 & 0.0 \\ \hline
% fc7 & 7 & 0 & 7 & 10 & 1 & 0 & 50.2 & 0.0 \\ \hline
% fc8 & 8 & 0 & 6 & 9 & 1 & 0 & 59.7 & 0.0 \\ \hline
conv1\_1 & 9 & 7 & 1 & 5 & 1 & 0 & 173 & 74 \\ \hline
conv1\_2 & 8 & 4 & 3 & 11 & 1 & 0 & 125 & 2 \\ \hline
conv2\_1 & 8 & 4 & 3 & 9 & 1 & 0 & 121 & 2 \\ \hline
conv2\_2 & 8 & 3 & 2 & 8 & 1 & 0 & 119 & 1 \\ \hline
conv3\_1 & 8 & 3 & 2 & 8 & 1 & 0 & 115 & 1 \\ \hline
conv3\_2 & 8 & 1 & 4 & 10 & 1 & 0 & 109 & 0 \\ \hline
conv3\_3 & 8 & 1 & 4 & 10 & 1 & 0 & 109 & 0 \\ \hline
conv4\_1 & 8 & 1 & 6 & 8 & 1 & 0 & 104 & 0 \\ \hline
conv4\_2 & 7 & 1 & 6 & 8 & 1 & 0 & 97 & 0 \\ \hline
conv4\_3 & 7 & 1 & 6 & 8 & 1 & 0 & 98 & 0 \\ \hline
conv5\_1 & 7 & 1 & 6 & 8 & 1 & 0 & 102 & 0 \\ \hline
conv5\_2 & 8 & 1 & 6 & 8 & 1 & 0 & 102 & 0 \\ \hline
conv5\_3 & 7 & 1 & 6 & 8 & 1 & 0 & 102 & 0 \\ \hline
fc6 & 7 & 0 & 5 & 8 & 1 & 0 & 67 & 0 \\ \hline
fc7 & 7 & 0 & 5 & 8 & 1 & 0 & 86 & 0 \\ \hline
fc8 & 8 & 1 & 6 & 8 & 1 & 0 & 104 & 0 \\ \hline
\end{tabular}
}
\label{tab:ParAnalysisWtsErrMAC_1}
\end{table}
\begin{figure}[ht]
	\centering
% 	\scalebox{1}{\includegraphics[width=1 \columnwidth]{figures/exp_pareto_wtsErrMAC8.pdf}}
	\scalebox{1}{\includegraphics[width=0.99 \columnwidth]{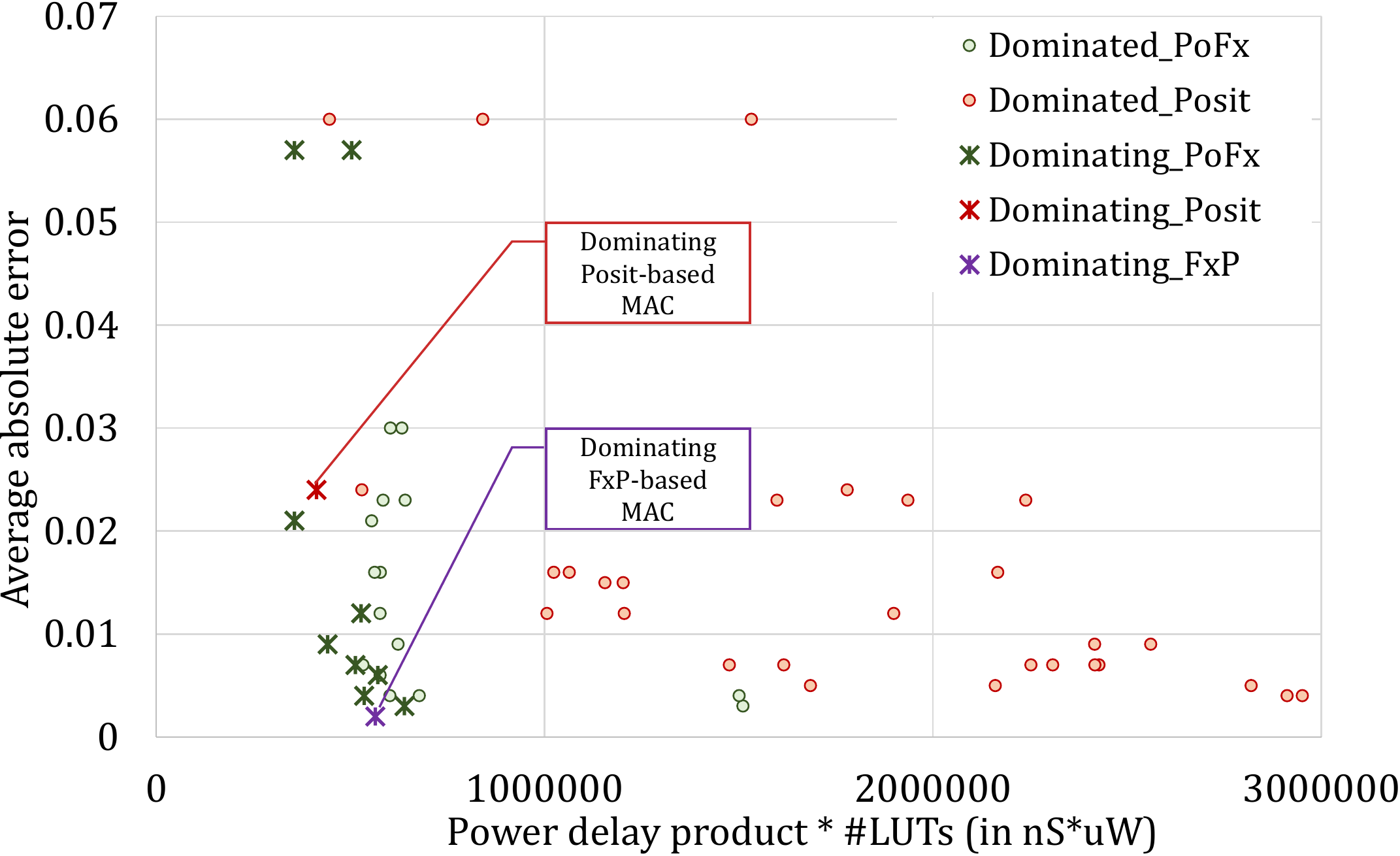}}
	\caption{Pareto analysis of $8$-bit \gls{mac} design along with errors induced in quantization of weights for Conv1\_1 layer of pre-trained VGG16.}
	\label{fig:exp_pareto_wtsErrMAC8_conv2_1_error}
\end{figure}

\begin{table}[t]
\caption{Pareto analysis of \gls{mac} designs with weights quantization error. Objectives: CPD, Power, Average Error, \#LUTs, \#Bit-width of parameters}
\centering
\def\arraystretch{1.0}
\resizebox{1.0 \columnwidth}{!}{
\begin{tabular}{|l|c|c|c|c|c|c|c|c|}
\hline
\multicolumn{1}{|c|}{\textbf{VGG16 Layer}} & \multicolumn{6}{c|}{\textbf{Number of points on   Pareto front}} & \multicolumn{2}{c|}{\multirow{3}{*}{\textbf{\begin{tabular}[c]{@{}c@{}}\%   Improvement in \\ hypevolume due to \\ PoFx-based MACs\end{tabular}}}} \\ \cline{1-7}
\multirow{2}{*}{\textbf{MAC Type}} & \multicolumn{2}{c|}{\multirow{2}{*}{\textbf{PoFx-based}}} & \multicolumn{2}{c|}{\multirow{2}{*}{\textbf{Posit-based}}} & \multicolumn{2}{c|}{\multirow{2}{*}{\textbf{FxP-based}}} & \multicolumn{2}{c|}{} \\
 & \multicolumn{2}{c|}{} & \multicolumn{2}{c|}{} & \multicolumn{2}{c|}{} & \multicolumn{2}{c|}{} \\ \hline
\textbf{Max Bits} & \multicolumn{1}{c|}{\textbf{8}} & \multicolumn{1}{c|}{\textbf{16}} & \multicolumn{1}{c|}{\textbf{8}} & \multicolumn{1}{c|}{\textbf{16}} & \multicolumn{1}{c|}{\textbf{8}} & \multicolumn{1}{c|}{\textbf{16}} & \textbf{8} & \multicolumn{1}{c|}{\textbf{16}} \\ \hline
% conv1\_1 & 21 & 25 & 12 & 42 & 0 & 0 & 37.4 & 35.6 \\ \hline
% conv1\_2 & 20 & 24 & 13 & 25 & 0 & 0 & 25.8 & 25.5 \\ \hline
% conv2\_1 & 20 & 21 & 12 & 25 & 0 & 0 & 25.4 & 25.3 \\ \hline
% conv2\_2 & 17 & 21 & 13 & 20 & 0 & 0 & 25.6 & 25.1 \\ \hline
% conv3\_1 & 17 & 20 & 12 & 19 & 0 & 0 & 25.2 & 24.9 \\ \hline
% conv3\_2 & 17 & 16 & 14 & 20 & 0 & 0 & 25.3 & 24.9 \\ \hline
% conv3\_3 & 17 & 16 & 14 & 20 & 0 & 0 & 25.2 & 24.9 \\ \hline
% conv4\_1 & 17 & 16 & 12 & 18 & 0 & 0 & 25.2 & 24.8 \\ \hline
% conv4\_2 & 17 & 16 & 12 & 18 & 0 & 0 & 25.2 & 24.8 \\ \hline
% conv4\_3 & 17 & 16 & 12 & 18 & 0 & 0 & 25.2 & 24.8 \\ \hline
% conv5\_1 & 17 & 16 & 12 & 18 & 0 & 0 & 25.2 & 24.8 \\ \hline
% conv5\_2 & 17 & 16 & 12 & 18 & 0 & 0 & 25.2 & 24.8 \\ \hline
% conv5\_3 & 17 & 16 & 12 & 18 & 0 & 0 & 25.2 & 24.8 \\ \hline
% fc6 & 17 & 15 & 13 & 17 & 0 & 0 & 26.6 & 24.8 \\ \hline
% fc7 & 17 & 16 & 13 & 17 & 0 & 0 & 25.7 & 24.7 \\ \hline
% fc8 & 17 & 16 & 12 & 18 & 0 & 0 & 25.2 & 24.8 \\ \hline
conv1\_1 & 21 & 31 & 10 & 41 & 0 & 0 & 40 & 74 \\ \hline
conv1\_2 & 20 & 32 & 11 & 23 & 0 & 0 & 27 & 50 \\ \hline
conv2\_1 & 20 & 31 & 10 & 23 & 0 & 0 & 27 & 48 \\ \hline
conv2\_2 & 17 & 30 & 11 & 18 & 0 & 0 & 27 & 48 \\ \hline
conv3\_1 & 17 & 29 & 10 & 17 & 0 & 0 & 27 & 47 \\ \hline
conv3\_2 & 17 & 26 & 12 & 18 & 0 & 0 & 27 & 46 \\ \hline
conv3\_3 & 17 & 26 & 12 & 18 & 0 & 0 & 26 & 46 \\ \hline
conv4\_1 & 17 & 26 & 10 & 16 & 0 & 0 & 27 & 46 \\ \hline
conv4\_2 & 17 & 26 & 10 & 16 & 0 & 0 & 27 & 46 \\ \hline
conv4\_3 & 17 & 26 & 10 & 16 & 0 & 0 & 27 & 46 \\ \hline
conv5\_1 & 17 & 26 & 10 & 16 & 0 & 0 & 27 & 46 \\ \hline
conv5\_2 & 17 & 26 & 10 & 16 & 0 & 0 & 27 & 46 \\ \hline
conv5\_3 & 17 & 26 & 10 & 16 & 0 & 0 & 27 & 46 \\ \hline
fc6 & 17 & 25 & 11 & 15 & 0 & 0 & 28 & 46 \\ \hline
fc7 & 17 & 26 & 11 & 15 & 0 & 0 & 27 & 46 \\ \hline
fc8 & 17 & 26 & 10 & 16 & 0 & 0 & 27 & 46 \\ \hline
\end{tabular}
}
\label{tab:ParAnalysisWtsErrMAC_2}
\end{table}
\par{\siva{In our current work we focus only on the quantization of weights and biases. The use of a specific quantized representation of the weights and biases will require the use of a compatible \gls{mac} design for inference. Hence, we performed a joint analysis of the performance of the various \gls{mac} designs and the errors induced in the parameters by the corresponding quantization scheme. The various \gls{mac} designs are grouped under three categories -- \gls{pofx}-based, Posit-based (that includes both multiply and adder combination and FMA-based designs) and \gls{fxp}-based. For the \gls{pofx}-based and Posit-based designs, \su{lower bit-width input designs were also considered}. For example, for $8$-bit quantization, $N$ was varied from $5$ to $8$. Similarly, for $16$-bit quantization, $N$ was varied from $5$ to $16$. \autoref{tab:ParAnalysisWtsErrMAC_1} shows the Pareto analysis results for $8$- and $16$-bit \gls{mac}s with the three objectives -- PDP, average quantization-induced error and the LUT utilization. We report the number of dominating points for each of the three types of quantization schemes used for the parameters of each layer of VGG16.
As shown in the table, using \gls{pofx}-based designs contribute significantly to the number of points on the Pareto-front for $8$-bit precision. We also report the percentage increase in the Pareto-front hypervolume due to the usage of \gls{pofx}-based designs over the collection of Posit and \gls{fxp}-based designs only. As seen in the table, using \gls{pofx}-based designs we report up to 173\% increase in the hypervolume for $8$-bits precision.~\autoref{fig:exp_pareto_wtsErrMAC8_conv2_1_error} shows the dominating and dominated points for each of the three categories in the corresponding design space for $8$-bit precision \gls{mac}s for the first layer (Conv1\_1) of VGG16. It can be observed that the Posit- and \gls{fxp}-based designs contribute one point each to the resulting Pareto-front, compared to 9 \gls{pofx}-based points.\footnote{Since we have used a 2D plot for showing the pareto-front for 3 objectives, some dominating points appear as dominated in~\autoref{fig:exp_pareto_wtsErrMAC8_conv2_1_error}.}
}}
\par{\siva{The improvements for $16$-bit precision are lower compared to $8$-bits. However, as shown in~\autoref{tab:ParAnalysisWtsErrMAC_2}, if we also consider the bits-width of the parameters as a design objective in the analysis, we report consistent improvements using \gls{pofx}-based designs for both $8$- and $16$-bits precision. Since the number of input bits is an indicator of the communication power dissipation (and energy consumption) for moving weights, using \gls{pofx}-based quantization can result in reducing the overall power dissipation during \gls{dnn} inference. 
}}

\subsubsection{Output Activation Error Analysis}
% \salim{In the second step of behavioral analysis, our framework utilizes the quantized parameters to evaluate each configuration's impact on the quantization of each layer's output activations. For this analysis, the input activations to a layer under consideration are kept at \gls{flp} precision. After computing the output activations, they are quantized using the quantization configuration employed to quantize the respective parameters.}
\begin{figure}[ht]
	\centering
% 	\scalebox{1}{\includegraphics[width=1 \columnwidth]{figures/exp_pareto_wtsErrMAC8.pdf}}
	\scalebox{1}{\includegraphics[width=0.99 \columnwidth]{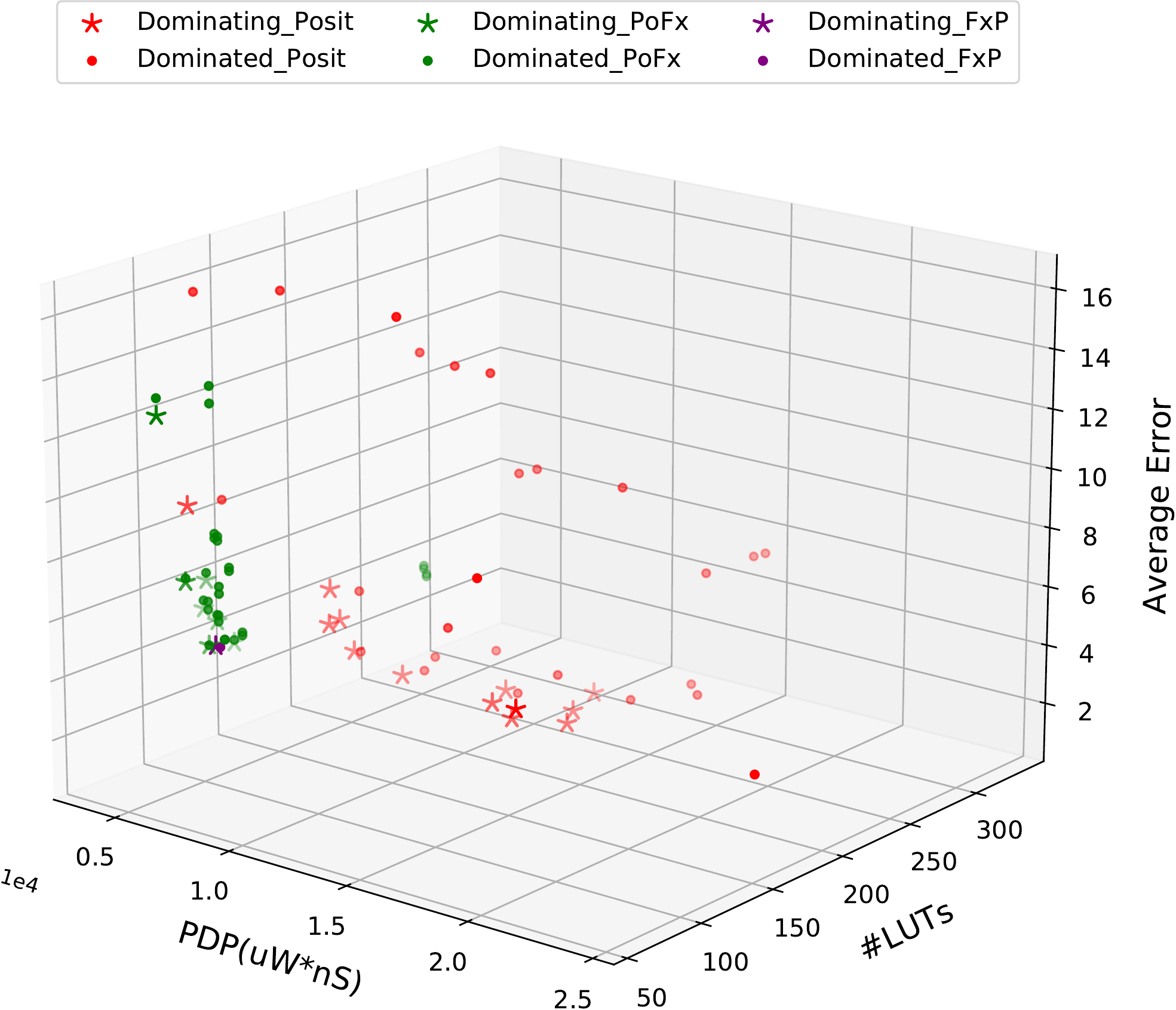}}
	\caption{Pareto analysis of $8$-bit MAC design along with errors induced in output activations for Conv1\_1 layer of pre-trained VGG16. 
	%12.4\% improvement in the Pareto-front hypervolume was observed due to the addition of PoFx-based designs. Number of Pareto-front points of each type: PoFx(7), Posit(13), FxP(1). 
	}
	\label{fig:exp_pareto_outActErrMAC8_conv1_1_error}
\end{figure}
%\subsubsection{Classification Error Analysis}
% Please add the following required packages to your document preamble:
% \usepackage{multirow}
% Please add the following required packages to your document preamble:
% \usepackage{multirow}
\par{\siva{
In the second step of behavioral analysis, our framework utilizes the quantized parameters to evaluate each configuration's impact on the output activations of each layer. The computation of the output activation involves using a \gls{mac} design that is compatible with the chosen quantization scheme. Similar to the analysis presented in~\autoref{fig:exp_pareto_wtsErrMAC8_conv2_1_error} for the errors induced in the parameters,~\autoref{fig:exp_pareto_outActErrMAC8_conv1_1_error} shows the design space while considering the errors in the output activations for the first layer--- Conv1\_1--- of VGG16. The 3D scatter plot shows the various design points corresponding to the three categories of \gls{mac} designs, \gls{pofx}-, Posit- and \gls{fxp}-based. \su{It can be observed from~\autoref{fig:exp_pareto_outActErrMAC8_conv1_1_error} that the \gls{pofx} and \gls{fxp}-based designs' contribution to the Pareto-front is mainly due to better hardware performance--- lower PDP and reduced number of utilized LUTs. Similarly, Posit-based designs' contribution is mainly due to lower average error, albeit at high hardware costs.} The resulting Pareto-front in~\autoref{fig:exp_pareto_outActErrMAC8_conv1_1_error} has $7$, $13$ and $1$ points from \gls{pofx}, Posit and \gls{fxp}-based designs
respectively, with 12.4\% improvement in the hypervolume over the collection of only Posit- and \gls{fxp}-based designs. 
It must be noted that since we focus on the quantization schemes for only the parameters, during the behavioral analysis, the input activations for each of the layers are kept at \gls{flp} precision. After computing the output activations, they are quantized using the configuration employed to quantize the respective parameters. This lets us evaluate the impact of the proposed methods and designs while other aspects are kept unchanged. 
}}

\begin{table}[t]
\tiny
\centering
\caption{{Classification} accuracy of VGG-16 network~\cite{simonyan2014deep} on ImageNet dataset~\cite{10.1007/s11263-015-0816-y} with quantization of weights and biases using different schemes}
{\renewcommand{\arraystretch}{1.0}
\resizebox{1.0\columnwidth}{!}{
\begin{tabular}{|c|c|c|c|c|c|}
\hline
\multicolumn{2}{|c|}{\multirow{2}{*}{\textbf{Configuration}}} & \multicolumn{2}{c|}{\textbf{\begin{tabular}[c]{@{}c@{}}Configuration \\ Parameters\end{tabular}}} &  \multirow{2}{*}{\textbf{Top-1 {[}\%{]}}} & \multirow{2}{*}{\textbf{Top-5 {[}\%{]}}} \\ \cline{3-4}
\multicolumn{2}{|c|}{} & \textbf{N} & \textbf{ES} &  &  \\ \hline
\multicolumn{2}{|c|}{\gls{flp}} & - & - & 69.72 & 89.09 \\ \hline \hline
\multicolumn{2}{|c|}{\gls{fxp}-16} & - & - & 69.66 & 89.02 \\ \hline
\rowcolor[HTML]{C0C0C0}\multicolumn{2}{|c|}{\gls{fxp}-8} & - & - & 64.71 & 86.26 \\ \hline
\multicolumn{2}{|c|}{\gls{fxp}-7} & - & - & 10.94 & 26.10 \\ \hline \hline
\multicolumn{2}{|c|}{Posit} & 7 & 1 & 68.88 & 88.50 \\ \hline
\multicolumn{2}{|c|}{Posit} & 8 & 1 & 69.59 & 89.00 \\ \hline
\multicolumn{2}{|c|}{Posit} & 6 & 2 & 66.32 & 86.99 \\ \hline
\multicolumn{2}{|c|}{Posit} & 7 & 2 & 68.77 & 88.54 \\ \hline
\rowcolor[HTML]{C0C0C0}\multicolumn{2}{|c|}{Posit} & 8 & 2 & 69.65 & 89.00 \\ \hline
\multicolumn{2}{|c|}{Posit} & 6 & 3 & 64.86 & 86.04 \\ \hline
\multicolumn{2}{|c|}{Posit} & 7 & 3 & 68.02 & 87.97 \\ \hline
\multicolumn{2}{|c|}{Posit} & 8 & 3 & 69.43 & 88.86 \\ \hline \hline
\multirow{8}{*}{\begin{tabular}[c]{@{}c@{}}PoFx\\ (N-1, ES)\end{tabular}} & Posit\_FxP & 6 & 1 & 46.05 & 71.12 \\ \cline{2-6} 
 & Posit\_FxP & 7 & 1 & 11.13 & 26.08 \\ \cline{2-6} 
 & Posit\_FxP & 5 & 2 & 43.59 & 69.08 \\ \cline{2-6} 
 & Posit\_FxP & 6 & 2 & 11.96 & 27.22 \\ \cline{2-6} 
 & Posit\_FxP & 7 & 2 & 1.92 & 6.31 \\ \cline{2-6} 
 & Posit\_FxP & 5 & 3 & 41.37 & 66.99 \\ \cline{2-6} 
 & Posit\_FxP & 6 & 3 & 11.67 & 26.84 \\ \cline{2-6} 
 & Posit\_FxP & 7 & 3 & 1.79 & 6.11 \\ \hline \hline
\multirow{8}{*}{\begin{tabular}[c]{@{}c@{}}PoFx\\ (N-1, ES)\end{tabular}}& \cellcolor[HTML]{C0C0C0}FxP\_Posit\_FxP & \cellcolor[HTML]{C0C0C0}6 & \cellcolor[HTML]{C0C0C0}1 & \cellcolor[HTML]{C0C0C0}64.38 & \cellcolor[HTML]{C0C0C0}85.94 \\ \cline{2-6} 
 & FxP\_Posit\_FxP & 7 & 1 & 64.48 & 86.15 \\ \cline{2-6} 
 & FxP\_Posit\_FxP & 5 & 2 & 58.27 & 81.99 \\ \cline{2-6} 
 & \cellcolor[HTML]{C0C0C0}FxP\_Posit\_FxP & \cellcolor[HTML]{C0C0C0}6 & \cellcolor[HTML]{C0C0C0}2 & \cellcolor[HTML]{C0C0C0}64.36 & \cellcolor[HTML]{C0C0C0}85.99 \\ \cline{2-6} 
 & FxP\_Posit\_FxP & 7 & 2 & 64.40 & 86.08 \\ \cline{2-6} 
 & FxP\_Posit\_FxP & 5 & 3 & 57.13 & 81.13 \\ \cline{2-6} 
 & FxP\_Posit\_FxP & 6 & 3 & 62.67 & 84.62 \\ \cline{2-6} 
 & FxP\_Posit\_FxP & 7 & 3 & 64.45 & 86.15 \\ \hline
\end{tabular}
}}
\label{tab:VGG_classification}
\end{table}
\subsubsection{Classification Error Analysis}
\siva{Finally, the behavioral analysis involves estimating the impact of the proposed methods on the classification accuracy.}
\salim{\autoref{tab:VGG_classification} shows the percentage Top-1 and Top-5 classification accuracies of the ImageNet validation dataset~\cite{10.1007/s11263-015-0816-y} using different quantization schemes. For this experiment, the activations have \gls{flp} precision, and the parameters (weights and biases) are quantized using various 8-bit schemes. For comparison, we also show the classification accuracy using 7-bit and 16-bit \gls{fxp}-based quantization techniques. The \gls{fxp}-16 and Posit$(N = 8, ES = 2)$ produce similar classification results by reducing the final output accuracy by only $0.06$ and $0.07$, respectively when compared with \gls{flp}-based results. The \gls{fxp}-8 based configuration reduces the Top-1 and Top-5 classification accuracy by $5.01$ and $2.83$, respectively. However, the \gls{fxp}-7-based quantization significantly drops the final classification accuracy. For the \gls{pofx}-based schemes, we consider the normalized \gls{pofx} technique and utilize \emph{Posit(N-1, ES)} configurations for N-bit Posit numbers. \autoref{tab:VGG_classification} reveals that the direct conversion of Posit numbers to \gls{fxp} scheme (\emph{Posit-FxP}) significantly diminishes the final output accuracy. However, utilizing \gls{fxp}$\rightarrow$ Posit$\rightarrow$ \gls{fxp} based conversion, the \gls{pofx} has an insignificant impact on the final output. For example, compared to the \gls{fxp}-8 based results, the \gls{fxp}-8$\rightarrow$ Posit(N-1 = 6,ES = 2)$\rightarrow$~\gls{fxp}-8 decreases the Top-1 and {Top-5} classification accuracy by only $0.35$ and $0.26$.
}

\su{\autoref{tab:class_MAC} shows the joint analysis of the ImageNet dataset classification accuracy and the corresponding MAC designs for a subset of the configurations. It contains only those configurations from~\autoref{tab:VGG_classification} that have comparable accuracy and having feasible hardware designs.  For instance, arithmetic blocks for Posit($N=6, ES=3$) could not be generated using SmallPosit~HDL~\cite{SmallPositHDL}. Similarly, as shown in~\autoref{tab:VGG_classification} the {Posit-FxP} modes have much lower accuracy than similar configurations for \gls{fxp}$\rightarrow$ Posit$\rightarrow$ \gls{fxp}, while requiring the same \gls{pofx}-based MAC, and are hence omitted from the analysis.
The PDP and LUT utilization values for each configuration in~\autoref{tab:class_MAC} are obtained from the lowest PDP design for that configuration. The PDP and LUT metrics shown in  the table correspond to values relative to the maximum shown in the table's top row. }
{\siva{
%\autoref{tab:class_MAC} shows the joint analysis of the classification accuracy and the corresponding \gls{mac} designs for inference. The PDP and LUT utilization values for each configuration are obtained from the \textit{lowest PDP} design for that configuration. The PDP and LUT metrics shown in the table correspond to values relative to the maximum shown in the top row of the table. \autoref{tab:class_MAC} contains a subset of the configurations shown in~\autoref{tab:VGG_classification}. Only those configurations that have comparable accuracy and having feasible hardware designs are shown in~\autoref{tab:class_MAC}. For instance, arithmetic blocks for Posit($N=6, E 5=3$) could not be generated using SmallPosit~HDL~\cite{SmallPositHDL}. Similarly, as shown in~\autoref{tab:VGG_classification} the {Posit-FxP} modes have much lower accuracy than similar configurations for \gls{fxp}--- Posit---~\gls{fxp}, while requiring the same \gls{pofx}-based MAC, and are hence omitted from the analysis. 
The highest value of PDP and LUT utilization occurs for the configurations Posit($N=8,ES=1$) and \gls{fxp}-16 respectively. The highest and lowest values of the performance metrics for each of the two categories -- Posit and \gls{pofx} are highlighted in bold text in~\autoref{tab:class_MAC}.
It can be observed that the Posit configuration for the highest Top-1 accuracy, Posit($N=8,ES=2$), corresponds to the \gls{mac} design with highest LUT utilization. Similarly the Posit configuration with highest Top-5 accuracy, Posit($N=8,ES=1$) (and Posit($N=8,ES=2$)), corresponds to highest (and relatively \textit{higher}) PDP value. The Posit configuration with the lowest accuracy, Posit($N=6,ES=2$) corresponds to the design with lowest PDP and LUT utilization among Posit-based \gls{mac}s.
}}
\begin{table}[t]
\caption{Joint analysis of classification accuracy and MAC hardware characteristics}
\centering
\def\arraystretch{1.0}
\resizebox{1.0 \columnwidth}{!}{
\begin{tabular}{|c|c|c|c|c|c|c|}
\hline
\multirow{2}{*}{\textbf{Configuration}} & \multirow{2}{*}{\textbf{N}} & \multirow{2}{*}{\textbf{ES}} & \multirow{2}{*}{\begin{tabular}[c]{@{}c@{}}\textbf{Top-1}\\ {[}\%{]}\end{tabular}} & \multirow{2}{*}{\begin{tabular}[c]{@{}c@{}}\textbf{Top-5}\\ {[}\%{]}\end{tabular}} & \multicolumn{2}{c|}{\textbf{Relative MAC Metrics}} \\ \cline{6-7} 
 &  &  &  &  & \begin{tabular}[c]{@{}c@{}}\textbf{PDP}\\ {[}Maximum:\\ 13616 uW*nS{]}\end{tabular} & \begin{tabular}[c]{@{}c@{}}\textbf{LUTs}\\ {[}Maximum:\\ 319{]}\end{tabular} \\ \hline \hline
\multirow{2}{*}{FxP} & 16 & - & 69.66 & 89.02 & 0.763 & \cellcolor[HTML]{C0C0C0}1.000 \\ \cline{2-7} 
 & 8 & - & 64.71 & 86.26 & 0.475 & 0.282 \\ \hline \hline
\multirow{7}{*}{\begin{tabular}[c]{@{}c@{}}Posit \\ (N,ES)\end{tabular}} & 7 & 1 & 68.88 & 88.5 & 0.578 & 0.671 \\ \cline{2-7} 
 & 8 & 1 & 69.59 & \textbf{89} & \cellcolor[HTML]{C0C0C0}\textbf{1.000} & 0.815 \\ \cline{2-7} 
 & 6 & 2 & \textbf{66.32} & \textbf{86.99} & \textbf{0.441} & \textbf{0.555} \\ \cline{2-7} 
 & 7 & 2 & 68.77 & 88.54 & 0.550 & 0.618 \\ \cline{2-7} 
 & 8 & 2 & \textbf{69.65} & \textbf{89} & 0.853 & \textbf{0.837} \\ \cline{2-7} 
 & 7 & 3 & 68.02 & 87.97 & 0.469 & 0.567 \\ \cline{2-7} 
 & 8 & 3 & 69.43 & 88.86 & 0.747 & 0.712 \\ \hline \hline
\multirow{8}{*}{\begin{tabular}[c]{@{}c@{}}PoFx\\ (N-1,ES)\end{tabular}} & 6 & 1 & 64.38 & 85.94 & 0.432 & \textbf{0.304} \\ \cline{2-7} 
 & 7 & 1 & \textbf{64.48} & \textbf{86.15} & 0.451 & 0.326 \\ \cline{2-7} 
 & 5 & 2 & 58.27 & 81.99 & 0.417 & 0.310 \\ \cline{2-7} 
 & 6 & 2 & 64.36 & 85.99 & \textbf{0.388} & \textbf{0.304 }\\ \cline{2-7} 
 & 7 & 2 & 64.4 & 86.08 & \textbf{0.478} & 0.326 \\ \cline{2-7} 
 & 5 & 3 & \textbf{57.13} & \textbf{81.13} & 0.446 & \textbf{0.304} \\ \cline{2-7} 
 & 6 & 3 & 62.67 & 84.62 & 0.418 & \textbf{0.304} \\ \cline{2-7} 
 & 7 & 3 & 64.45 & 86.15 & 0.413 & \textbf{0.361} \\ \hline
\end{tabular}
}
\label{tab:class_MAC}
\end{table}

\par{\siva{Similar correlations were also observed in the case of \gls{pofx}-based designs. \su{Designs with higher PDP usually result in better accuracy. Compared to \gls{fxp}-8 based designs the \gls{pofx}($N-1=7,ES=1$) achieves similar accuracy with lower PDP ($\approx5\%$) and slightly higher LUT overhead ($\approx15\%$). Similarly, \gls{pofx}($N-1=6,ES=2$) achieves comparable accuracy with even lower PDP ($\approx18\%$) and less LUT overheads ($\approx8\%$). Additionally, these \gls{pofx}-based designs requires less bits for representing the parameters of a network.} This can result in lower communication and storage overheads in the accelerator design for each layer of the network.
}}
\subsection{Accelerator Design Analysis}
\label{subsec:expRsrcEst}
\par{\siva{In order to estimate the system-level impact of using the proposed \gls{pofx} methodology, we integrated the candidate solutions in the design of an accelerator for a fully-connected layer of a \gls{dnn}. The accelerator was designed using C++ and synthesized using Xilinx's Vivado HLS. To keep the design generic, we implemented a matrix-vector multiplication. The matrix, representing the weights of the fully-connected layer, was of size $64 \times 10$. Each vector, representing an input activation, is of size $1 \times 64$. One thousand input activations were used to estimate the switching activity in order to compute the power dissipation. The implemented accelerator uses ReLU activation function.
}}
\subsubsection{Accelerator DSE}
\begin{table}[t]
\caption{Accelerator implementations used for analysis}
\centering
\def\arraystretch{1.0}
\resizebox{0.9 \columnwidth}{!}{
\begin{tabular}{|c|c|c|c|}
\hline
\textbf{Design} & \textbf{\begin{tabular}[c]{@{}c@{}}Optimization\\ type\end{tabular}} & \textbf{\begin{tabular}[c]{@{}c@{}}Memory type\\ for weights and \\ activation arrays\end{tabular}} & \textbf{\begin{tabular}[c]{@{}c@{}}Partitioning of\\ weights matrix\end{tabular}} \\ \hline \hline
base & None & Default & Default \\ \hline
dotOpt & \begin{tabular}[c]{@{}c@{}}Unroll loop for \\ dot product\end{tabular} & Default & Row-wise \\ \hline
fullOpt & \begin{tabular}[c]{@{}c@{}}Unroll loops for\\ dot product and \\ outer loop\end{tabular} & Default & \begin{tabular}[c]{@{}c@{}}Complete \\ partitioning\\ (Row and Column)\end{tabular} \\ \hline \hline
dotOpt\_LRAM & \begin{tabular}[c]{@{}c@{}}Unroll loop for \\ dot product\end{tabular} & LUTRAMs & Row-wise \\ \hline
fullOpt\_LRAM & \begin{tabular}[c]{@{}c@{}}Unroll loops for\\ dot product and \\ outer loop\end{tabular} & LUTRAMs & \begin{tabular}[c]{@{}c@{}}Complete \\ partitioning\end{tabular} \\ \hline \hline
dotOpt\_BRAM & \begin{tabular}[c]{@{}c@{}}Unroll loop for \\ dot product\end{tabular} & BlockRAMs & Row-wise \\ \hline
fullOpt\_BRAM & \begin{tabular}[c]{@{}c@{}}Unroll loops for\\ dot product and \\ outer loop\end{tabular} & BlockRAMs & \begin{tabular}[c]{@{}c@{}}Complete \\ partitioning\end{tabular} \\ \hline
\end{tabular}
}
\label{tab:accDesigns}
\end{table}
\par
\begin{figure}[t]
	\centering
	\scalebox{1}{\includegraphics[width=0.99 \columnwidth]{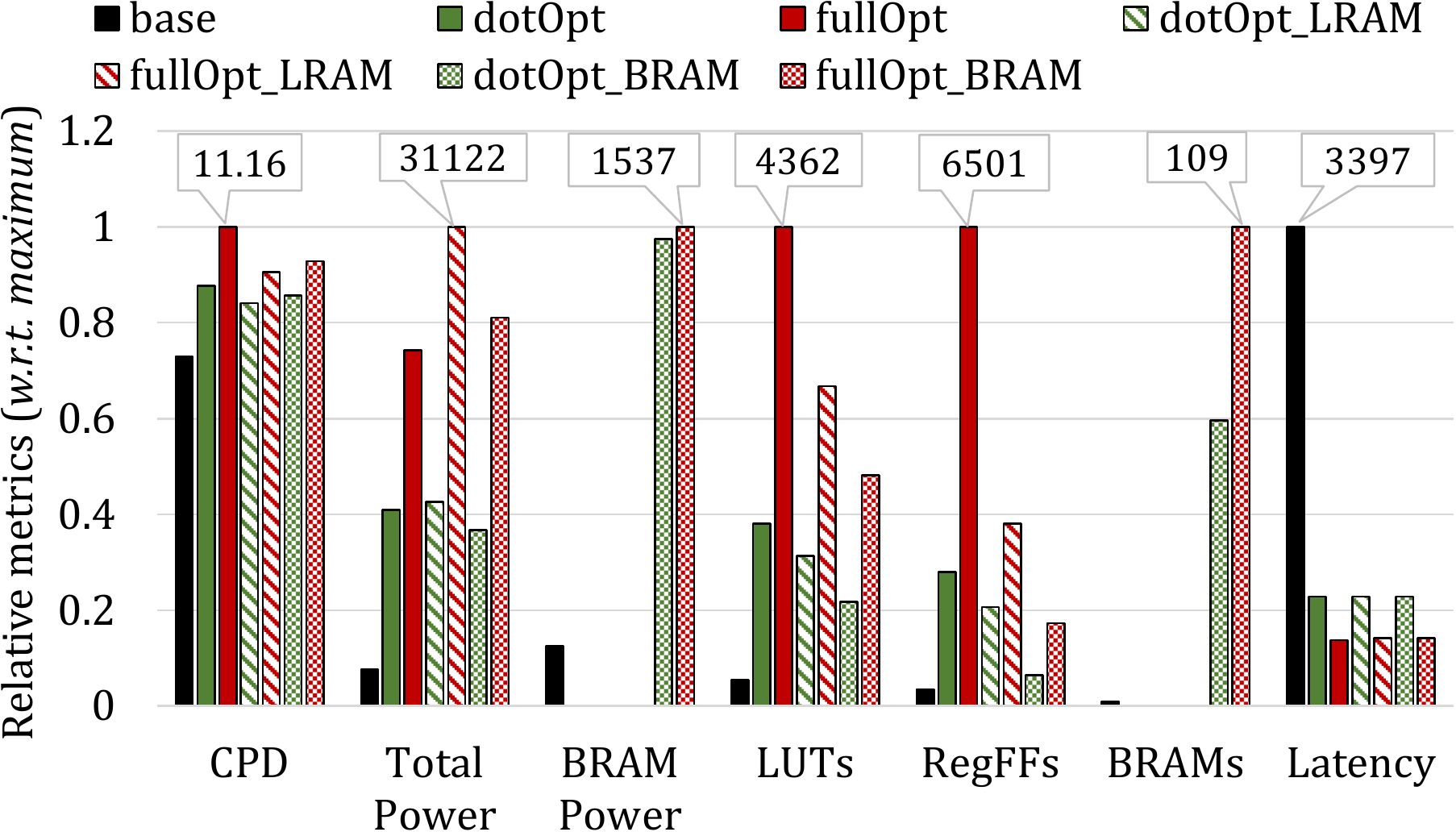}}
	\caption{Hardware metrics for accelerator design implementing FxP-based MAC using different optimization modes.}
	\label{fig:exp_accDsgn_8bits_FxP_top}
\end{figure}

{\siva{In Section~\ref{subsec:rsrcMeth}, we presented some of the degrees of freedom in the design of an accelerator that can result in varying performance metrics and resource requirements. \autoref{fig:exp_accDsgn_8bits_FxP_top} shows the resulting metrics from seven different implementations of the accelerator under test using $8$-bit \gls{fxp} parameters and activations. The \textit{base} implementation refers to the basic design without using any HLS directives. As shown in~\autoref{tab:accDesigns}, the other implementations vary in terms of the optimization goals, and the type and partitioning of the memory used for storing weights and activations. As seen in~\autoref{fig:exp_accDsgn_8bits_FxP_top}, the optimizations of \textit{fullOpt} result in the lowest latency implementations. However, it also results in the highest power dissipation and maximum resource utilization. Also it can be noted from the figure that using BRAMs results in lower total power dissipation than using LUTRAMs for both \textit{fullOpt} and \textit{dotOpt} implementations.}
\par{Further, it can be observed that the CPD varies with the optimization modes, even with the usage of the same arithmetic hardware, as the implementation of multiple parallel operators spreads the designs spatially and the routing delays tend to increase accordingly. It must be noted that~\autoref{fig:exp_accDsgn_8bits_FxP_top} shows only a subset of the possible implementations using the various degrees of freedom. While it is possible to generate many more design points using various types of array partitioning and loop-related HLS directives, we limit our evaluation to these seven types of implementations for showing the impact of our proposed designs.
}}
\subsubsection{Accelerator Resource Requirements}
\begin{figure*}[ht]
	\centering
	\scalebox{0.9}{\includegraphics[width=2 \columnwidth]{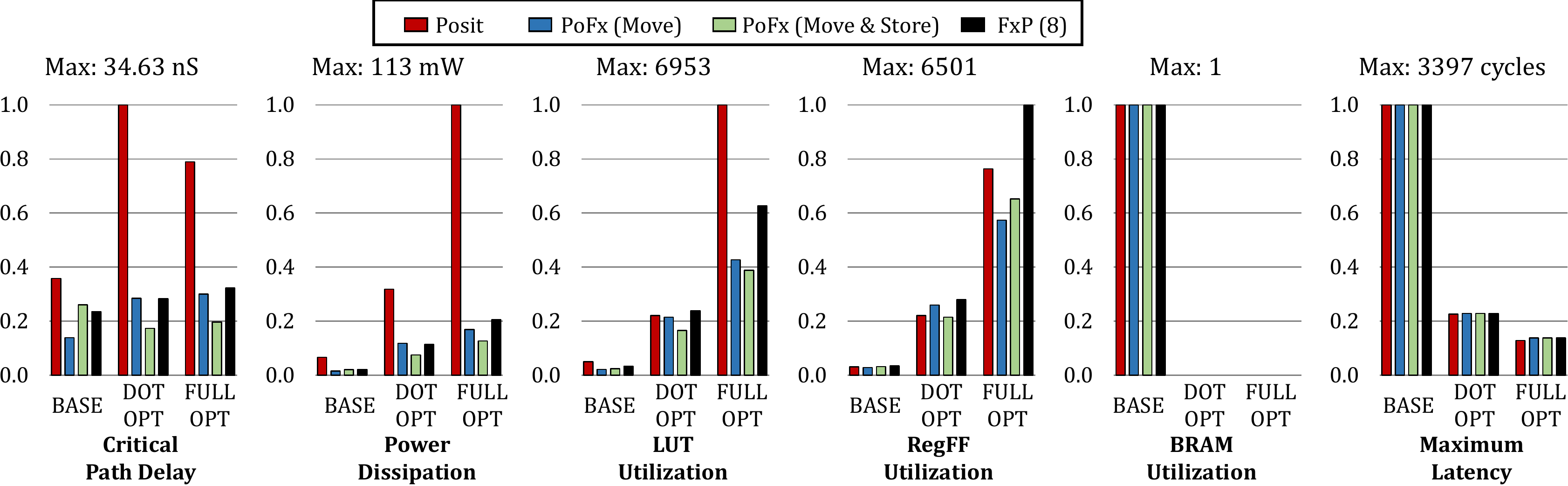}}
	\caption{Relative hardware metrics for accelerator using \textit{base}, \textit{dotOpt} and \textit{fullOpt} implementations for four types of designs. The quantization configurations used are Posit($N=6,ES=0$), \gls{pofx}($N-1=5, ES=0$) and \gls{fxp}($M=8$). }
	\label{fig:exp_accDsgn_base_dotOpt_fullOpt_6_5_5_8}
\end{figure*}
\begin{figure*}[ht]
	\centering
	\scalebox{0.85}{\includegraphics[width=2 \columnwidth]{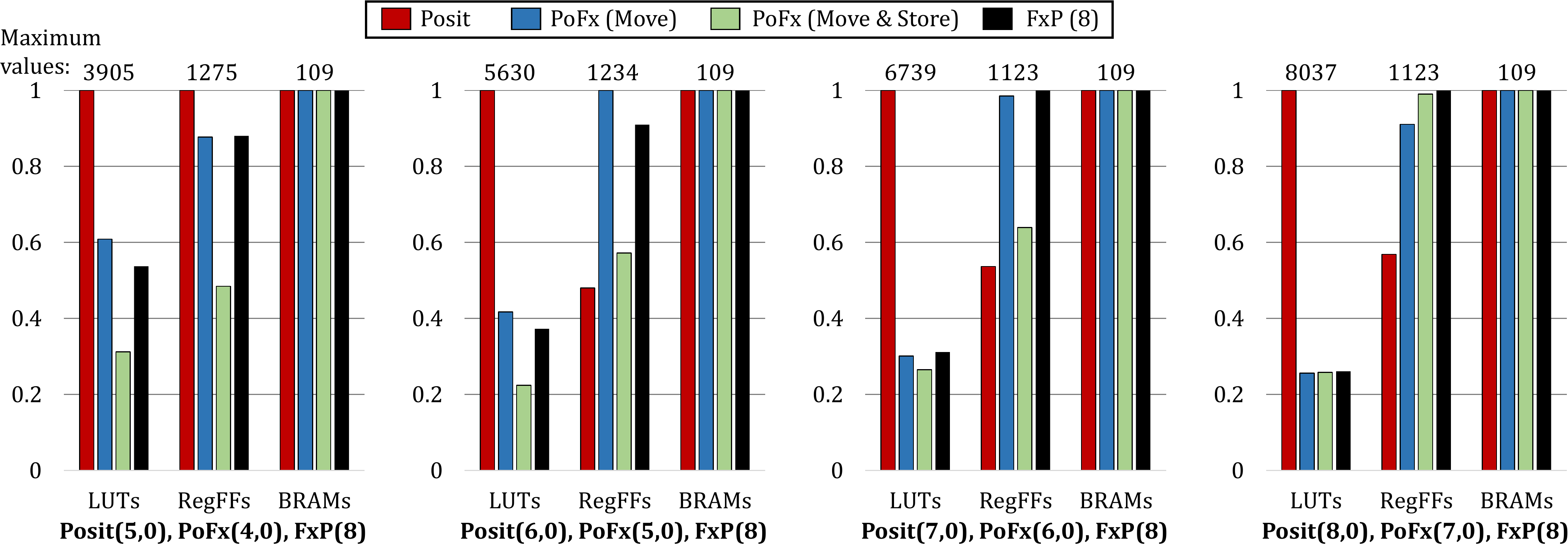}}
	\caption{Variation in the relative resource utilization of BRAM-based accelerator (\textit{fullOpt\_BRAM}) implemented with varying Posit and PoFx designs compared to FxP(8)-based designs.}
	\label{fig:exp_accDsgn_BRAM}
\end{figure*}
\begin{figure*}[ht]
	\centering
	\scalebox{0.85}{\includegraphics[width=2 \columnwidth]{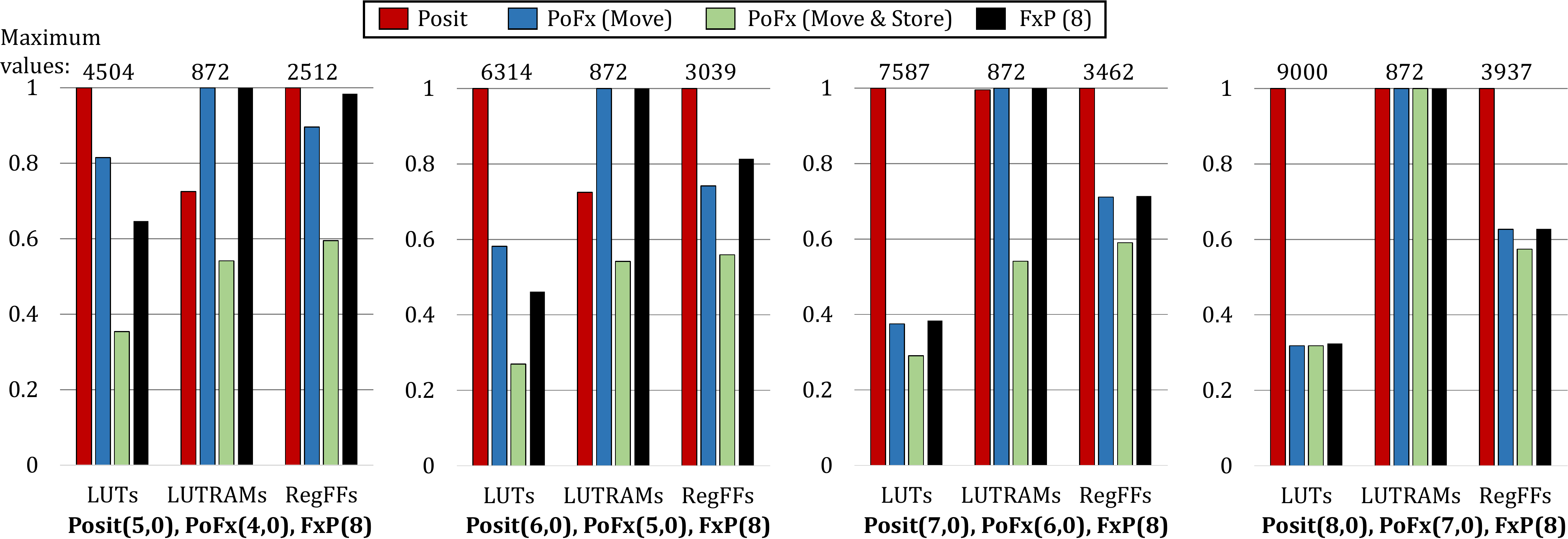}}
	\caption{Variation in the relative resource utilization of LUTRAM-based accelerator (\textit{fullOpt\_LRAM}) implemented with varying Posit and PoFx designs compared to FxP(8)-based designs.}
	\label{fig:exp_accDsgn_LRAM}
	%\vspace{-4pt}
\end{figure*}
% \begin{figure*}[t] 
%   \centering
%   	\subfloat[Posit(N=5,ES=0), PoFx(N-1=5,ES=0)\label{exp_accDsgn_BRAM_5}]{%
%       \includegraphics[width=0.32\textwidth]{figures/exp_accDsgn_BRAM_5.pdf}}
%     \hspace*{0.2em}
%     \subfloat[Posit(N=6,ES=0), PoFx(N-1=6,ES=0)\label{exp_accDsgn_BRAM_6}]{%
%         \includegraphics[width=0.32\textwidth]{figures/exp_accDsgn_BRAM_6.pdf}}
% \hspace*{0.2em}
%     \subfloat[Posit(N=7,ES=0), PoFx(N-1=7,ES=0)\label{exp_accDsgn_BRAM_7}]{%
%         \includegraphics[width=0.32\textwidth]{figures/exp_accDsgn_BRAM_7.pdf}}
%   \caption{Variation in the hardware performance metrics of BRAM-based accelerator implemented with varying Posit and PoFx designs compared to FxP(8)-based designs.}
%   \label{fig:exp_accDsgn_BRAM} 
% \end{figure*}
% \begin{figure*}[t] 
%   \centering
%   	\subfloat[Posit(N=5,ES=0), PoFx(N-1=5,ES=0)\label{exp_accDsgn_LRAM_5}]{%
%       \includegraphics[width=0.32\textwidth]{figures/exp_accDsgn_LRAM_5.pdf}}
%     \hspace*{0.2em}
%     \subfloat[Posit(N=6,ES=0), PoFx(N-1=6,ES=0)\label{exp_accDsgn_LRAM_6}]{%
%         \includegraphics[width=0.32\textwidth]{figures/exp_accDsgn_LRAM_6.pdf}}
% \hspace*{0.2em}
%     \subfloat[Posit(N=7,ES=0), PoFx(N-1=7,ES=0)\label{exp_accDsgn_LRAM_7}]{%
%         \includegraphics[width=0.32\textwidth]{figures/exp_accDsgn_LRAM_7.pdf}}
%   \caption{Variation in the hardware performance metrics of LRAM-based accelerator implemented with varying Posit and PoFx designs compared to FxP(8)-based designs.}
%   \label{fig:exp_accDsgn_LRAM} 
% \end{figure*}
\par{\siva{
In order to compare the effect of using Posit-based, \gls{pofx}-based and \gls{fxp}-based \gls{mac} units, we implemented the following four accelerator designs:
\begin{enumerate}
    \item \textit{Posit}: The accelerator stores and computes all operations in Posit($N,ES$) format.
    \item \textit{\gls{pofx}(Move)}: The weights are moved to the accelerator in normalized \gls{pofx}($N-1,ES$) representation, converted to \gls{fxp} and stored as \gls{fxp}($M=8$) numbers. During computations, the \gls{fxp}($M=8$) weights are fetched from local memory and used directly for arithmetic.
    \item \textit{\gls{pofx}(Move \& Store)}: The weights are moved from main memory and stored in local memory in normalized \gls{pofx}($N-1,ES$) format. During computation, the weights are fetched from local memory, converted to \gls{fxp}($M=8$) and used in the computation of the output activation values.
    \item \textit{\gls{fxp}(8)}: The weights are moved from main memory to accelerator and stored in the local memory of the accelerator as \gls{fxp}($M=8$) numbers. Similar to \gls{pofx}(Move), the computation stage does not involve any conversions between number representations.
\end{enumerate}
}}

\par{\siva{
\autoref{fig:exp_accDsgn_base_dotOpt_fullOpt_6_5_5_8} shows the comparison of the performance metrics of
these four designs for \textit{base}, \textit{dotOpt} and \textit{fullOpt} implementations. The configurations used for representing the weights in these designs are Posit($N = 6, ES=0$), \gls{pofx}($N-1=5, ES=0$) and \gls{fxp}($M=8$). It can be observed that the Posit-based design has higher CPD, power dissipation and LUT utilization for almost every implementation. For instance, we report $\approx80\%$ reduction in the CPD with the \gls{pofx}(Move \& Store)-based design for the \textit{fullOpt} implementation. The latency metric is similar in case of each implementation type across the four design types. It can be observed that the \gls{pofx}(Move \& Store) has higher CPD than \gls{pofx}(Move) for \textit{base} implementation, but lower CPD for \textit{dotOpt} and \textit{fullOpt} implementations. }
\par{In the base implementation of \gls{pofx}(Move \& Store), the additional over head of the \gls{pofx} conversion during computation increases the CPD. However, in the \textit{dotOpt} and \textit{fullOpt} implementations, the additional interfaces of the partitioned, higher bit-width \gls{fxp} weights-array dominates the low-cost \gls{pofx} converter delay. Since in the absence of any memory-specific HLS directives, the LUTs are used to store the weights, the LUT utilization of \gls{pofx}(Move \& Store) design is the lowest for each implementation type. We report $\approx60\%$ reduction in the LUTs utilization with the \gls{pofx}(Move \& Store)-based design over the Posit-based design for the \textit{fullOpt} implementation. Further, BRAMs are not instantiated in the \textit{dotOpt} and \textit{fullOpt} implementations in any of the designs. However, as shown in~\autoref{fig:exp_accDsgn_8bits_FxP_top}, BRAMs present a more power-efficient alternative.
}}

\par{\siva{
We explored the impact of using memory-related HLS directives in the four designs for the \textit{fullOpt} and \textit{dotOpt} implementations. \autoref{fig:exp_accDsgn_BRAM} shows the accelerators' relative resource requirements for the \textit{fullOpt\_BRAM} implementation of the four designs with varying configurations of Posit($N,ES$) and \gls{pofx}($N-1, ES$). It can be observed that the LUT utilization of Posit is much higher in all cases. This can be attributed to the high hardware costs of the Posit arithmetic blocks. Similarly the RegFF utilization of \gls{pofx}(Move \& Store) is lower than that of \gls{pofx}(Move) designs for most cases. However, the BRAM utilization remains constant for all the designs across all configurations. This is due to the granularity of the BRAM memory. Even if weights are stored as values lower than $8$-bits, equal number of instances of BRAMs are used. \su{However, as shown in~\autoref{fig:exp_accDsgn_LRAM}, if LUTRAMs are used for storing the weights and activations (\textit{fullOpt\_LRAM} implementation), lower LUTRAM utilization is observed in \gls{pofx}(Move \& Store) than \gls{pofx}(Move). For instance, compared to the Posit($N=7$, $ES=0$) we report $\approx46\%$ reduction in LUTRAMs utilization with the \gls{pofx}($N-1=6$, $ES=0$) design. This difference is reduced to zero in \gls{pofx}($N-1=7,ES=0$). Similarly, the difference in RegFF utilization of \gls{pofx}(Move) and \gls{pofx}(Move \& Store) designs reduces with increasing value of $N-1$. Therefore, the proposed \gls{pofx} representation results in reduction in the accelerator's overall resource consumption. }
}}
% \textcolor{red}{Update text with same numbers as reported in abstract, introduction and conclusion}

%\clearpage
\section{Conclusion}
\label{sec:conc}
%\lipsum[1-3]
\su{To implement machine learning applications on resource- and energy-constrained embedded systems with limited computational power, it is imperative to consider the unique features of various optimization techniques together. This paper proposes the \textit{ExPAN(N)D} framework for analyzing and combining the number representation efficacy of the Posit scheme and the resource- and compute-efficiency of \gls{fxp}-based schemes. ExPAN(N)D utilizes a modified and novel representation of Posit numbers systems to represent the trained parameters of \gls{dnn}s. Using the proposed scheme, we use $N-1$ bits for an $N$-bit Posit configuration to reduce the storage requirements. For performing arithmetic operations on trained parameters, stored in Posit format, ExPAN(N)D proposes and utilizes a resource-efficient Posit to \gls{fxp} converter \gls{pofx}. Using \gls{pofx}, all arithmetic operations are performed using \gls{fxp}-based arithmetic operators. }

\su{Compared to $6$-bit Posit-based implementation, our proposed $8$-bit \gls{pofx}-based \gls{ann} accelerator provides up to 80\% and 60\% reduction in overall resource utilization and critical path delay, respectively for the highest throughput design. 
Further, compared to an $8$-bit \gls{fxp}-based implementation, the $8$-bit \gls{pofx}-based accelerator provides up to $46\%$ reduction in storing the trained parameters. ExPAN(N)D utilizes a TensorFlow-based behavioral framework to evaluate the impact of different quantization configurations on the final output accuracy of \gls{ann}s. We intend to extend the proposed framework by incorporating other networks' optimization techniques such as approximate arithmetic operators and various other quantization schemes.}

% Resource Utilization:
% Post(6,0): 6953 x 1.0 = 6953
% PoFx(5,0): 6953 x 0.4 = 

% Critical Path Delay:
% Post(6,0): 34.63 x 1.0 = 6953
% PoFx(5,0): 34.63 x 0.2 = 

% \begin{thebibliography}{00}
% \bibitem{b1} G. Eason, B. Noble, and I. N. Sneddon, ``On certain integrals of Lipschitz-Hankel type involving products of Bessel functions,'' Phil. Trans. Roy. Soc. London, vol. A247, pp. 529--551, April 1955.
% \bibitem{b2} J. Clerk Maxwell, A Treatise on Electricity and Magnetism, 3rd ed., vol. 2. Oxford: Clarendon, 1892, pp.68--73.
% \bibitem{b3} I. S. Jacobs and C. P. Bean, ``Fine particles, thin films and exchange anisotropy,'' in Magnetism, vol. III, G. T. Rado and H. Suhl, Eds. New York: Academic, 1963, pp. 271--350.
% \bibitem{b4} K. Elissa, ``Title of paper if known,'' unpublished.
% \bibitem{b5} R. Nicole, ``Title of paper with only first word capitalized,'' J. Name Stand. Abbrev., in press.
% \bibitem{b6} Y. Yorozu, M. Hirano, K. Oka, and Y. Tagawa, ``Electron spectroscopy studies on magneto-optical media and plastic substrate interface,'' IEEE Transl. J. Magn. Japan, vol. 2, pp. 740--741, August 1987 [Digests 9th Annual Conf. Magnetics Japan, p. 301, 1982].
% \bibitem{b7} M. Young, The Technical Writer's Handbook. Mill Valley, CA: University Science, 1989.
% \end{thebibliography}
% \scriptsize
% \clearpage
\renewcommand{\baselinestretch}{0.95}
\bibliographystyle{IEEEtran}
% argument is your BibTeX string definitions and bibliography database(s)
\bibliography{references}

% Generated by IEEEtran.bst, version: 1.14 (2015/08/26)
\begin{thebibliography}{10}
\providecommand{\url}[1]{#1}
\csname url@samestyle\endcsname
\providecommand{\newblock}{\relax}
\providecommand{\bibinfo}[2]{#2}
\providecommand{\BIBentrySTDinterwordspacing}{\spaceskip=0pt\relax}
\providecommand{\BIBentryALTinterwordstretchfactor}{4}
\providecommand{\BIBentryALTinterwordspacing}{\spaceskip=\fontdimen2\font plus
\BIBentryALTinterwordstretchfactor\fontdimen3\font minus
  \fontdimen4\font\relax}
\providecommand{\BIBforeignlanguage}[2]{{%
\expandafter\ifx\csname l@#1\endcsname\relax
\typeout{** WARNING: IEEEtran.bst: No hyphenation pattern has been}%
\typeout{** loaded for the language `#1'. Using the pattern for}%
\typeout{** the default language instead.}%
\else
\language=\csname l@#1\endcsname
\fi
#2}}
\providecommand{\BIBdecl}{\relax}
\BIBdecl

\bibitem{DBLP:journals/corr/abs-1708-02709}
\BIBentryALTinterwordspacing
T.~Young, D.~Hazarika, S.~Poria, and E.~Cambria, ``Recent trends in deep
  learning based natural language processing,'' \emph{CoRR}, vol.
  abs/1708.02709, 2017. [Online]. Available:
  \url{http://arxiv.org/abs/1708.02709}
\BIBentrySTDinterwordspacing

\bibitem{DBLP:journals/corr/HeZR015}
K.~He, X.~Zhang, S.~Ren, and J.~Sun, ``Delving deep into rectifiers: Surpassing
  human-level performance on imagenet classification.''

\bibitem{6639344}
L.~{Deng}, G.~{Hinton}, and B.~{Kingsbury}, ``New types of deep neural network
  learning for speech recognition and related applications: an overview,'' in
  \emph{2013 IEEE International Conference on Acoustics, Speech and Signal
  Processing}, 2013, pp. 8599--8603.

\bibitem{simonyan2014deep}
K.~Simonyan and A.~Zisserman, ``Very deep convolutional networks for
  large-scale image recognition,'' 2014.

\bibitem{han2015deep}
S.~Han, H.~Mao, and W.~J. Dally, ``{Deep compression: Compressing deep neural
  networks with pruning, trained quantization and huffman coding},''
  \emph{arXiv preprint arXiv:1510.00149}, 2015.

\bibitem{Liu_2015_CVPR}
B.~Liu, M.~Wang, H.~Foroosh, M.~Tappen, and M.~Pensky, ``{Sparse Convolutional
  Neural Networks},'' in \emph{Proceedings of the IEEE Conference on Computer
  Vision and Pattern Recognition (CVPR)}, June 2015.

\bibitem{8877390}
N.~{Burgess}, J.~{Milanovic}, N.~{Stephens}, K.~{Monachopoulos}, and
  D.~{Mansell}, ``{Bfloat16 Processing for Neural Networks},'' in \emph{2019
  IEEE 26th Symposium on Computer Arithmetic (ARITH)}, 2019, pp. 88--91.

\bibitem{langroudi2019cheetah}
H.~F. Langroudi, Z.~Carmichael, D.~Pastuch, and D.~Kudithipudi, ``{Cheetah:
  Mixed Low-Precision Hardware \& Software Co-Design Framework for DNNs on the
  Edge},'' 2019.

\bibitem{10.5555/3045390.3045690}
D.~D. Lin, S.~S. Talathi, and V.~S. Annapureddy, ``{Fixed Point Quantization of
  Deep Convolutional Networks},'' in \emph{Proceedings of the 33rd
  International Conference on International Conference on Machine Learning -
  Volume 48}, ser. ICML’16.\hskip 1em plus 0.5em minus 0.4em\relax JMLR.org,
  2016, p. 2849–2858.

\bibitem{PositHardwareGenerator}
R.~{Chaurasiya}, J.~{Gustafson}, R.~{Shrestha}, J.~{Neudorfer}, S.~{Nambiar},
  K.~{Niyogi}, F.~{Merchant}, and R.~{Leupers}, ``{Parameterized Posit
  Arithmetic Hardware Generator},'' in \emph{2018 IEEE 36th International
  Conference on Computer Design (ICCD)}, 2018, pp. 334--341.

\bibitem{PositArithmetic}
M.~K. {Jaiswal} and H.~K.~. {So}, ``{Universal number posit arithmetic
  generator on FPGA},'' in \emph{2018 Design, Automation Test in Europe
  Conference Exhibition (DATE)}, 2018, pp. 1159--1162.

\bibitem{PACoGen}
------, ``{PACoGen: A Hardware Posit Arithmetic Core Generator},'' \emph{IEEE
  Access}, vol.~7, pp. 74\,586--74\,601, 2019.

\bibitem{PositsFPGA}
A.~{Podobas} and S.~{Matsuoka}, ``{Hardware Implementation of POSITs and Their
  Application in FPGAs},'' in \emph{2018 IEEE International Parallel and
  Distributed Processing Symposium Workshops (IPDPSW)}, 2018, pp. 138--145.

\bibitem{DeepPositron}
Z.~{Carmichael}, H.~F. {Langroudi}, C.~{Khazanov}, J.~{Lillie}, J.~L.
  {Gustafson}, and D.~{Kudithipudi}, ``{Deep Positron: A Deep Neural Network
  Using the Posit Number System},'' in \emph{2019 Design, Automation Test in
  Europe Conference Exhibition (DATE)}, 2019, pp. 1421--1426.

\bibitem{AdaptvePosit}
S.~H. Fatemi~Langroudi, V.~Karia, J.~Gustafson, and D.~Kudithipudi, ``{Adaptive
  Posit: Parameter aware numerical format for deep learning inference on the
  edge},'' 06 2020, pp. 3123--3131.

\bibitem{FastDNNusingPosit}
M.~Cococcioni, F.~Rossi, E.~Ruffaldi, and S.~Saponara, ``{Fast deep neural
  networks for image processing using posits and ARM scalable vector
  extension},'' \emph{Journal of Real-Time Image Processing}, 05 2020.

\bibitem{DeepPeNSieve}
R.~Murillo, A.~Del~Barrio, and G.~Botella, ``{Deep PeNSieve: A deep learning
  framework based on the posit number system},'' \emph{Digital Signal
  Processing}, vol. 102, p. 102762, 05 2020.

\bibitem{PositNNFramework}
S.~H. Fatemi~Langroudi, Z.~Carmichael, J.~Gustafson, and D.~Kudithipudi,
  ``{PositNN Framework: Tapered Precision Deep Learning Inference for the
  Edge},'' 07 2019, pp. 53--59.

\bibitem{InferenceOnEmbeddedDevices}
S.~H. {Fatemi Langroudi}, T.~{Pandit}, and D.~{Kudithipudi}, ``Deep learning
  inference on embedded devices: Fixed-point vs posit,'' in \emph{2018 1st
  Workshop on Energy Efficient Machine Learning and Cognitive Computing for
  Embedded Applications (EMC2)}, 2018, pp. 19--23.

\bibitem{zangPositMAC}
H.~Zhang, J.~He, and S.-B. Ko, ``Efficient posit multiply-accumulate unit
  generator for deep learning applications,'' 05 2019, pp. 1--5.

\bibitem{jain2020clarinet}
R.~Jain, N.~Sharma, F.~Merchant, S.~Patkar, and R.~Leupers, ``{CLARINET: A
  RISC-V Based Framework for Posit Arithmetic Empiricism},'' 2020.

\bibitem{DBLP:journals/corr/ZhouNZWWZ16}
\BIBentryALTinterwordspacing
S.~Zhou, Z.~Ni, X.~Zhou, H.~Wen, Y.~Wu, and Y.~Zou, ``{DoReFa-Net: Training Low
  Bitwidth Convolutional Neural Networks with Low Bitwidth Gradients},''
  \emph{CoRR}, vol. abs/1606.06160, 2016. [Online]. Available:
  \url{http://arxiv.org/abs/1606.06160}
\BIBentrySTDinterwordspacing

\bibitem{8318896}
P.~{Gysel}, J.~{Pimentel}, M.~{Motamedi}, and S.~{Ghiasi}, ``{Ristretto: A
  Framework for Empirical Study of Resource-Efficient Inference in
  Convolutional Neural Networks},'' \emph{IEEE Transactions on Neural Networks
  and Learning Systems}, vol.~29, no.~11, pp. 5784--5789, 2018.

\bibitem{DBLP:journals/corr/RastegariORF16}
\BIBentryALTinterwordspacing
M.~Rastegari, V.~Ordonez, J.~Redmon, and A.~Farhadi, ``{XNOR-Net: ImageNet
  Classification Using Binary Convolutional Neural Networks},'' \emph{CoRR},
  vol. abs/1603.05279, 2016. [Online]. Available:
  \url{http://arxiv.org/abs/1603.05279}
\BIBentrySTDinterwordspacing

\bibitem{DBLP:journals/corr/CourbariauxBD15}
\BIBentryALTinterwordspacing
M.~Courbariaux, Y.~Bengio, and J.~David, ``{BinaryConnect: Training Deep Neural
  Networks with binary weights during propagations},'' \emph{CoRR}, vol.
  abs/1511.00363, 2015. [Online]. Available:
  \url{http://arxiv.org/abs/1511.00363}
\BIBentrySTDinterwordspacing

\bibitem{DBLP:journals/corr/abs-1811-05896}
\BIBentryALTinterwordspacing
M.~de~Prado, M.~Denna, L.~Benini, and N.~Pazos, ``{QUENN:} quantization engine
  for low-power neural networks,'' \emph{CoRR}, vol. abs/1811.05896, 2018.
  [Online]. Available: \url{http://arxiv.org/abs/1811.05896}
\BIBentrySTDinterwordspacing

\bibitem{9126777}
S.~{Gupta}, S.~{Ullah}, K.~{Ahuja}, A.~{Tiwari}, and A.~{Kumar}, ``{ALigN: A
  Highly Accurate Adaptive Layerwise Log\_2\_Lead Quantization of Pre-Trained
  Neural Networks},'' \emph{IEEE Access}, vol.~8, pp. 118\,899--118\,911, 2020.

\bibitem{8587722}
S.~{Vogel}, M.~{Liang}, A.~{Guntoro}, W.~{Stechele}, and G.~{Ascheid},
  ``{Efficient Hardware Acceleration of CNNs using Logarithmic Data
  Representation with Arbitrary log-base},'' in \emph{2018 IEEE/ACM
  International Conference on Computer-Aided Design (ICCAD)}, 2018, pp. 1--8.

\bibitem{9116476}
C.~{De la Parra}, A.~{Guntoro}, and A.~{Kumar}, ``{ProxSim: GPU-based
  Simulation Framework for Cross-Layer Approximate DNN Optimization},'' in
  \emph{2020 Design, Automation Test in Europe Conference Exhibition (DATE)},
  2020, pp. 1193--1198.

\bibitem{10.1145/2966986.2967021}
\BIBentryALTinterwordspacing
V.~Mrazek, S.~S. Sarwar, L.~Sekanina, Z.~Vasicek, and K.~Roy, ``{Design of
  Power-Efficient Approximate Multipliers for Approximate Artificial Neural
  Networks},'' in \emph{Proceedings of the 35th International Conference on
  Computer-Aided Design}, ser. ICCAD ’16.\hskip 1em plus 0.5em minus
  0.4em\relax New York, NY, USA: Association for Computing Machinery, 2016.
  [Online]. Available: \url{https://doi.org/10.1145/2966986.2967021}
\BIBentrySTDinterwordspacing

\bibitem{8863138}
M.~S. {Ansari}, V.~{Mrazek}, B.~F. {Cockburn}, L.~{Sekanina}, Z.~{Vasicek}, and
  J.~{Han}, ``{Improving the Accuracy and Hardware Efficiency of Neural
  Networks Using Approximate Multipliers},'' \emph{IEEE Transactions on Very
  Large Scale Integration (VLSI) Systems}, vol.~28, no.~2, pp. 317--328, 2020.

\bibitem{8342140}
B.~S. {Prabakaran}, S.~{Rehman}, M.~A. {Hanif}, S.~{Ullah}, G.~{Mazaheri},
  A.~{Kumar}, and M.~{Shafique}, ``{DeMAS: An efficient design methodology for
  building approximate adders for FPGA-based systems},'' in \emph{2018 Design,
  Automation Test in Europe Conference Exhibition (DATE)}, 2018, pp. 917--920.

\bibitem{ullah2018smapproxlib}
S.~Ullah, S.~S. Murthy, and A.~Kumar, ``{SMApproxlib: library of FPGA-based
  approximate multipliers},'' in \emph{2018 55th ACM/ESDA/IEEE Design
  Automation Conference (DAC)}.\hskip 1em plus 0.5em minus 0.4em\relax IEEE,
  2018, pp. 1--6.

\bibitem{ullah2020area}
S.~Ullah, H.~Schmidl, S.~S. Sahoo, S.~Rehman, and A.~Kumar, ``{Area-optimized
  Accurate and Approximate Softcore Signed Multiplier Architectures},''
  \emph{IEEE Transactions on Computers}, 2020.

\bibitem{10.1109/ASP-DAC47756.2020.9045171}
\BIBentryALTinterwordspacing
Z.~Ebrahimi, S.~Ullah, and A.~Kumar, ``{LeAp: Leading-One Detection-Based
  Softcore Approximate Multipliers with Tunable Accuracy},'' in \emph{2020 25th
  Asia and South Pacific Design Automation Conference (ASP-DAC)}.\hskip 1em
  plus 0.5em minus 0.4em\relax IEEE Press, 2020, p. 605–610. [Online].
  Available: \url{https://doi.org/10.1109/ASP-DAC47756.2020.9045171}
\BIBentrySTDinterwordspacing

\bibitem{rajagopalan2011xilinx}
V.~Rajagopalan, V.~Boppana, S.~Dutta, B.~Taylor, and R.~Wittig, ``{Xilinx
  Zynq-7000 EPP: An extensible processing platform family},'' in \emph{2011
  IEEE Hot Chips 23 Symposium (HCS)}.\hskip 1em plus 0.5em minus 0.4em\relax
  IEEE, 2011, pp. 1--24.

\bibitem{logicore2017axi}
{LogiCORE, IP}, ``{AXI Interconnect (v2.1)},'' 2017.

\bibitem{ultra96}
\BIBentryALTinterwordspacing
{ Avnet, Inc}, ``{ULTRA96-V2},'' 2019. [Online]. Available:
  \url{https://www.avnet.com/opasdata/d120001/medias/docus/193/5365-pb-ultra96-v2-v4a.pdf}
\BIBentrySTDinterwordspacing

\bibitem{PositsPrimer}
\BIBentryALTinterwordspacing
Gustafson and Yonemoto, ``{Beating Floating Point at Its Own Game: Posit
  Arithmetic},'' \emph{Supercomput. Front. Innov.: Int. J.}, vol.~4, no.~2, p.
  71–86, Jun. 2017. [Online]. Available:
  \url{https://doi.org/10.14529/jsfi170206}
\BIBentrySTDinterwordspacing

\bibitem{eyeriss2016}
Y.~{Chen}, J.~{Emer}, and V.~{Sze}, ``{Eyeriss: A Spatial Architecture for
  Energy-Efficient Dataflow for Convolutional Neural Networks},'' in \emph{2016
  ACM/IEEE 43rd Annual International Symposium on Computer Architecture
  (ISCA)}, 2016, pp. 367--379.

\bibitem{SmallPositHDL}
B.~Wu., ``{SmallPositHDL},''
  \url{https://github.com/starbrilliance/SmallPositHDL}, 2020.

\bibitem{abadi2016tensorflow}
M.~Abadi, P.~Barham, J.~Chen, Z.~Chen, A.~Davis, J.~Dean, M.~Devin,
  S.~Ghemawat, G.~Irving, M.~Isard \emph{et~al.}, ``{Tensorflow: A system for
  large-scale machine learning},'' in \emph{12th $\{$USENIX$\}$ Symposium on
  Operating Systems Design and Implementation ($\{$OSDI$\}$ 16)}, 2016, pp.
  265--283.

\bibitem{10.1007/s11263-015-0816-y}
\BIBentryALTinterwordspacing
O.~Russakovsky, J.~Deng, H.~Su, J.~Krause, S.~Satheesh, S.~Ma, Z.~Huang,
  A.~Karpathy, A.~Khosla, M.~Bernstein, A.~C. Berg, and L.~Fei-Fei, ``{ImageNet
  Large Scale Visual Recognition Challenge},'' \emph{Int. J. Comput. Vision},
  vol. 115, no.~3, p. 211–252, Dec. 2015. [Online]. Available:
  \url{https://doi.org/10.1007/s11263-015-0816-y}
\BIBentrySTDinterwordspacing

\end{thebibliography}

\vspace{-10pt}
\renewcommand{\baselinestretch}{0.95}
%\vspace{95pt}
\begin{IEEEbiography}[{\includegraphics[width=1in,height=1.25in,clip,keepaspectratio]{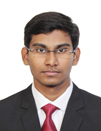}}]{Suresh Nambi} is currently working as a Guest Researcher with the Chair for Processor Design at TU Dresden. He received his undergraduate B.E. degree in Electrical and Electronics from BITS Pilani, India in the year 2020. He worked as a Mitacs Globalink Research scholar at EAM Lab, York University (Summer 2019). His current research involves exploring quantization and approximation methods for designing energy-efficient Deep Neural Networks for edge computing. His research interests include Reconfigurable computing, Printed electronics and Digital VLSI.
\end{IEEEbiography}
\begin{IEEEbiography}[{\includegraphics[width=1in,height=1.25in,clip,keepaspectratio]{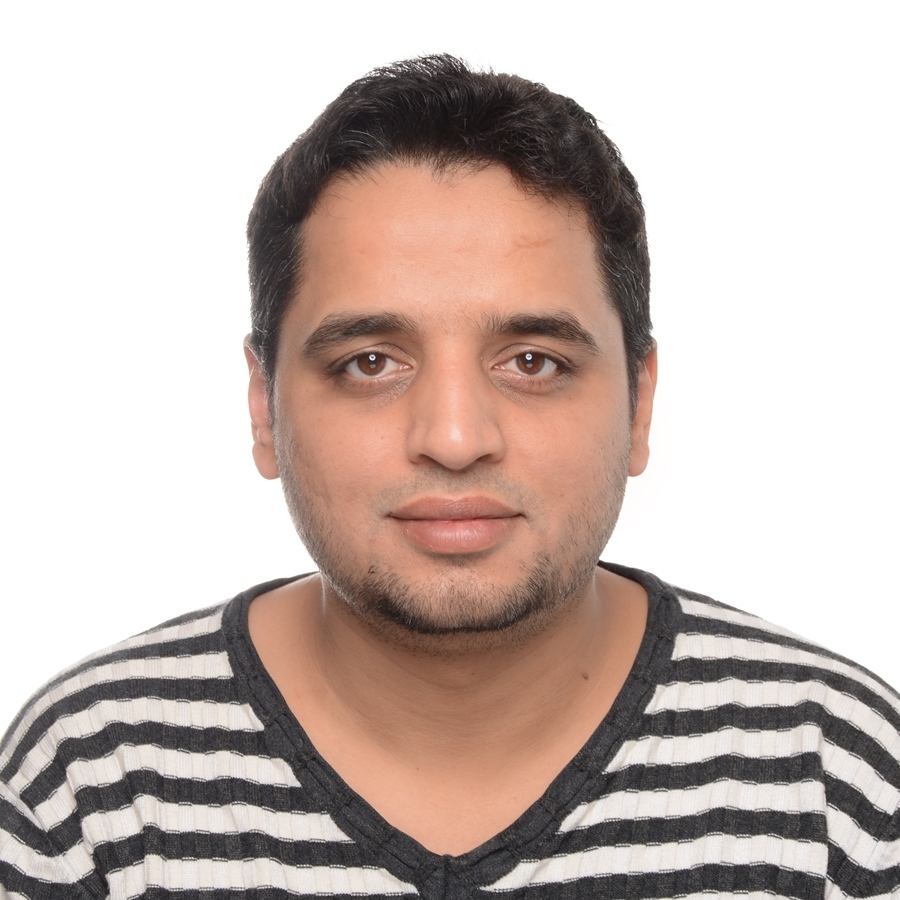}}]{Salim Ullah} is a Ph.D. student at the Chair for Processor Design, Technische Universit{\"a}t Dresden. He has completed his BSc and MSc in Computer Systems Engineering from the University of Engineering and Technology Peshawar, Pakistan. His current research interests include the Design of Approximate Arithmetic Units, Approximate Caches, and Hardware Accelerators for Deep Neural Networks.
\end{IEEEbiography}
\begin{IEEEbiography}[{\includegraphics[width=1.0in,height=1.0in,clip]{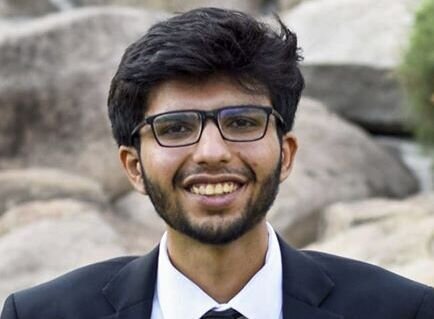}}]{Aditya Lohana} is currently working as a Software Engineer for Microsoft India. He completed his undergraduate studies with a B.E. (2016-2020) in Computer Science from BITS Pilani. During this time, he spent a semester as Guest Researcher at TU Dresden with the Chair for Processor Design and worked on privacy-aware distributed Machine Learning. His research interests include Deep Learning, Natural Language Processing and Distributed Systems.
\end{IEEEbiography}
\begin{IEEEbiography}[{\includegraphics[width=1in,height=1.25in,clip,keepaspectratio]{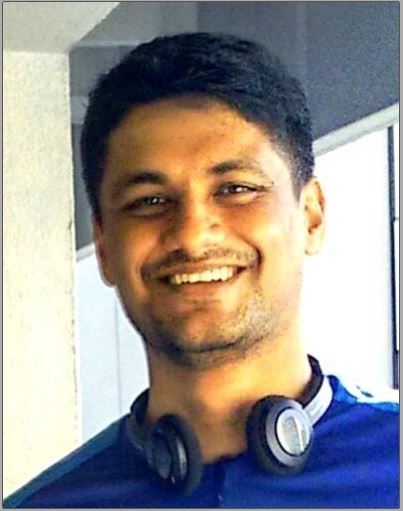}}]{Siva Satyendra Sahoo} is currently working as a Postdoctoral Researcher with the Chair for Processor Design at TU Dresden. He received his doctoral degree (Ph.D., 2015-2019) in the field of reliability in heterogeneous embedded systems from the National University of Singapore, Singapore. He completed his masters (M.Tech, 2010-2012) from the Indian Institute of Science, Bangalore in the specialization Electronics Design Technology. He has also worked with Intel India, Bangalore in the domain of Physical Design. His research interests include Embedded Systems, Machine Learning, Approximate Computing, Reconfigurable Computing, Reliability-aware Computing Systems, and System-level Design.
\end{IEEEbiography}
\begin{IEEEbiography}[{\includegraphics[width=1in,height=1.25in,clip,keepaspectratio]{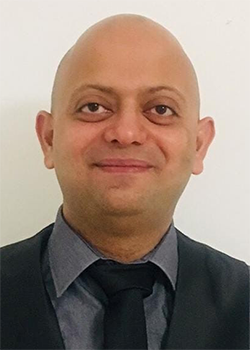}}]{Farhad Merchant} received his Ph.D. from the Indian Institute of Science, Bangalore (India), in 2016. His Ph.D. thesis title was "Algorithm-Architecture Co-design for Dense Linear Algebra Computations." He was a receipt of the DAAD fellowship during his PhD. He worked as a postdoctoral research fellow at Nanyang Technological University (NTU), Singapore, from March 2016 to December 2016. In December 2016, he moved to Corporate Research in Robert Bosch in Bangalore as a Researcher, where he worked on numerical methods for ordinary differential equations. He joined Institute for Communication Technologies and Embedded Systems, RWTH Aachen University, in December 2017 as a postdoctoral research fellow in the Chair for Software for Systems on Silicon. Farhad is receipt of the HiPEAC technology transfer award in 2019.
\end{IEEEbiography}
\begin{IEEEbiography}[{\includegraphics[width=1in,height=1.25in,clip,keepaspectratio]{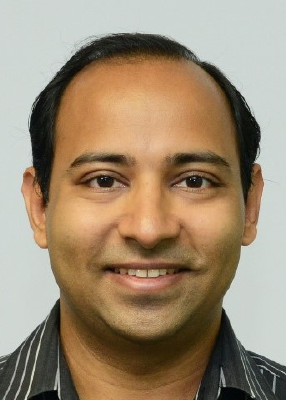}}]{Akash Kumar} (SM'13) received the joint Ph.D. degree in electrical engineering and embedded systems from the Eindhoven University of Technology, Eindhoven, The Netherlands, and the National University of Singapore (NUS), Singapore, in 2009. From 2009 to 2015, he was with NUS. He is currently a Professor with Technische Universit{\"a}t Dresden, Dresden, Germany, where he is directing the Chair for Processor Design. His current research interests include the Design, Analysis, and Resource Management of Low-Power and Fault-Tolerant Embedded Multiprocessor Systems.
\end{IEEEbiography}
\end{document}